\documentclass{aastex62}

\received{xxx}
\revised{yyy}
\accepted{zzz}

\begin{document}

\title{Cosmological evolution of quasar radio emission in the view of multifractality}

\correspondingauthor{A. Bewketu Belete}
\email{asnakew@fisica.ufrn.br}

\author{A. Bewketu Belete}
\altaffiliation{Departamento de F\'isica Te\'orica e Experimental, Universidade Federal do Rio Grande do Norte, Natal, RN 59078-970, Brazil}
\affiliation{Departamento de F\'isica Te\'orica e Experimental, Universidade Federal do Rio Grande do Norte, Natal, RN 59078-970, Brazil}
\author{Smain Femmam}
\affiliation{UHA University France, 2 Rue des Fr\`eres Lumi\`ere, 68100 Mulhouse, France}

\author{Merja Tornikosk}
\affiliation{Aalto University Mets\"ahovi Radio Observatory, Mets\"ahovintie 114,\\
02540 Kylm\"al\"a, Finland}

\author{Anne L\"ahteenm\"aki}
\affiliation{Aalto University Mets\"ahovi Radio Observatory, Mets\"ahovintie 114,\\
02540 Kylm\"al\"a, Finland}
\affiliation{Aalto University Department of Electronics and Nanoengineering,\\
P.O. BOX 15500, FI-00076 AALTO, Finland }

\author{Joni Tammi}
\affiliation{Aalto University Mets\"ahovi Radio Observatory, Mets\"ahovintie 114,\\
02540 Kylm\"al\"a, Finland}

\author{I. C. Le\~ao}
\affiliation{Departamento de F\'isica Te\'orica e Experimental, Universidade Federal do Rio Grande do Norte, Natal, RN 59078-970, Brazil}
\author{B. L. Canto Martins}
\affiliation{Departamento de F\'isica Te\'orica e Experimental, Universidade Federal do Rio Grande do Norte, Natal, RN 59078-970, Brazil}
\author{J. R. De Medeiros}
\affiliation{Departamento de F\'isica Te\'orica e Experimental, Universidade Federal do Rio Grande do Norte, Natal, RN 59078-970, Brazil}



\begin{abstract}

Variations in scaling behavior in the flux and emissions of distant astronomical sources with respect to their cosmic time are important l phenomena that can provide valuable information about the dynamics within the sources and their cosmological evolution with time. Different studies have been applying linear analysis to understand and model quasars' light curves. Here, we study the multifractal behavior of selected quasars' radio emissions in their observed frame (at 22 and 37 GHz bands) and and their rest frame. To this end, we apply the wavelet transform-based multifractal analysis formalism called wavelet transform modulus maxima. In addition, we verify whether the autoregressive integrated moving average (ARIMA) models fit our data or not. In our work, we observe strong multifractal behavior for all the sources. Additionally, we find that the degree of multifractality is strongly similar for each source and significantly different between sources at 22 and 37 GHz. This similarity implies that the two frequencies have the same radiation region and mechanism, whereas the difference indicates that the sources have intrinsically different dynamics. Furthermore, we show that the degree of multifractality is the same in the observed and rest frames of the quasars, i.e., multifractality is an intrinsic property of radio quasars. Finally, we show that the ARIMA models fit the 3C 345 quasar at 22 GHz and partially fit most of the time series with the exception of the 3C 273 and 3C 279 quasars at 37 GHz, for which the models are found to be inadequate.
 
\end{abstract}



\section{Introduction} \label{sec:intro}

Radio astronomy remains one of the most significant research areas in astronomy and cosmology to date; this is because a wide variety of astronomical objects that are not detectable in the optical wavelengths as well as thermal and nonthermal radiation mechanisms and propagation phenomena can be studied at radio wavelengths. Radio observations have played a significant role in the discovery, subsequent observation, and classification of active galactic nuclei (hereinafter AGNs), radio galaxies and other radio sources (e.g., \citet{2016ApJ...826..132S, 2016MNRAS.459.1310K, 2015MNRAS.453.2438T, 2009BASI...37...63S, 2007MNRAS.382.1019B, 2006MNRAS.371..945H, 1997ApJ...487L.135R, 1964ApJ...140....1G}). Moreover, as an application of radio astronomy, studying the behavior of the radio luminosity function (e.g., \citet{2001MNRAS.322..536W, 1990MNRAS.247...19D}) plays a significant role in understanding the formation and evolution of radio galaxies and the cosmological evolution of radio sources, which in turn help to uncover the physics of the early universe. \\
Quasars, an extremely luminous compact region thought to reside at the center of galaxies, are distant astronomical objects that belong to a subclass of AGNs (e.g., \citet{2017A&ARv..25....2P, 2017FrASS...4...47P, 2017AAS...22925029F, 1987A&A...176..197C}). Flat spectrum radio quasars (hereinafter FSRQs), which include both high- and low-polarization quasars (hereinafter HPQs and LPQs, respectively), along with BL Lac objects, form a subset of AGNs known as blazars. The former are known for their high luminosity and strong and broad emission features in their spectra, whereas the latter are known for their low luminosity and very weak or even absent emission features in their spectra \citep{2001MNRAS.326.1455M}. Radio-loud quasars/blazars show some of the most violent high-energy phenomena observed in AGNs to date. They emit radiation across the electromagnetic spectrum, from the radio to X-rays and/or gamma rays (e.g., \citet{2017A&ARv..25....2P, 2017FrASS...4...47P, 2017AAS...22925029F, 2009A&A...504L...9V, 1987A&A...176..197C}) and have the characteristics of very high luminosity, nonthermal radiation, strong radio emission,  and large flux fluctuations and have been known as highly variable energetic sources across the electromagnetic spectrum though most (if not all) of them are strongly variable in a very short time scale at higher frequencies \citep{1999MNRAS.306..247L, 1988Natur.335..330C, 1998ApJ...497..178W, 1997ARA&A..35..445U, 1996MNRAS.279..429N, 1995MNRAS.273..583D, 1979ApJ...229L.115C}; moreover, high and variable polarization and superluminal motion are characteristics of blazars. The low-energy blazar emission is thought to be the result of electron synchrotron radiation, with the peak frequency reflecting the maximum energy at which electrons can be accelerated \citep{2015MNRAS.452.1303P}.\\
The timescale variability and other properties of AGNs in general, and of quasars in particular, have been extensively studied at several radio frequencies using different time series analyses approaches. Here, we study the multifractal behavior of selected radio-loud quasars such as 3C 273, 3C 279, 3C 345, and 3C 454.3. The flat spectrum radio-loud quasar 3C 273 is a low polarization quasar \citep{2011A&A...532A.146L, 1991AJ....102.1946V} at a distance of z = 0.158 \citep{1995PhLA..203..161A}. This radio quasar has been monitored by Mets$\ddot{a}$hovi Radio Observatory at 22 and 37 GHz for 24 and 39 years, respectively, and it is among AGNs whose emission has been studied at all wavelengths \citep{1990A&A...234...73C, 1987A&A...176..197C}. Particularly, the radio variability of 3C 273 has been studied by \citep{2000A&A...361..850T, 1999A&A...349...45T}. 
The HPQ quasar 3C 279  \citep{2011A&A...532A.146L} at z = 0.536 \citep{1996MNRAS.279..429N} has been observed by Mets$\ddot{a}$hovi Radio Observatory at 22 and 37 GHz for 24 and 39 years, respectively. The flat spectrum radio quasar 3C 279 has been known as one of the brightest quasars at all wavelengths, and its multiwavelength behavior and jet structure have been studied by different authors (e.g. \citet{2016MNRAS.457.3535Z, 2008ApJ...689...79C, 2001ApJ...558..583H}).   
The HPQ quasar 3C 345 \citep{2011A&A...532A.146L}  at a distance of z =  0.595 \citep{1996MNRAS.279..429N, 1984ApJ...279..465M, 1965ApJ...142.1674B} is one of the most studied radio-loud flat-spectrum quasars whose radio spectrum is dominated by a flat-spectrum core \citep{1989AJ.....97.1550K}. 3C 345 has been monitored by Mets$\ddot{a}$hovi Radio Observatory at 22 and 37 GHz for 24 and 38 years, respectively. The radio flux observations of 3C 345 have shown that variations are common in the radio frequency range \citep{1985ApJS...59..513A}. \citet{1991ApJ...380..351V} indicated that the light-curve of 3C 345 is nonlinear and stochastic. The quasar 3C 454.3 is an HPQ \citep{2011A&A...532A.146L} at z = 0.859 \citep{1996MNRAS.279..429N} with a flat spectrum that belongs to the blazar class of active galactic nuclei \citep{2006A&A...453..817V}, which has been studied across the electromagnetic spectrum \citep{2010ApJ...715..362J} and known as variable source at all wavelengths. This quasar has been monitored by Mets$\ddot{a}$hovi Radio Observatory at 22 and 37 GHz for 24 and 38 years, respectively.\\ 
Using the structure function, the discrete correlation, and Lomb-Scargle periodogram analysis approach, \citet{2007A&A...469..899H} has studied the variability time scales of a sample of AGN including our candidates at several frequencies between 4.8 and 230 GHz, has shown that the slopes between frequencies were different and did not find a significant difference between most of the sources. Similarly, \citet{1992A&A...254...80V} studied variability timescales of a sample of AGNs including our sources at 22 and 37 GHz and found that the spectral and variability characteristics between most of the sources were very similar, except for the difference observed between HPQs and LPQs. Additionally, \citet{2008A&A...488..897H} has studied the variability behavior of 80 AGNs including our candidates at 22, 37 and 90 GHz using wavelet analysis and has shown that the timescales at 22 and 37 GHz did not differ significantly from source to source. Using the structure function analysis, \citet{2010ChA&A..34..343W} studied the variability characteristics of 3C 273 and 3C 345 at 22 and 37 GHz and found that each source has similar slopes at these two frequencies, indicating that for both sources, the two frequencies have the same region and mechanism.
 Moreover, \citet{1992A&A...254...71V} studied the flux variations of extragalactic radio sources at 37, 22, 14.5, 8, and 4.8 GHz and suggested that all outbursts in AGN have similar evolutionary tracks, defined by the motion of the turnover peak of the shock spectrum in time, and most of the observed differences result from variations of the frequency at which the outburst reaches its maximum development. Similarly, \citet{1999ApJS..120...95V} studied the total flux density variations of a group of AGNs at 22 and 37 GHz and has shown that the variations were adequately decomposed into flares having an exponential rise.\\ See \citet{2018RAA....18...40Z,  2018A&A...614A.148B, 2017MNRAS.472..788G,  2013ApJ...773..147J, 2011A&A...528L..10B, 2010ChA&A..34..357D, 2010arXiv1012.2820S, 2009A&A...504L...9V, 2008ApJ...689...79C, 2008A&A...491..755R, 2007A&A...469..899H, 2006A&A...456..895L, 2005A&A...440..845L, 1999MNRAS.306..247L, 1999ApJS..120...95V, 1998ApJ...497..178W, 1994ApJ...437...91S, 1992A&A...254...71V, 1991AJ....102.1946V, 1990A&A...234...73C, 1987A&A...176..197C}) for more studies on the variability timescale and other physical properties of our candidates across the electromagnetic spectrum.\\
Most (if not all) of the approaches used in the abovementioned works were linear analyses. Here, we apply a  nonlinear analysis approach to reveal the nonlinear characteristics about the sources considered, and this is the main motivation for investigating the new methods and emphasizes the importance of the present work. In this work, we apply a wavelet transform-based multifractality analysis approach called wavelet transform modulus maxima (hereinafter WTMM) and study the multiscaling or multifractal behavior of the quasars 3C 273, 3C 279, 3C 345, and 3C 454.3 radio observations at 22 and 37 GHz. 
Why multifractality analysis? Most astrophysical objects are possibly associated with continuous nonlinear stochastic systems due to their complexity in nature. A fractal behavior can be observed in the time series of complex systems \citep{2016JAsGe...5..301M}. It has been shown that quasars, in general, are among complex systems that have nonlinear time series characterized by fractal behavior \citep{1991ApJ...380..351V} and also by sudden bursts of very large amplitude \citep{1990ApJ...359...63B, 1989A&A...226....9K}, which implies that the dynamical evolution of quasars is nonlinear (i.e., described by nonlinear stochastic differential equations) \citep{1991ApJ...380..351V}. Additionally, there is a suggestion that extra-galactic radio sources are intermittent on timescales of $\sim$ $10^{4} - 10^{5}$ yr \citep{1997ApJ...487L.135R}. Traditionally, different classical approaches, such as power spectrum function (PS), structure function (SF), PS-periodogram, covariance analyses, and others, which are suitable only for addressing the signals characteristic of linear systems \citep{1992ApJ...391..518V}, have been in use to analyze and study nonlinear signals due to the unavailability of better approaches that are necessary to gain detailed information about the dynamics of complex systems. The multifractal formalism was introduced in the mid-1980s to provide a statistical description of the fluctuations of regularity of singular measures found in chaotic dynamical systems \citep{1986PhRvA..33.1141H}. Currently, multifractal analysis is being used in several fields of science to characterize nonlinearity or detect singularity in various types of signals from complex systems and study correlations between different physical parameters  \citep{2012Trevino}. Currently, the multifractal analysis approach has been applied to a large number of empirical as well as theoretical studies of a wide variety of problems. As an example, using multifractality analysis, it has been shown that the X-ray light curve of the BL Lac object PKS 2155-304 is monofractal, and the optical light curve of  the quasar 3C 345 has a multifractal nature - nonlinear behavior \citep{1992ApJ...391..518V}. In addition, using multifractality analysis, it has been shown that the distribution of The Infrared Astronomical Satellite (IRAS) galaxies is homogeneous at the large scale \citep{2000MNRAS.318L..51P}, which is in agreement with the cosmological principle. Additionally, \citet{2018MNRAS.tmp.1264B} has studied the fractal nature of the light curves of 3C 273 at specific frequencies across its electromagnetic spectrum and shown that most of the light curves have presented a multifractal signature, confirming the nonlinear and intermittent nature of the source. For more studies on multifractality analysis by different authors in different science cases, see  \citet{2017AdSpR..60.1363M, 2016JAsGe...5..301M, 2016cosp...41E.946K, 2016AGUFM.P11A1846A, 2016ApJ...831...87D, 2013EPSC....8...30A, 2012SPIE.8222E..0FJ, Ouahabi:2011:WMA:2036797.2036809, 2009PhPl...16j2307N, 2007EPJB...60..483L, 1998adap.org..8004D}.\\
The aims of our work are as follows: (i) to study the multifractal behavior (if any) of selected radio-loud quasars located at different redshifts, (ii) to verify how fractal signatures of each source, and between sources, behave at 22 and 37 GHz (any similarity or difference? If yes/no, what can we learn about them?), and (iii) to determine any possible correlation between multifractal behavior and redshift at those two frequencies, i.e., how the fractal signature of these radiations change with redshift (the cosmological evolution of quasars' radio emissions from the view of multifractality analysis). The first aim helps to draw a conclusion regarding whether radio-loud quasars are multifractal/intermittent and nonlinear systems. From the second aim, we can roughly understand whether the selected radio frequencies have the same radiation region and mechanism or not. Furthermore, knowledge of the correlation between multifractal behavior and redshift is essential not only for robustly understanding the cosmological evolution of quasars' multifractal signature at specific radio frequencies and for interpretation of the behavior of the relativistic plasma and nonthermal radiation associated with the jet out flowing from the black hole/accretion disk systems but also for supporting the claim that quasars' redshift is cosmological in nature. In all our discussions, a flat $\Lambda$CDM cosmology with $\Omega_{\lambda}$ = 0.70, $\Omega_{m}$ = 0.3, and $H_o = 70 kms^{-1}$ Mp$c^{-1}$ is used, unless otherwise specified. Our work is structured as follows: in section \ref{data}, we present the light curves of the data used, the method and procedures. The results obtained and a discussion are presented in section \ref{res}, and the summary and conclusions are included in section \ref{concl}.
\section{Data Collection, Method and Procedures}\label{data}
\subsection{Data Collection: Light-curves}
\begin{figure*}
\centering
\includegraphics[scale=0.33855]{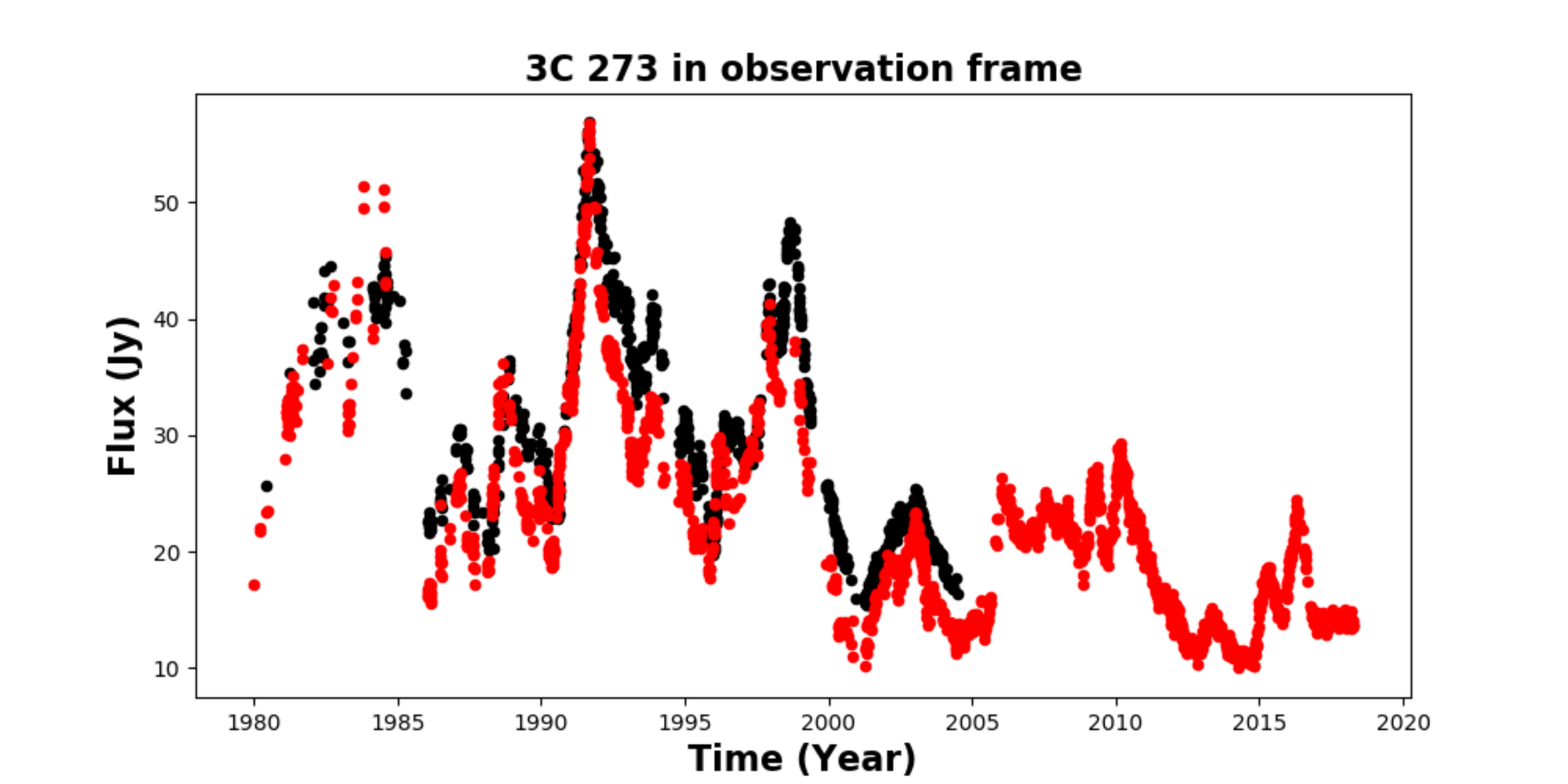}
\includegraphics[scale=0.33855]{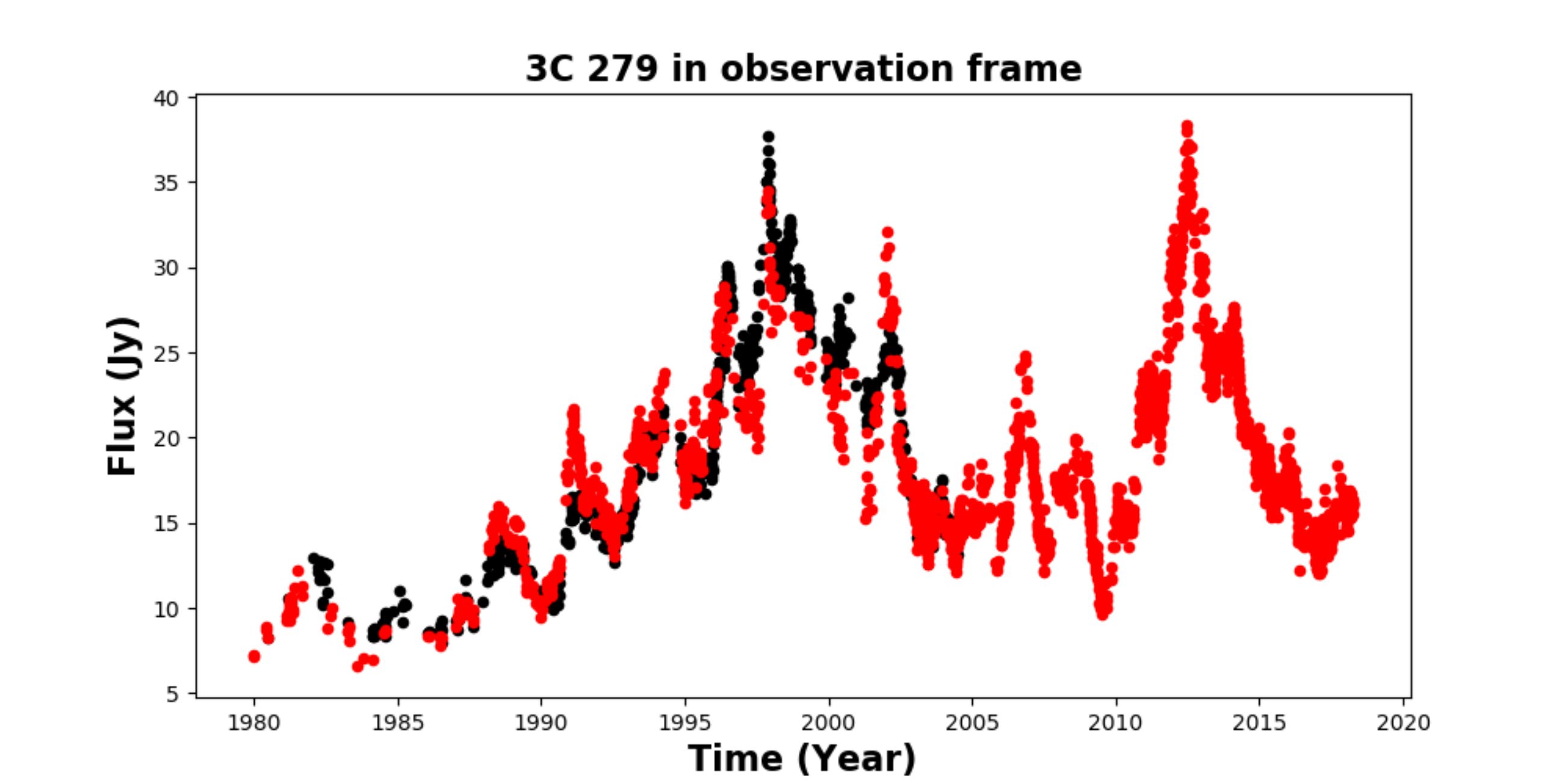}
\includegraphics[scale=0.33855]{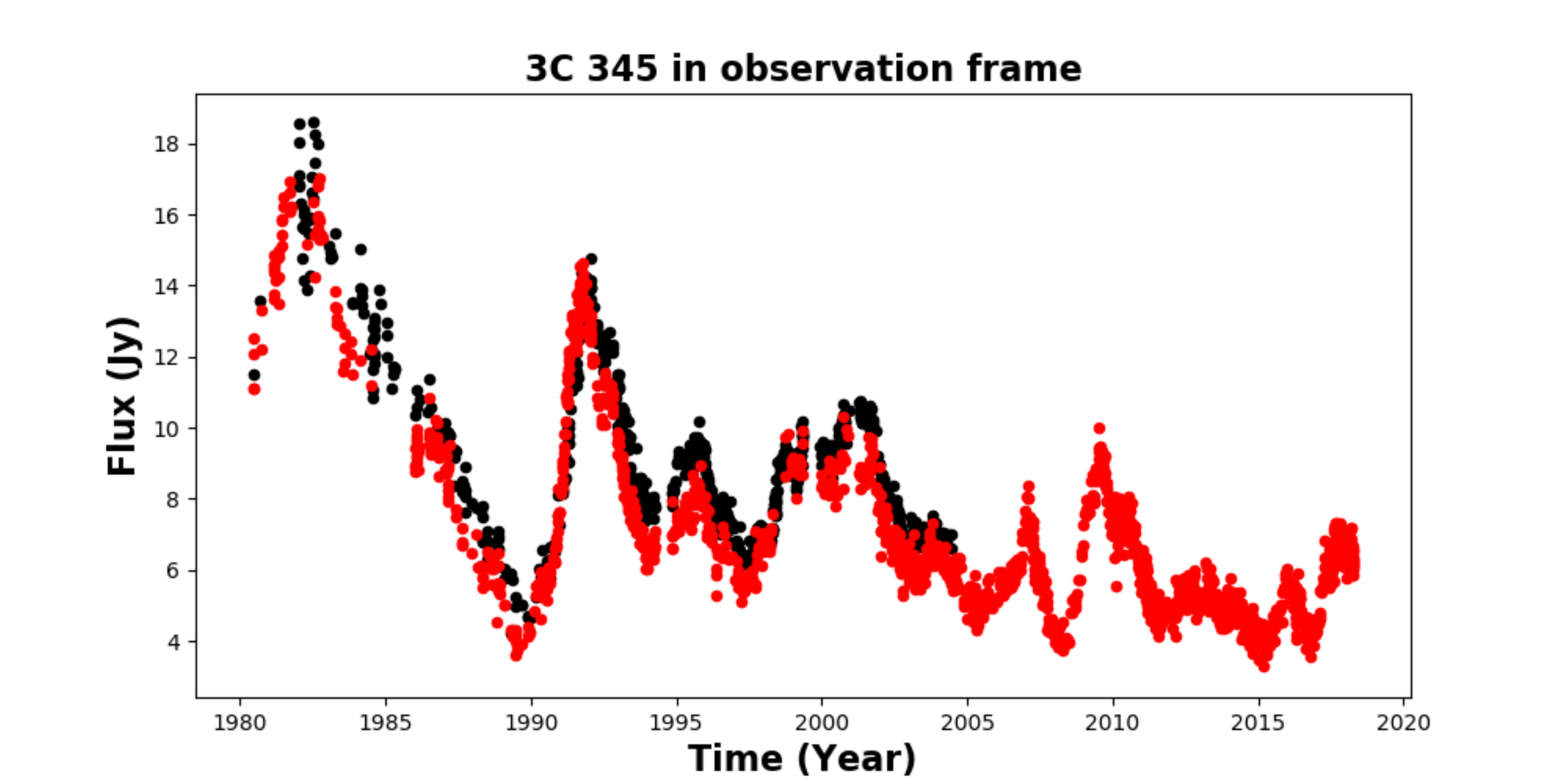}
\includegraphics[scale=0.33855]{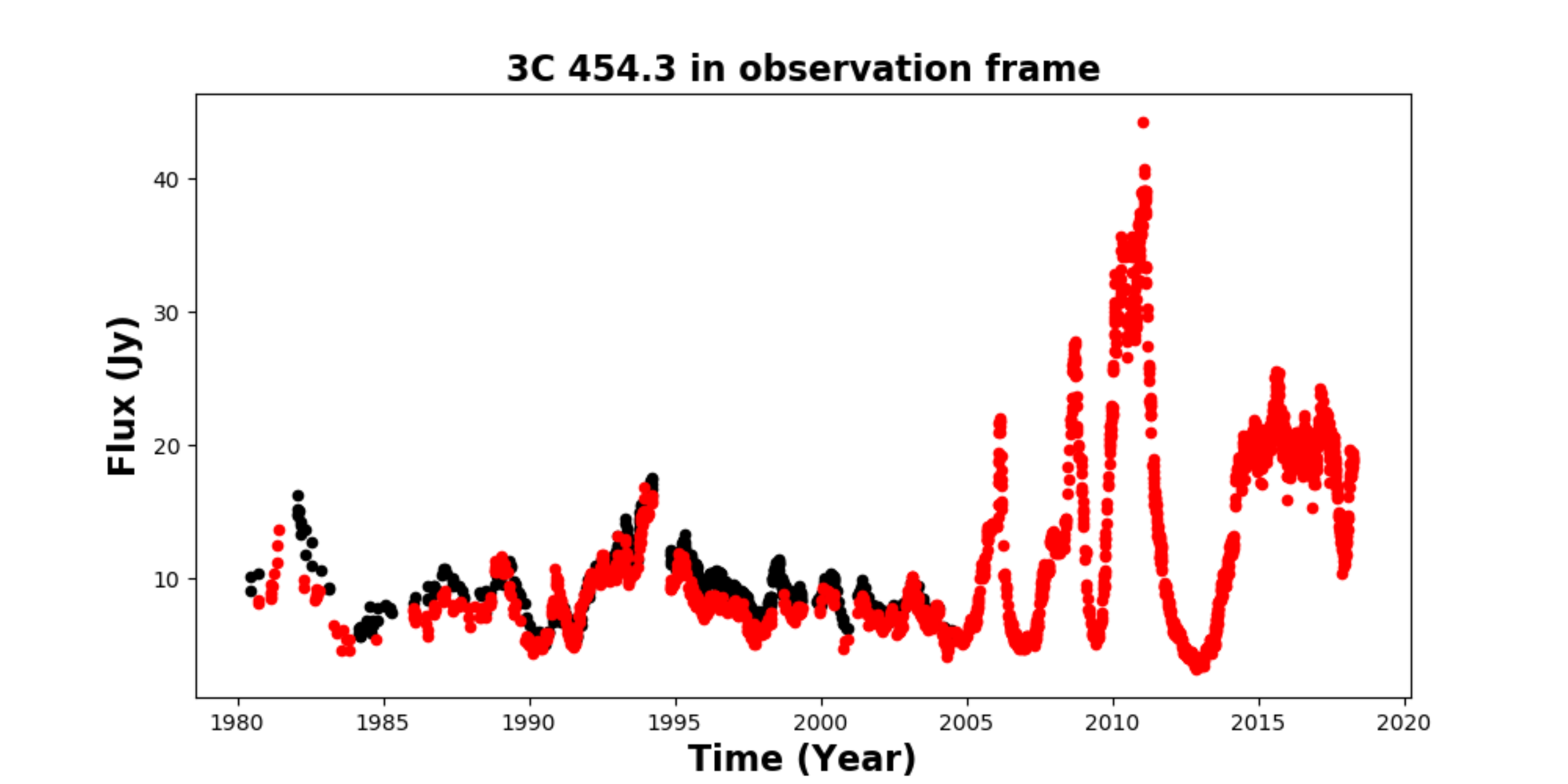}
\caption{Light curves of the sources 3C 273, 3C 279, 3C 345, and 3C 454.3 at 22 GHz (black) and 37 GHz (red) in the observation frame.}
\label{fig1}
\end{figure*}

\begin{figure*}
\centering
\includegraphics[scale=0.33855]{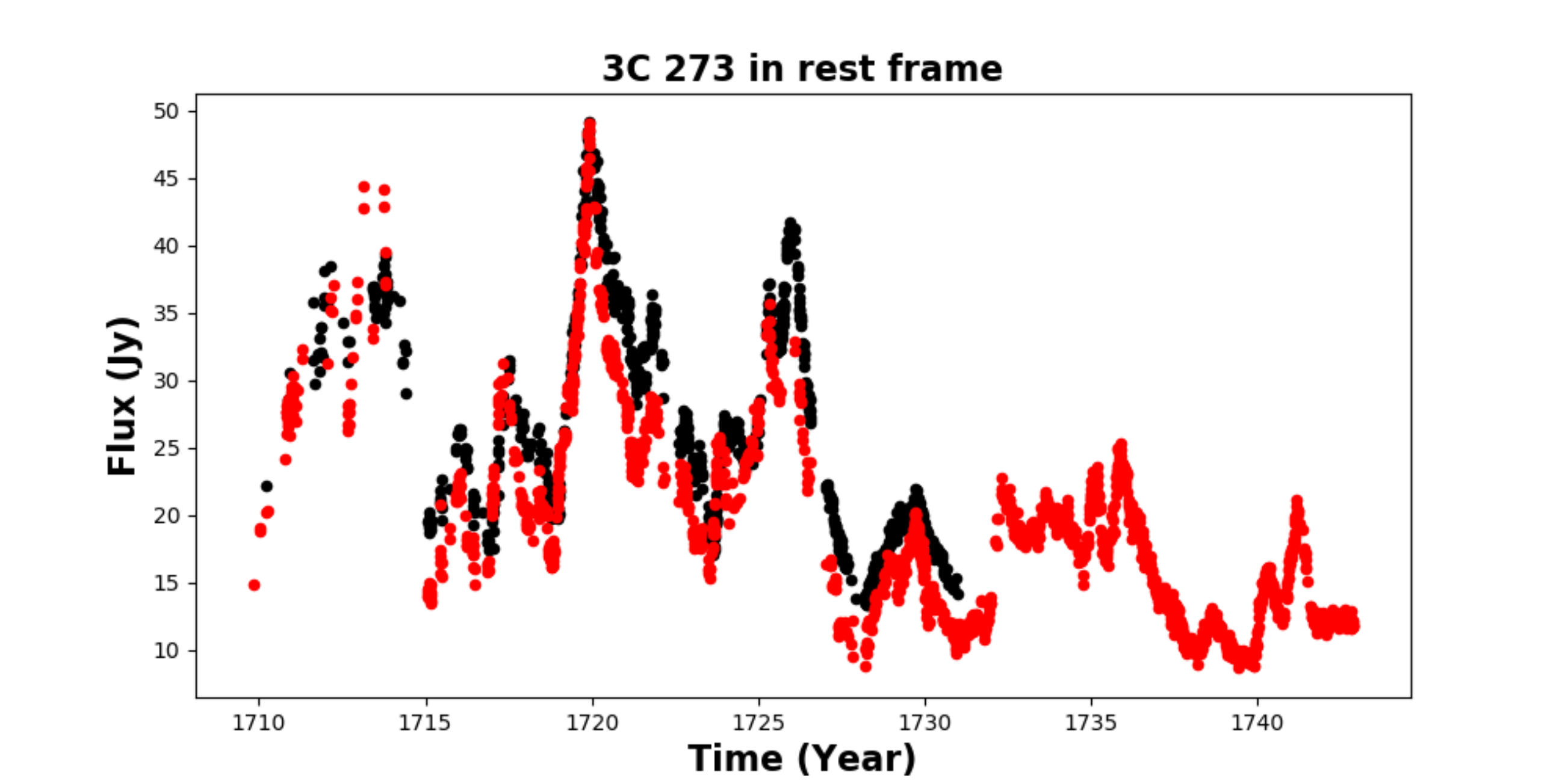}
\includegraphics[scale=0.33855]{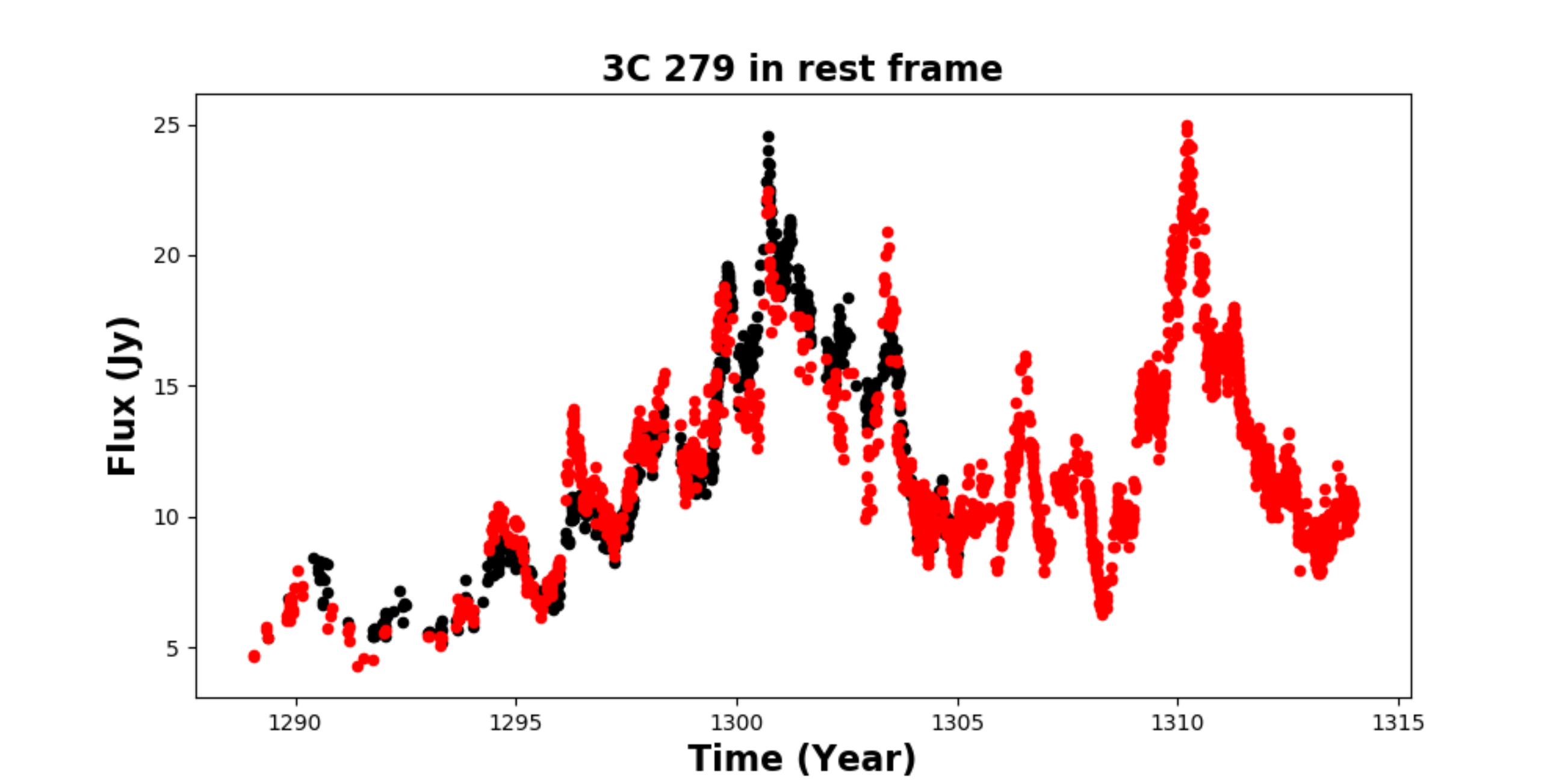}
\includegraphics[scale=0.33855]{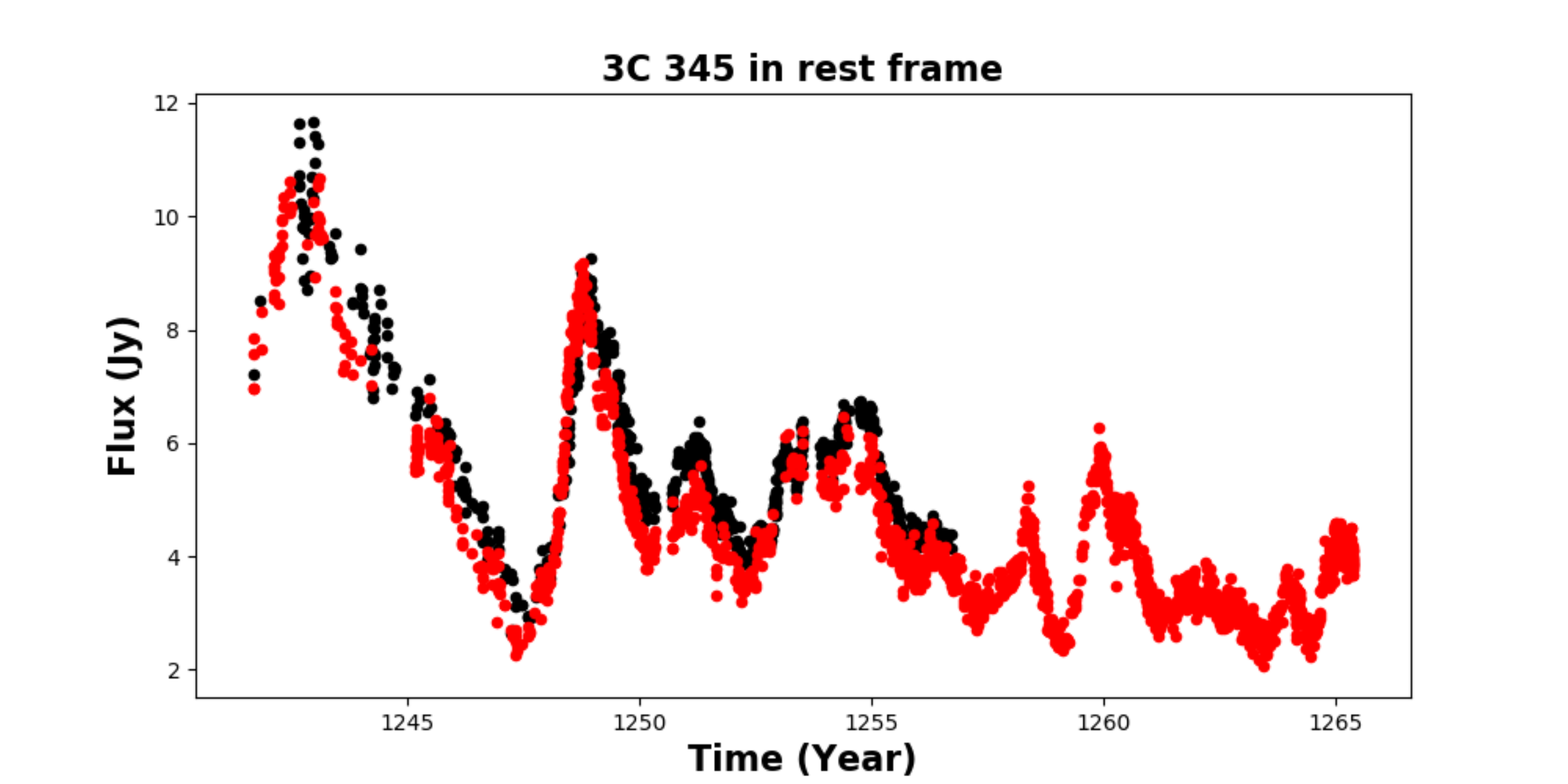}
\includegraphics[scale=0.33855]{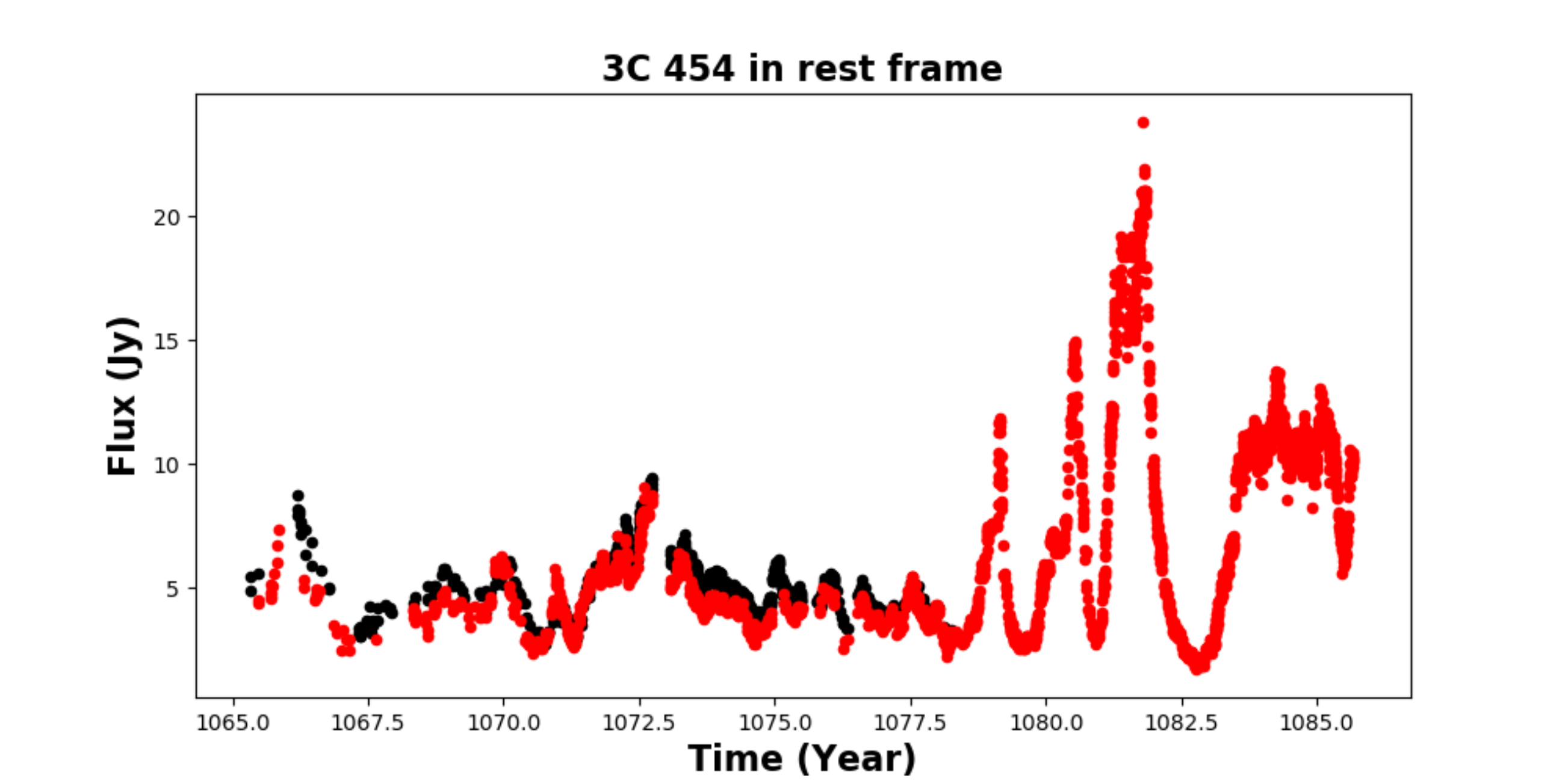}
\caption{Light curves of the sources 3C 273 at 25.4760 GHz (black) and 42.8460 GHz (red), 3C 279 at 33.7920 GHz (black) and 56.8320 GHz (red), 3C 345 at 35.0900 GHz (black) and 59.0150 GHz (red), and 3C 454.3 at 40.8980 GHz (black) and 68.7830 GHz (red) in the rest frame.}
\label{fig2}
\end{figure*}

We have collected the radio flux data of the sources 3C 273, 3C 279, 3C 345, and 3C 454.3, at 22 and 37 GHz frequencies from Aalto University Mets$\ddot{a}$hovi Radio Observatory in Finland. We have full light curves for 37 GHz and only up to 2004 for 22 GHz. Some of the most recent data have not been published yet, and the light curves up to 2004 are published in \citet{2005A&A...440..409T}. Our selection of these two frequencies takes advantage of the very long history of radio flux observations of those quasars at these bands. The light curves of the sources in the observation and rest frames are given in Figs. \ref{fig1} and  \ref{fig2}, respectively. We have corrected our data as $f_{rest} = f_{obs}*(1+z)$ , $t_{rest} = t_{obs}/(1+z)$, and $S_{rest} = S_{obs}/(1+Z)^{(1+\alpha)}$, where $f$ is frequency, $t$ is time, $S$ is flux, $z$ is redshift, and $\alpha$ is the spectral index.  In the radio domain, we assume that $\alpha = 0 $ as the spectrum is more or less flat.

\medskip

\medskip

\begin{table}
\caption{The time range of observation and number of data points for each source at 22 and 37 GHz.} 
\centering 
\begin{tabular}{l c c rrrrrrr} 
\hline\hline 
Source name && Frequency & Time range &Number of data\\ &&&&points
\\ [0.5ex]
\hline 
 & &22 GHz   &  1980-2004 & 938 \\[-1ex]
\raisebox{1.5ex}{3C 273} & \raisebox{1.5ex}{}&37 GHz
 &   1979-2018 & 1954 \\[1ex]
& &22 GHz & 1980-2004 & 757  \\[-1ex]
\raisebox{1.5ex}{3C 279} & \raisebox{1.5ex}{}& 37 GHz
& 1979-2018 & 1961\\[1ex]
& &22 GHz& 1980-2004 & 805 \\[-1ex]
\raisebox{1.5ex}{3C 345} & \raisebox{1.5ex}{}& 37 GHz
& 1980-2018 & 1660  \\[1ex]
& &22 GHz& 1980-2004 & 760  \\[-1ex]
\raisebox{1.5ex}{3C 454.3} & \raisebox{1.5ex}{}& 37 GHz
& 1980-2018 & 2292 \\[1ex]
\hline 
\end{tabular}
\label{tab1}
\end{table}
For all our candidates, we have an unequal length of data streams at both frequencies (Table \ref{tab1}). In WTMM-based multifractaltiy analysis, though the choice of the scale parameter is dependent on the length of a time series, it mainly depends on the absence and presence of local maxima lines at the scale considered, i.e., whether the scale at which the calculated wavelet coefficients hold maxima lines or not. At different scales, we can have different numbers of local maxima lines that carry local information about any singularity contained in that part of the time series. In general, the data length determines the choice of scale parameter. It is one of the parameters we use to calculate wavelet coefficients, from which we calculate the local maxima lines that we use in the multifractality analysis part. Obviously, the scale parameters at which we obtain wavelet coefficients that hold maxima lines for the current light curves change when the length of the time series changes, which in turn affects the multifractality strength to be calculated. For our case, we have chosen the most informative scale parameter, i.e., the scale parameter at which we have better maxima lines, at each frequency for all the time series, and therefore, our discussion of results takes this into consideration.
\subsection{Method and Procedures}
Signals can be efficiently represented by decomposing them in different frequencies using the Fourier analysis method. However, the most valuable information in a complex system signal is contained by its irregular structures and transient phenomena called singularities, and particularly in physics, it is important to analyze irregular structures in a signal to deduce properties about the underlying physical phenomena \citep{1988PhRvL..61.2281A, 1983JBAA...93..238M}, which is beyond the capability of Fourier analysis because it only decomposes a signal into its frequency domain. Therefore, the Fourier transform is not powerful and preferable for multifractality analysis, which requires a special technique of decomposing a signal into time and frequency domains. The continuous wavelet transform is an excellent tool for mapping the changing properties of nonstationary signals. Because of its capability of decomposing a signal into small fractions that are well localized in time and frequency and of detecting local regularities of a signal (areas on the signal where a particular derivative is not continuous) such as nonstationarity, oscillatory behavior, breakdown, discontinuity in higher derivatives, the presence of long-range dependence, and other trends, wavelet analysis remains one of the most preferable signal analysis techniques to date \citep{2016JAsGe...5..301M, WaveletTransformModulusMaximaApproachforWorldStockIndexMultifractalAnalysis}. These strengths of wavelet transform make it preferable to other traditional singularity analysis techniques, and there is a claim that it is suitable for multifractal analysis and allows for reliable multifractal analysis to be performed \citet{1991Muzy}. Therefore, due to these and other reasons not mentioned here, we have chosen to apply a multifractality analysis approach that requires the continuous wavelet transform known as the wavelet transform modulus maxima. The WTMM was originally introduced by \citep{1991Muzy}. Basically, WTMM-based multifractality analysis consists of two statistically connected parts: the wavelet 
transform part and multifractality analysis part. Each part is discussed below.\vspace{5mm}\\
\textbf{I. Wavelet transform formalism}\\
\textit{A. Continuous wavelet transform}:\\
The direct continuous wavelet transform of a given signal $X$($t$) can be represented by:
\begin{equation}\label{eq1}
W(s,a)=  \frac{1}{\sqrt{s}}\int_{0}^{T}\left(\Psi\left(\frac{t-a}{s}\right). X(t)\right)dt
\end{equation}
where $W$ are the wavelet coefficients; $\Psi$($s$,$a$,$t$) - the mother wavelet function;
$s$ - the scaling parameter;
$a$ - the shift parameter;
$X$ - the signal;
$t$ - the time at which the signal is recorded; and
$T$ - maximal time value or signal length.
The analyzing  wavelet $\Psi$($t$) is generally chosen to be well localized
in space and frequency. Usually, $\Psi$($t$) is only required to be of zero mean, but in addition to these requirements, for the particular purpose of multifractal analysis, $\Psi$($t$) is also required to be orthogonal to some low-order polynomials, up to the degree $n$-1 (i.e., to have $n$ vanishing
moments) \citet{2006GeoJI.164...63E}:
\begin{equation}\label{eq2}
\int_{-\infty}^{+\infty} t^{m}\Psi(t)dt= 0,      \forall m, 0 \leq m \le n
\end{equation} 
A class of commonly used real-valued analyzing wavelets, which satisfies the condition given by eq. \ref{eq2}, is given by the successive derivatives of the Gaussian function \citet{2006GeoJI.164...63E}.

\begin{equation}\label{eq3}
\Psi^{(N)}(t)= \frac{d^{N}}{dt^{N}}e^{-\frac{t^{2}} {2}},     
\end{equation} 
for which $n$ = $N$.
Our analyzing wavelet is the Mexican Hat (MHAT) wavelet (second-order Gaussian wavelet), which is one of the wavelets that has been applied for WTMM-based analysis, and represented by the relation:
\begin{equation}\label{eq4}
\Psi(t)= (1-t^{2}).e^{-\frac{t^{2}} {2}}
\end{equation} 
where $\Psi$($t$) - the mother wavelet function; $t$ - the time at which the signal is recorded. 
At lower scales $s \approx 0$, the number of local maxima lines (hereinafter LcMx) tends to infinity. Since it is the maxima line that points toward each regularity or carries information about any singularity or nonlinearity in a signal \citep{muzy1994multifractal, mallat1992singularity, mallat1992characterization},  it is unnecessary to calculate wavelet coefficients that do not contain maxima line(s). Though there is a suggestion that the scaling parameter $s$ used in the WTMM approach is limited to $s \leq$ [128], it should be in the interval [1, $\frac{T}{2}]$ and can also be in the interval [1, $\frac{T}{4}$], which is still informative, mainly to reduce computation time \citep{WaveletTransformModulusMaximaApproachforWorldStockIndexMultifractalAnalysis}. The shifting parameter $a$ cannot be greater than the signal length $T$, and therefore, $a$  $\leq$ T.\\
The calculated wavelet coefficients $W_{s},{a}$ can be written in a matrix form given by \citep{WaveletTransformModulusMaximaApproachforWorldStockIndexMultifractalAnalysis}:
\begin{equation}\label{eq5}
W_{s,a}=W(s,a)|(s,a\in N)\wedge(s \in [1, s_{max}])\wedge(a \in [1, T]),
\end{equation}
where $W_{s,a}$ - the wavelet coefficients; $s_{max}$- the maximal scaling parameter;
$s$ - the scaling parameter; $a$ - the shifting parameter; and $T$ - the signal length. 
Additionally, one can calculate the absolute wavelet coefficients in matrix form as:
\begin{equation}\label{eq6}
W_{s,a}^{sq}=(W(s,a))^{2}|(s,a\in N)\wedge(s \in [1, s_{max}])\wedge(a \in [1, T]),
\end{equation}
where $W^{sq}$ - the squared wavelet coefficients matrix. Other parameters are as explained above. In the wavelet transform output plot, wavelet coefficients are colored by their absolute values. \\
\textit{B. Skeleton function construction}: The skeleton function is nothing but a collection of maxima lines at each scale of the calculated wavelet coefficient matrix, i.e., it is a scope of all local maxima lines that exist on each scale $s$. In other words, the skeleton matrix construction is a technique of excluding coefficients in the absolute wavelet coefficients matrix that are not maximal. As a result, in the skeleton matrix, only absolute wavelet coefficients that belong to local maxima lines exist. The need to collect all the maxima lines at each scale together in matrix form, the skeleton function, is from the fact that it is the maxima lines that carry valuable information about the signals, i.e, maxima lines point toward regularity in the signal. We construct the skeleton function as follows \citep{WaveletTransformModulusMaximaApproachforWorldStockIndexMultifractalAnalysis}:

\begin{equation}\label{eq7}
LcMx_{s,a}= \begin{cases}
 1 |\frac{\partial (W(s,a))^2}{\partial a}=0 ,\\
0| \neg(\frac{\partial (W(s,a))^2}{\partial a}=0)
  \end{cases}
\end{equation}
under the conditions ($s$,$a\in N$) $\wedge$ ($s \in [1, s_{max}$]) $\wedge$ ($a \in [1, T]$), where $LcMx$ - the wavelet skeleton function; $W$($s$, $a$) - the wavelet coefficeints;
$s$ - the scaling parameter; $s_{max}$ - the maximal scaling parameter; $a$ - the shift parameter; and $T$ - the signal length.
The wavelet skeleton function can also be calculated from squared wavelet coefficients matrix and expressed in a matrix form as follows:
\begin{equation}\label{eq8}
LcMx_{s,a}=LcMx(s,a)
\end{equation}
under the same conditions as eq. \ref{eq7}.  Since wavelet coefficients on corners provide little or no information, and consequently, local maxima lines (LML) on corners also provide no significant information about the singularity in the signal, we therefore take the edge effect into consideration by removing the local maxima lines on corners using the following formula \citep{WaveletTransformModulusMaximaApproachforWorldStockIndexMultifractalAnalysis}:

\begin{equation}\label{eq9}
LcMx_{s,a}= \begin{cases}
 1 |(\frac{\partial (W(s,a))^2}{\partial a}=0)\wedge(s < a < T-s) ,\\
0| \neg(\frac{\partial (W(s,a))^2}{\partial a}=0)\lor(T-s \leq a) \lor (a\leq s)
  \end{cases} 
\end{equation}
the skeleton function is a logical function that has only two variables, 0 and 1; one is used if the skeleton matrix element is a local maximum, and zero is used otherwise.\\
There is one problem in WTMM-based multifractality analysis: the skeleton function does not contain local maxima lines as required for fractal analysis but has disconnected broken lines, gaps, and single points in the LcMx matrix. Therefore, it is mandatory to apply some technique to fix these limitations. To that end, we applied an algorithm called the supremum algorithm, which consists of the following steps.\citep{WaveletTransformModulusMaximaApproachforWorldStockIndexMultifractalAnalysis}:\\
1) Define matches (relations between single local maxima points);
2) Define match conflicting (one cell to more) and nonconflicting cases;
3) Create chains from pairs;
4) Chain interpolation (to fill missing wavelet coefficients on WTMM line);
5) Add points to LcMc map (on line gaps);
6) Add single points to LcMc map; and
7) Change variables (at the end).\vspace{5mm}\\
\textbf{II. Multifractality analysis part}\\
\textit{C. Fractal partition function estimation}: At this time, the multifractality analysis starts. Using the wavelet modulus maxima coefficients obtained in the wavelet transform part, we calculate the thermodynamic partition function, a function that connects the wavelet transform and multifractality analysis part. \citet{1991Muzy} defined the thermodynamic partition function $Zq$($s$) as the
sum of the $q$-th powers of the local maxima of the modulus to avoid division by zero:
\begin{equation}\label{eq10}
Zq(s)=\sum_{a=1}^{T-1}(C(s).WTMM)^{q}|(LcMx_{s,a}=1),
\end{equation} 
where $Zq(s)$ - the thermodynamic partition function;
$WTMM$ - the wavelet modulus maxima coefficients;
$C$($s$) - the constant depending on the scaling parameter $s$;
$s$ - the scaling parameter;
$q$ - the moment, which takes any interval with zero mean, for our case $q$ $\in$ [-5, 5]; and
$LcMx$ - the wavelet skeleton function (aggregate of local maxima lines in matrix form).
The thermodynamic partition function is a function of two arguments - the scaling parameter $s$ and the power argument $q$. The moment $q$ discovers different regions of the singularity measurement in the signal, i.e., it indicates the presence of wavelet modulus maxima coefficients of different values. The condition 
$LcMx_{s,a}$ = 1 is to inform that only modulus maxima coefficients are used. The thermodynamic partition function is finite if the wavelet modulus maxima coefficients are not equal to zero ($WTMM_{s,a}\neq 0$). To satisfy the condition, all zero coefficients should be neglected in the wavelet modulus maxima
matrix WTMM. The origin of zero coefficients in wavelet modulus maxima matrix is in the LcMx wavelet skeleton function - all elements in the skeleton matrix that are not local maxima are zero-valued elements. What is important here is the relationship between the $Zq$($s$) and $s$, which determines the scalability of the signal under consideration. We investigate the changes of $Zq$($s$) in the time series at a different scales $s$ for each $q$. A plot of the logarithm of $Zq$($s$) against the logarithm of the time scale $s$ was created. This plot shows how $Zq$($s$) scales with $s$ and reveals the strength and nature of local fluctuations at each scale in the signal. In the WTMM approach, the wavelet transform maxima are used to define a partition function whose power-law behavior is used for an estimation of the local exponents. At small scales, the following relation is expected: 
\begin{equation}\label{eq11}
Zq(s)\sim s^{\tau(q)},
\end{equation}
where $\tau(q)$ - the scaling exponent function, which is the slope of the linear fitted line on the log-log plot of $Zq$($s$) and  $s$ for each $q$.\\
\textit{D. The scaling exponent function $\tau(q)$}: This is a function of one argument $q$ that is determined from the slope of the linear fitted line on the log-log plot of $Zq$($s$) against the logarithm of the time scale $s$ for each $q$, which means that the behavior of the scaling function $\tau$($q$) is completely dependent on the nature of the thermodynamic partition function, or in other words, the behavior of $\tau$($q$) is dependent on the scaling relationship between $Zq$($s$) and $s$. The mathematical relation of calculating $\tau$($q$) is given by the relation:
\begin{equation}\label{eq12}
\tau(q)= \lim_{s\to 0}\frac{\ln(Zq(s))}{ln(a)},
\end{equation}
where $Zq$($s$)- the thermodynamic partition function;
$\tau$ - the local scaling exponent;
$s$ - the scaling parameter; and
$q$ - the moment.
The condition $\tau$($q$ = 0) $+$ 1 = 0 is important for the multifractal spectrum calculation\citep{WaveletTransformModulusMaximaApproachforWorldStockIndexMultifractalAnalysis}.
We define monofractal and multifractal as follows: the time series is said to be monofractal  if $\tau$($q$) is linear with respect to $q$, and if $\tau$($q$) is nonlinear with respect to $q$, then the time series considered is classified as multifractal (Frish and Parisi, 1985).\\
\textit{E. The multifractal spectrum function, $f$($\alpha$)}:
Once we determine the scaling function $\tau$($q$), it is necessary to estimate the multifractal spectrum $f$($\alpha$) to be able to fully draw conclusions about the multifractal behavior of the signal considered. Using the calculated scaling function, we estimate the multifractal function via Legendre transformation as \citep{1986PhRvA..33.1141H}:
\begin{equation}\label{eq13}
\alpha = \alpha(q) = \frac{\partial \tau(q)}{\partial q},
\end{equation}
where $\alpha$ is the singularity exponent or Holder exponent.
\begin{equation}\label{eq14}
f(\alpha) = q.\alpha - \tau(q),
\end{equation}
where $f$($\alpha$) is the multifractal spectrum function. 
We extract two important items of information from $f$($\alpha$) against the $\alpha$ plot: the width ($\Delta\alpha$ = $\alpha_{max}$ - $\alpha_{min}$) and the symmetry in the shape of $\alpha$ defined as $A = (\alpha_{max}-\alpha_{0})/(\alpha_{0}-\alpha_{min})$ , where  $\alpha_{0}$ is the value of  $\alpha$ when $f(\alpha)$ assumes its maximum value. \citet{2003GeoRL..30.2146A} and \citet{2002SHIMIZU} have proposed that the width of a multifractal spectrum is the measure of the degree of multifractality. Smaller values of $\Delta\alpha$ (i.e., $\Delta\alpha$ becomes close to zero) indicate the monofractal limit, whereas larger values indicate the strength of the multifractal behavior in the signal \citet{2004PCE....29..295T}.  For the symmetry in the shape of $\alpha$, the asymmetry presents three shapes: asymmetry to the right-truncated ($A >1$), left-truncated ($0< A <1$), or symmetric ($A = 1$). \citet{2012Ihlen} presented that the symmetric spectrum originated from the leveling of the $q^{th}$-order generalized Hurst exponent for both positive and negative $q$ values. The leveling of $q^{th}$-order Hurst exponent reflects that the $q^{th}$-order fluctuation is insensitive to the magnitude of the local fluctuation. When the multifractal structure is sensitive to the small-scale fluctuation with large magnitudes, the spectrum will be found with a right truncation, whereas the multifractal spectrum will be found with left-side truncation when the time series has a multifractal structure that is sensitive to the local fluctuations with small magnitudes. Therefore, the width and shape of a multifractal spectrum are able to classify small and large magnitude (intermittency) fluctuations and determine the degree of the multifractality signature in a given signal.
\subsubsection{Autoregressive Integrated Moving Average (ARIMA) Model}
Quasars are known to be extremely variable in a short time scale from hours to months due to the rapid change in their accretion rate (e.g., \citet{doi:10.1146/annurev.aa.33.090195.001115, 1993JAVSO..22...55Q}). Thus, a time series model that is suitable for a nonstationary time series with short memory may be preferable to model a quasar time series. The autoregressive integrated moving average (hereinafter ARIMA) model is the extension of autoregressive moving average model (ARMA), which in turn is a combination of autoregressive (AR) and moving average (MA) linear time-series models. The AR, MA, and ARMA models are only applied for stationary time series, whereas ARIMA includes the case of nonstationarity as well. In addition, ARIMA models are capable of describing short-memory autocorrelations that exist in a time series. The ARIMA model is one of the autoregressive time-series models rarely used to understand the nature of astronomical times series \citep{10.3389/fphy.2018.00080}. Here, we apply the ARIMA model for the light curves in Fig. \ref{fig1}. The aim is only to check whether ARIMA fits our data or not, from which we can learn the type of correlations that exist in quasars' time series and their behavior. Since seasonality is not common in quasar time series but trends are, we use the nonseasonal ARIMA(p,d,q) model discussed in \citet{2014Hyndman}:

\begin{equation}\label{eq15}
y^{\prime}_{t}=c + \phi_{1} y^{\prime}_{t-1}+...+ \phi_{p}y^{\prime}_{t-p} + \theta_{1}\varepsilon_{t-1} + ... + \theta_{q}\varepsilon_{t-q} + \varepsilon_{t},
\end{equation}  

where $y^{\prime}_{t}$ is the differenced time series; p is the order of the autoregressive part- AR(p); q is the order of the moving average part- MA(q); and $\varepsilon_{t}$ is white noise. The ARIMA(p,d,q) model contains two predictors: a linear combination of lagged values of the variable- AR(p) and the linear combination of lagged errors- MA(q) model. The reference used for this model can be used to understand the details.\\
In fitting the model, we follow the procedures discussed in \citet{2014Hyndman}:
(1) We check for stationarity - if nonstationary, we apply differencing. Usually the autocorrelation function (hereinafter ACF) of a time series decays exponentially to zero if it is stationary and slowly to zero if it is nonstationary; 
(2) we plot the ACF and partial autocorrelation function (hereinafter PACF) of the already differenced time series;
(3) we first estimate the parameters $p$ and $q$  based on ACF and PACF in (2);
(4) we fit the model ARIMA(p,d,q) using the $p$ and $q$ estimated in (3); and
(5) we calculate and plot the corresponding residuals. We accept the model only if its residual is white noise. If the residual of the fitted model is not white noise, we repeat the steps from 3 to 5. In estimating the parameters $p$ and $q$, we use the function auto.arima() that provides us with a better model, though not always.
\section{Results and Discussion}\label{res}
In our work, we apply WTMM based-multifractality analysis and study the multifractal or nonlinear behavior of radio emissions of selected radio-loud quasars 3C 273, 3C 279, 3C 345, and 3C 454.3 in the observation frame (at 22 and 37GHz ) and in the rest frames ($f_{rest} = f_{obs}*(1+z)$). The aim is to search for the presence of a multifractal signature in the light curves of the sources at each band and verify whether there is any similarity or difference in the degree of multifractaltiy between the bands and between the sources both in their observed and rest frames.
\subsection {Analysis of light curves in the observation frame:} 
\begin{figure*}
\centering
\includegraphics[scale=0.5]{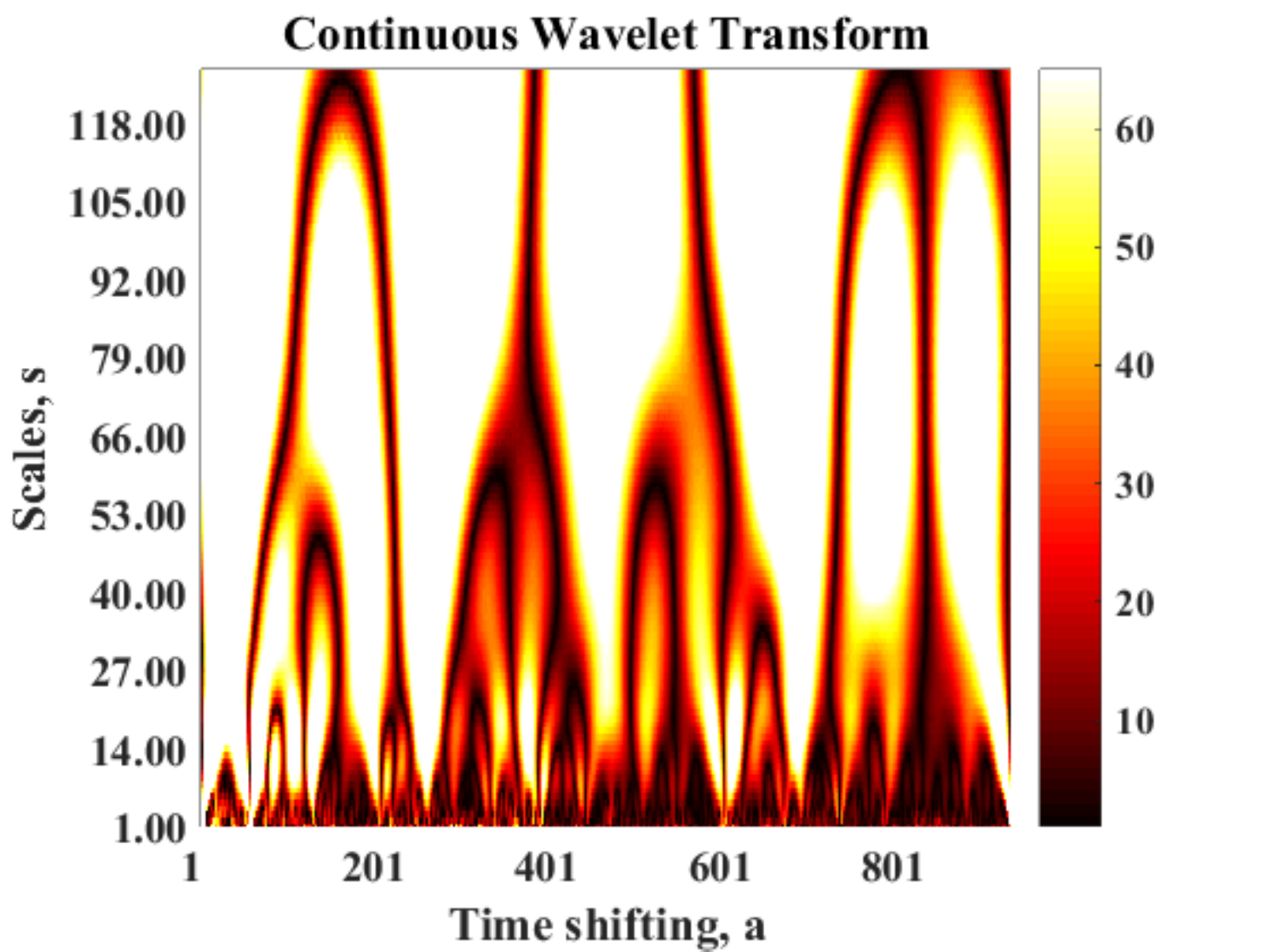}  
\includegraphics[scale=0.5]{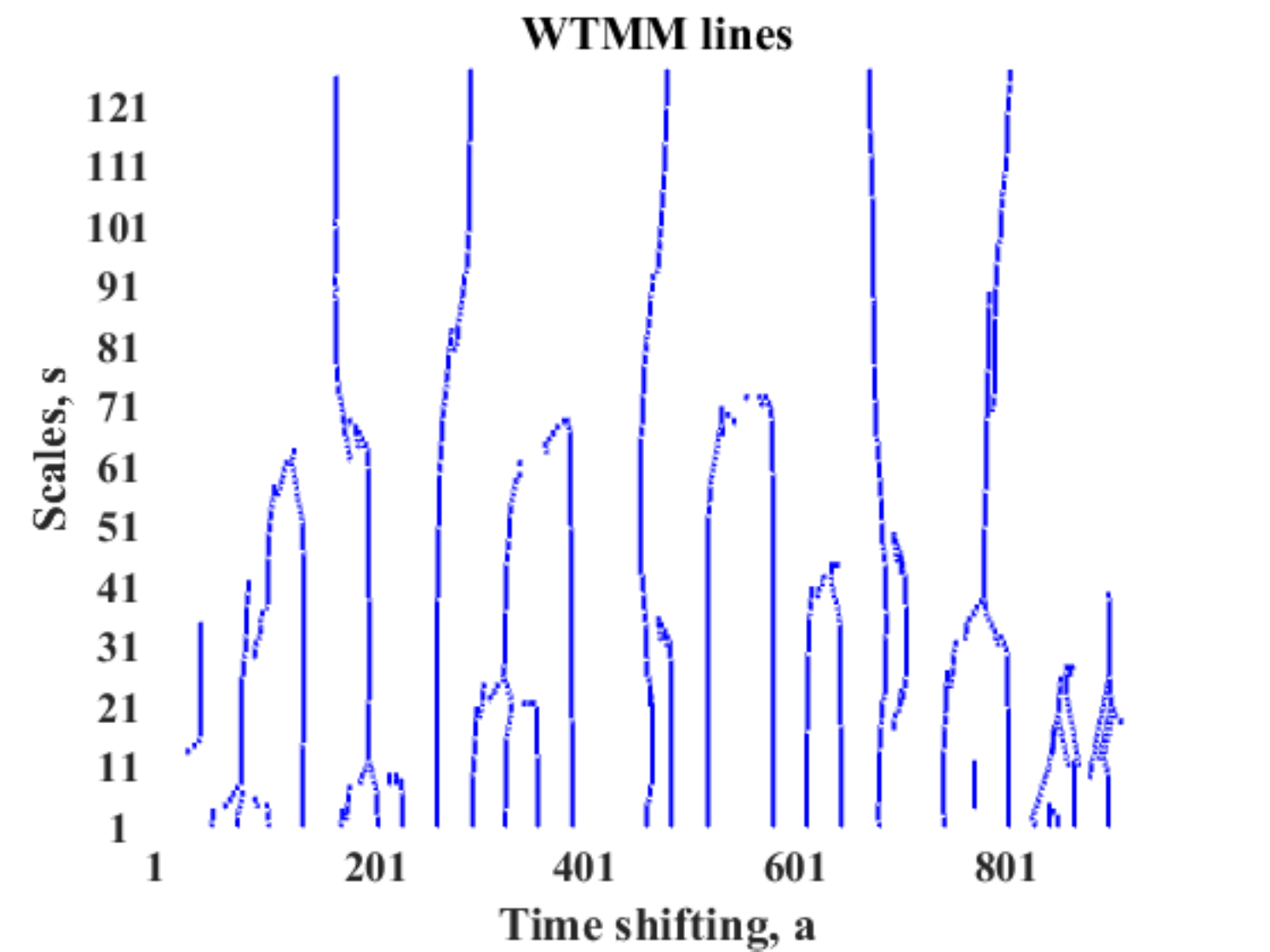}
\includegraphics[scale=0.5]{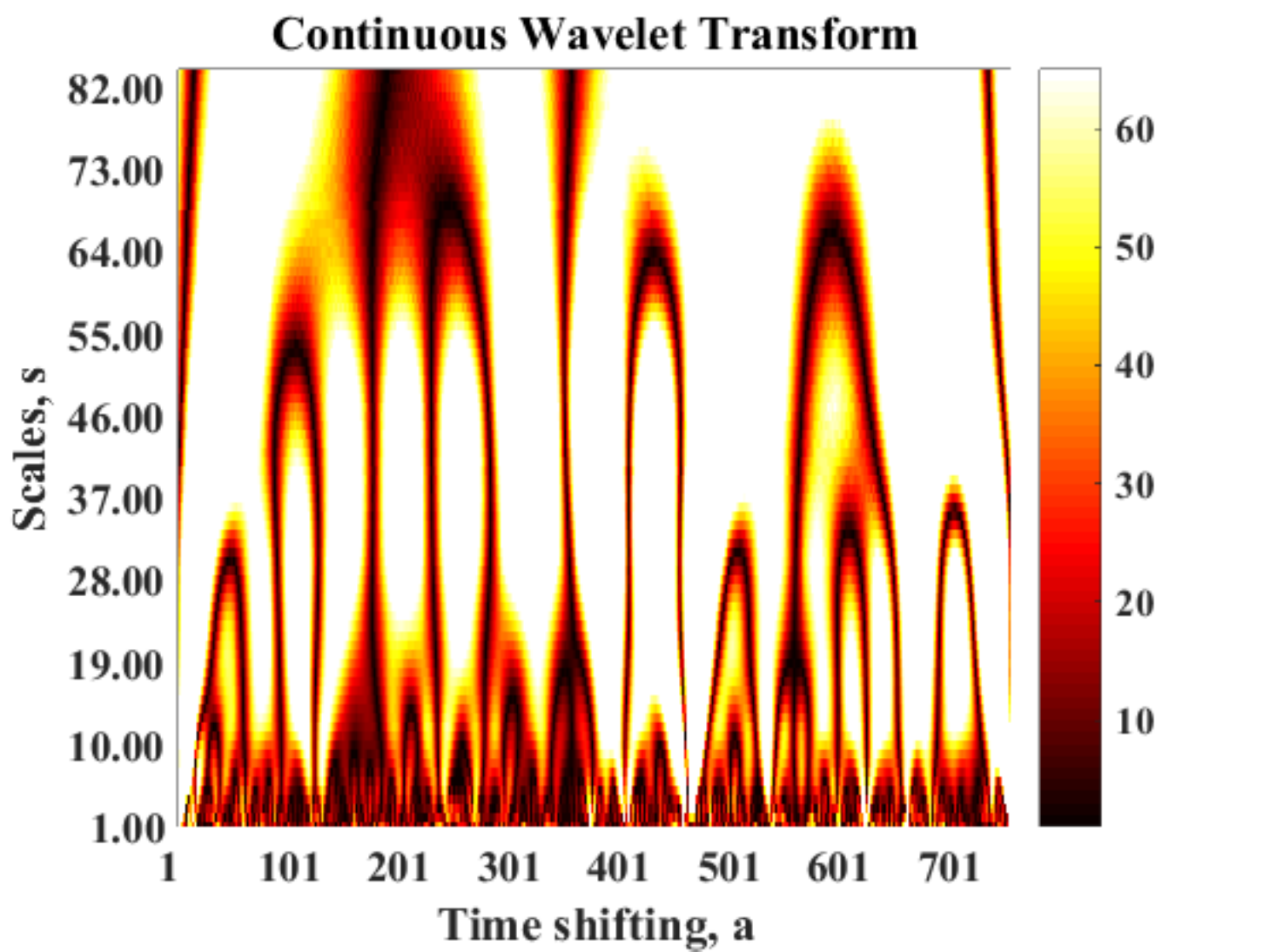}  
\includegraphics[scale=0.5]{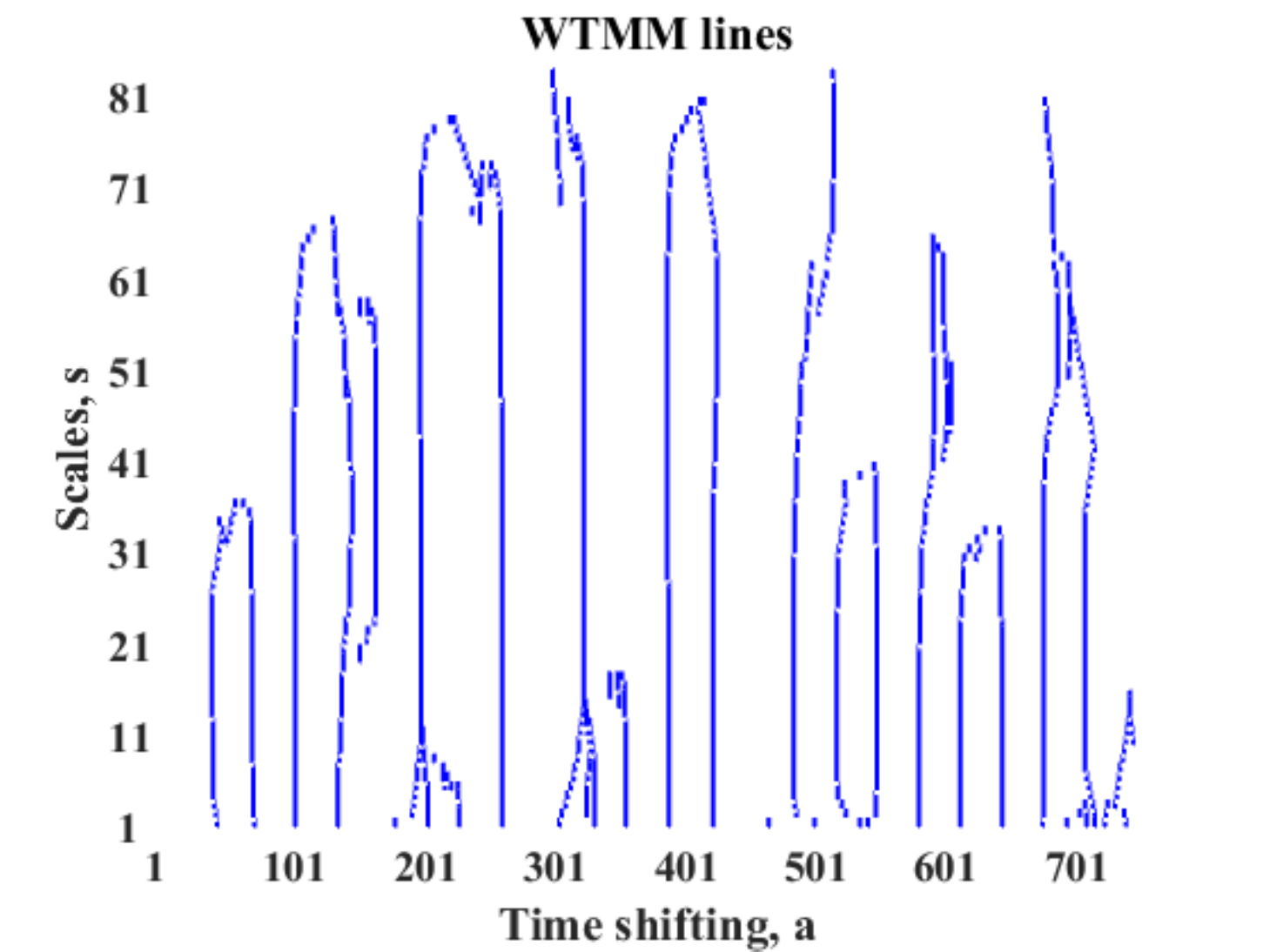}
\includegraphics[scale=0.5]{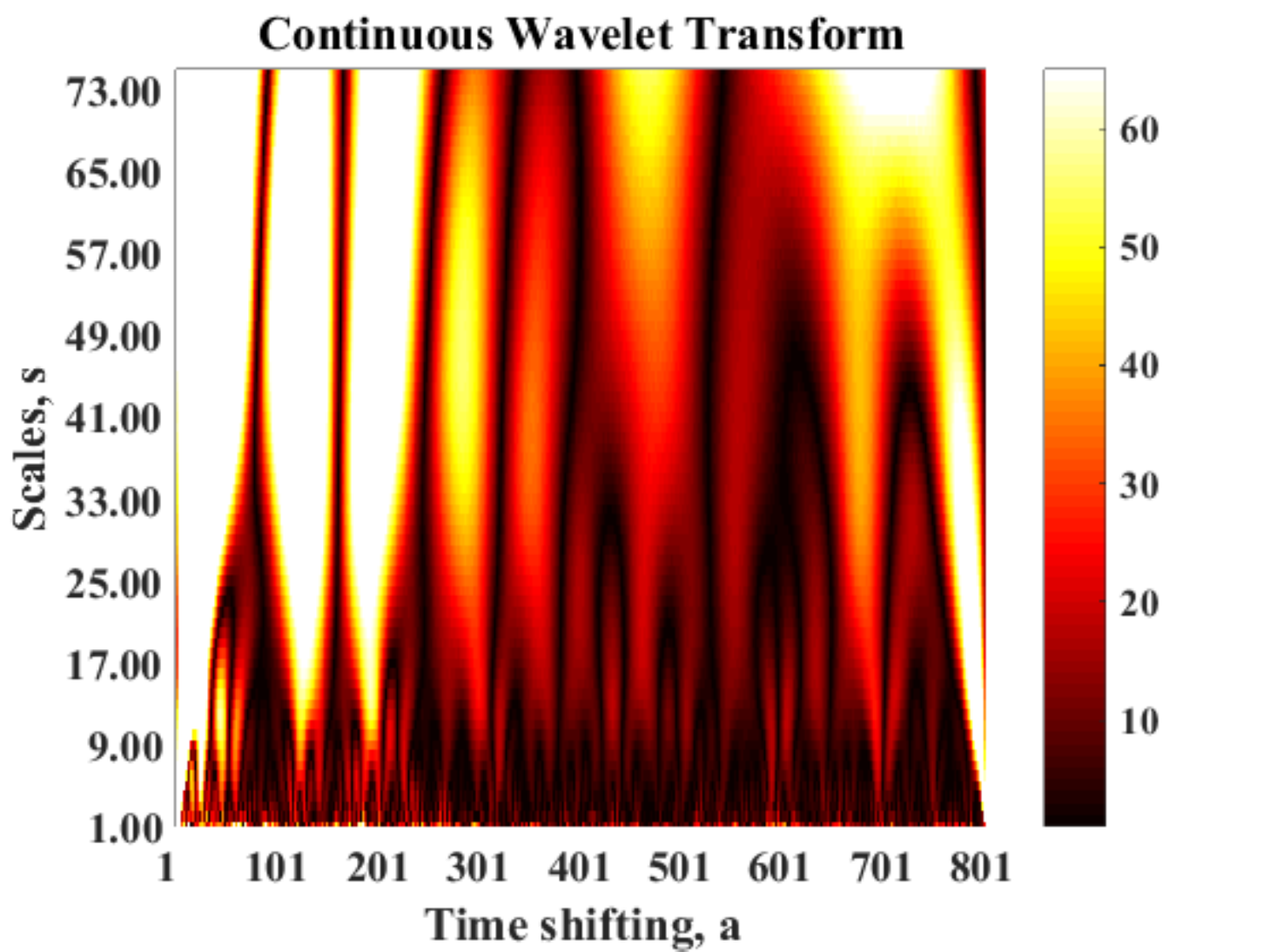} 
\includegraphics[scale=0.5]{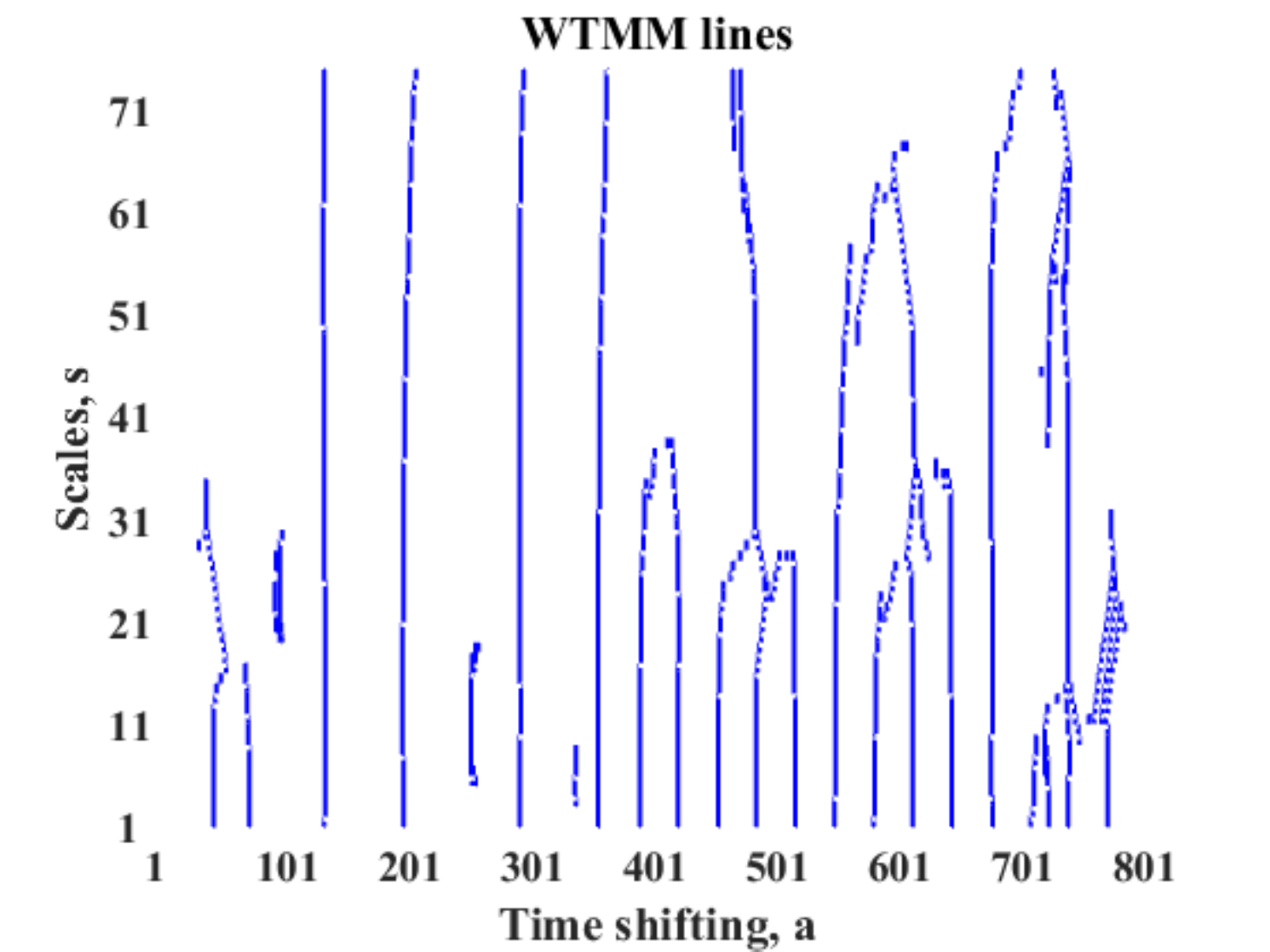}
\includegraphics[scale=0.5]{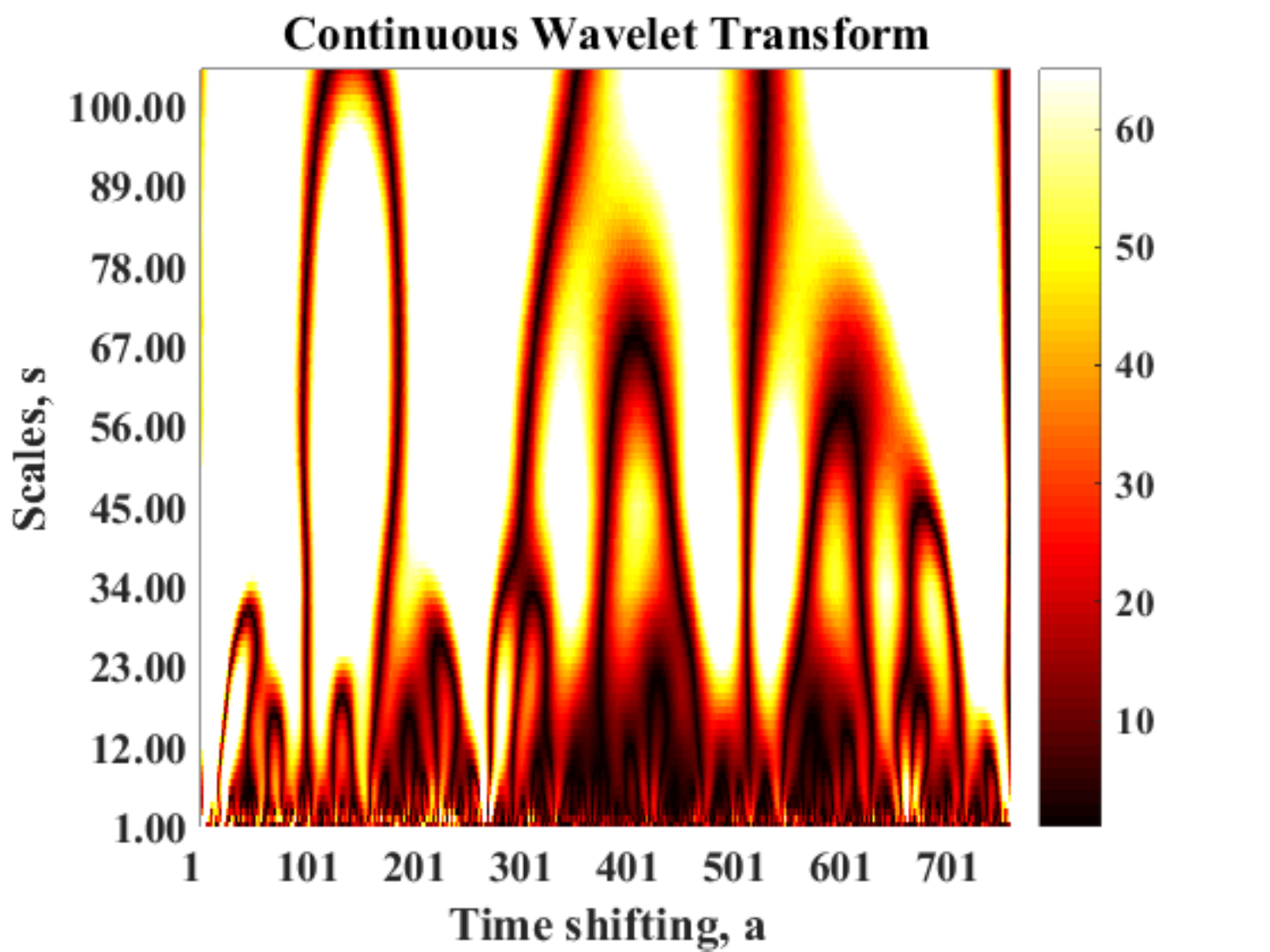}  
\includegraphics[scale=0.5]{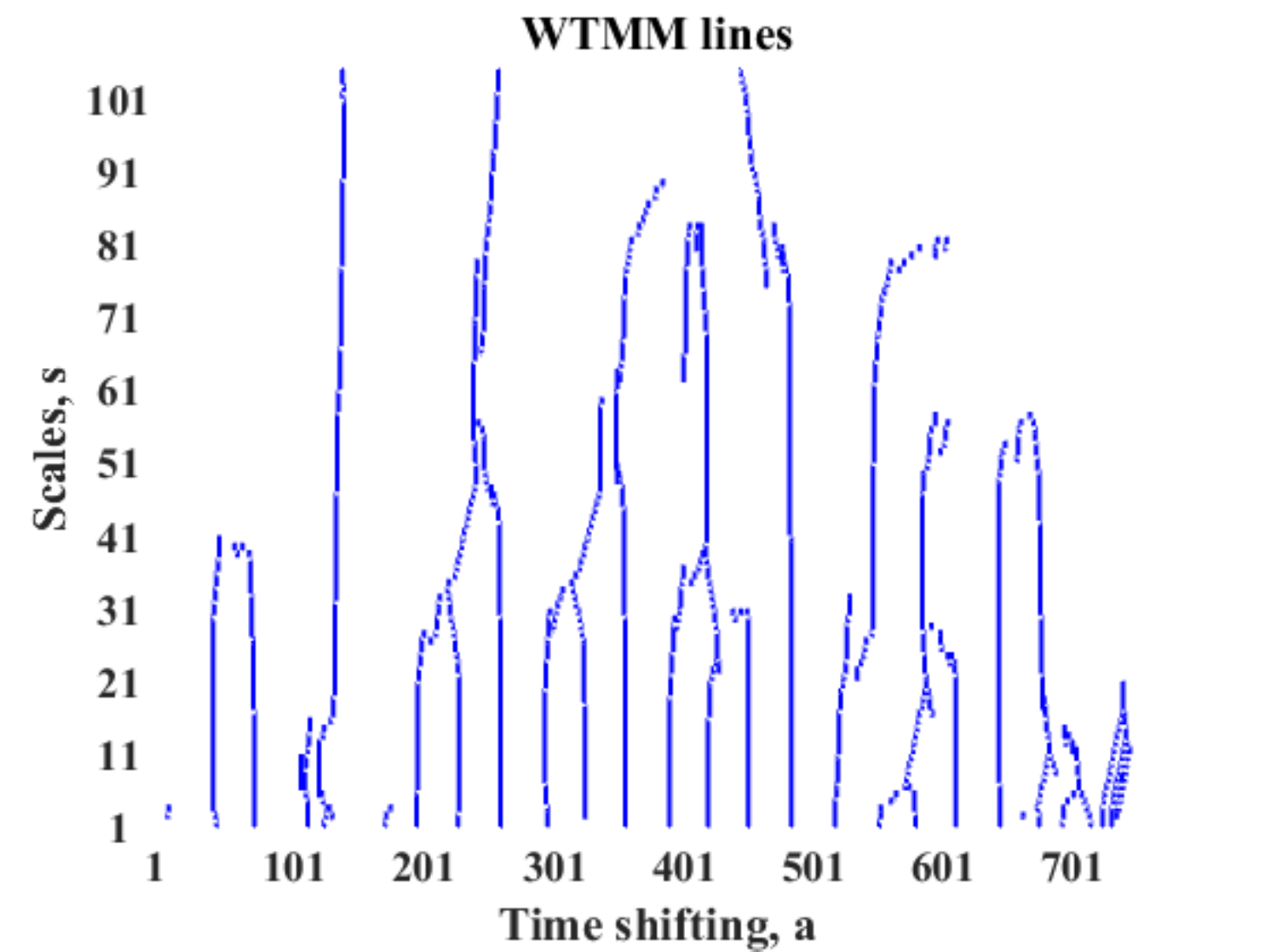}
\caption{The wavelet map and skeleton function (left to right) at 22 GHz for the sources 3C 273, 3C 279, 3C 345, and 3C 454.3 (top to bottom).}
\label{fig3}
\end{figure*}

\begin{figure*}
\centering
\includegraphics[scale=0.5]{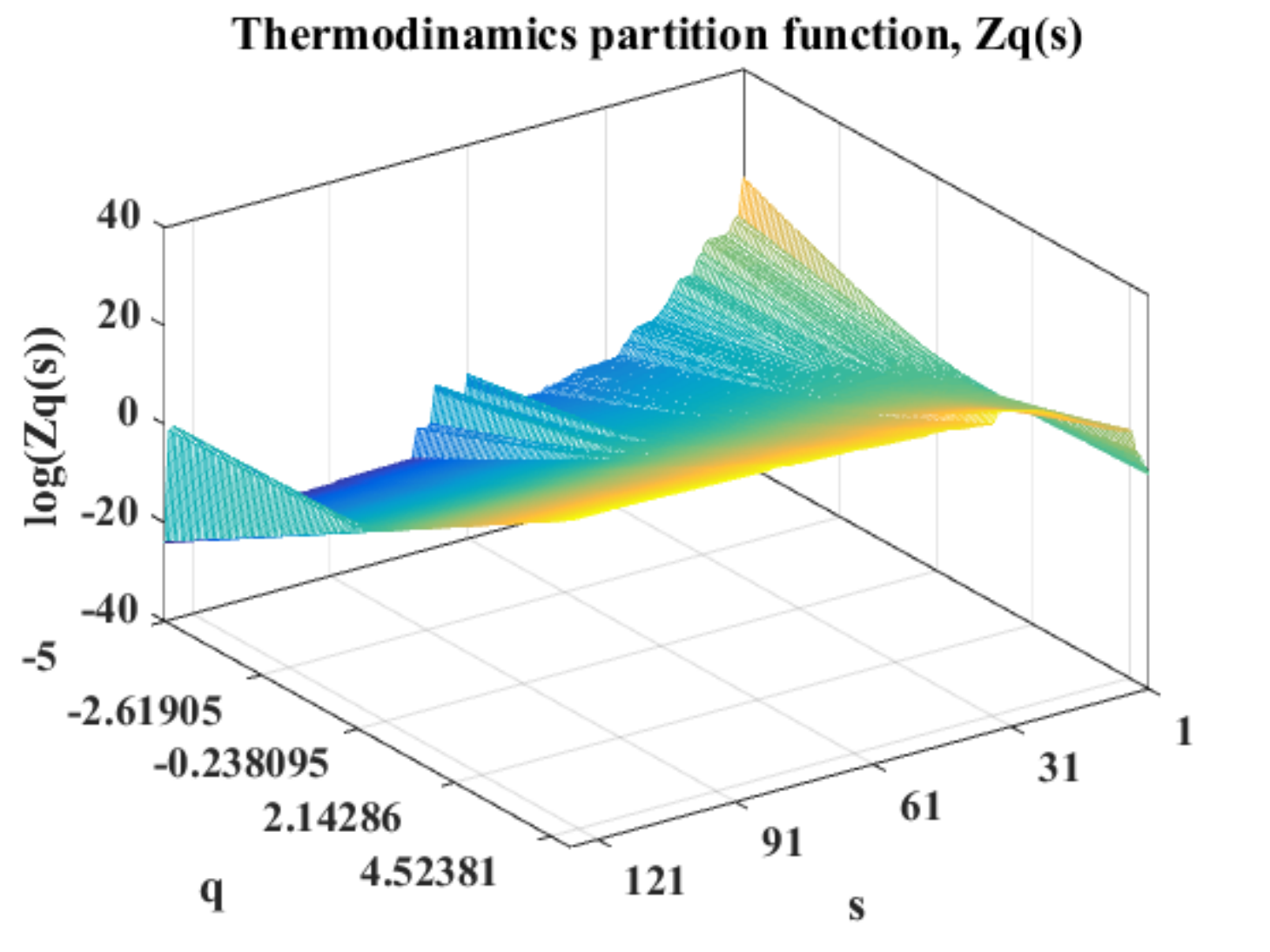} 
\includegraphics[scale=0.5]{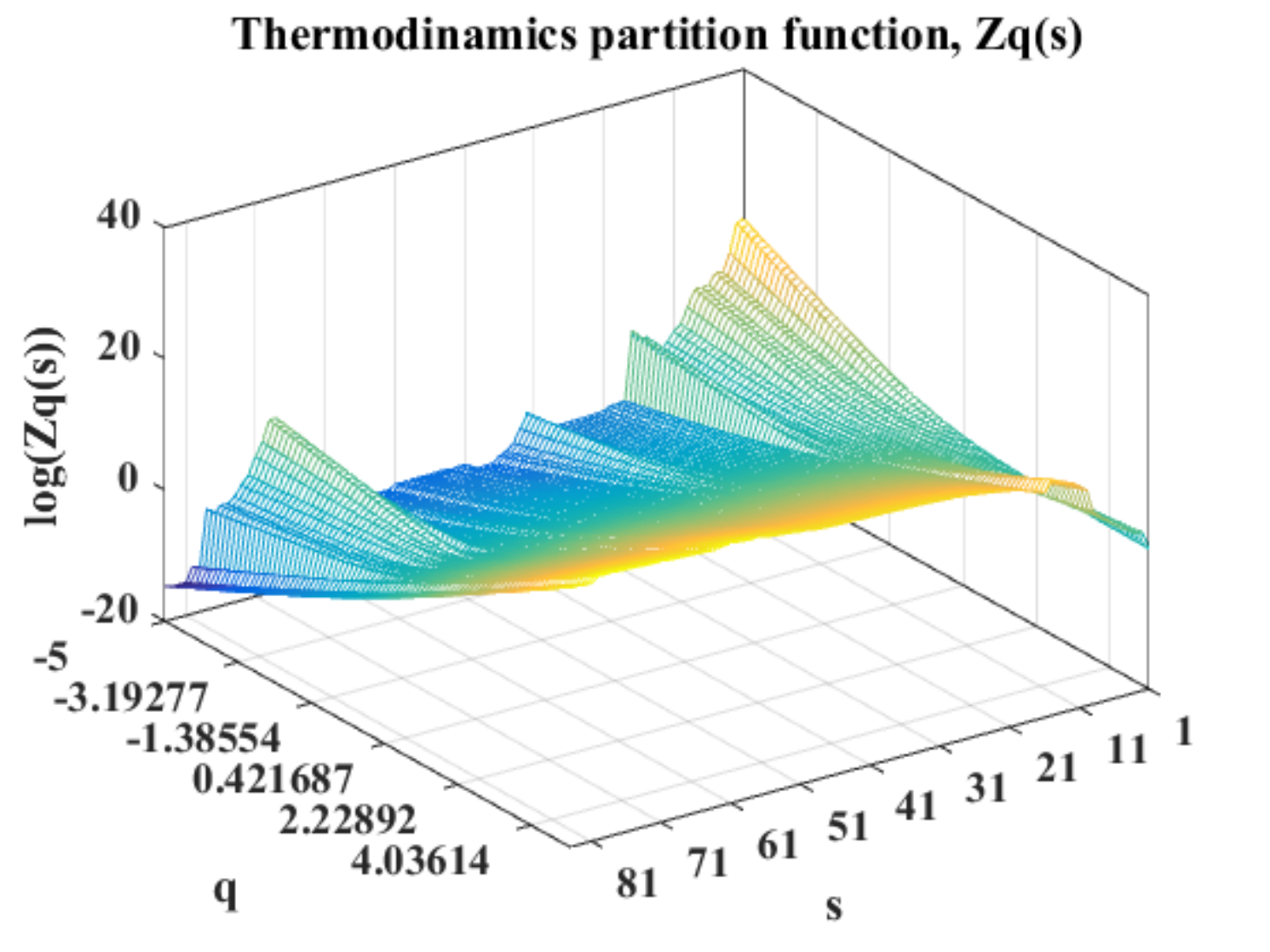} 
\includegraphics[scale=0.5]{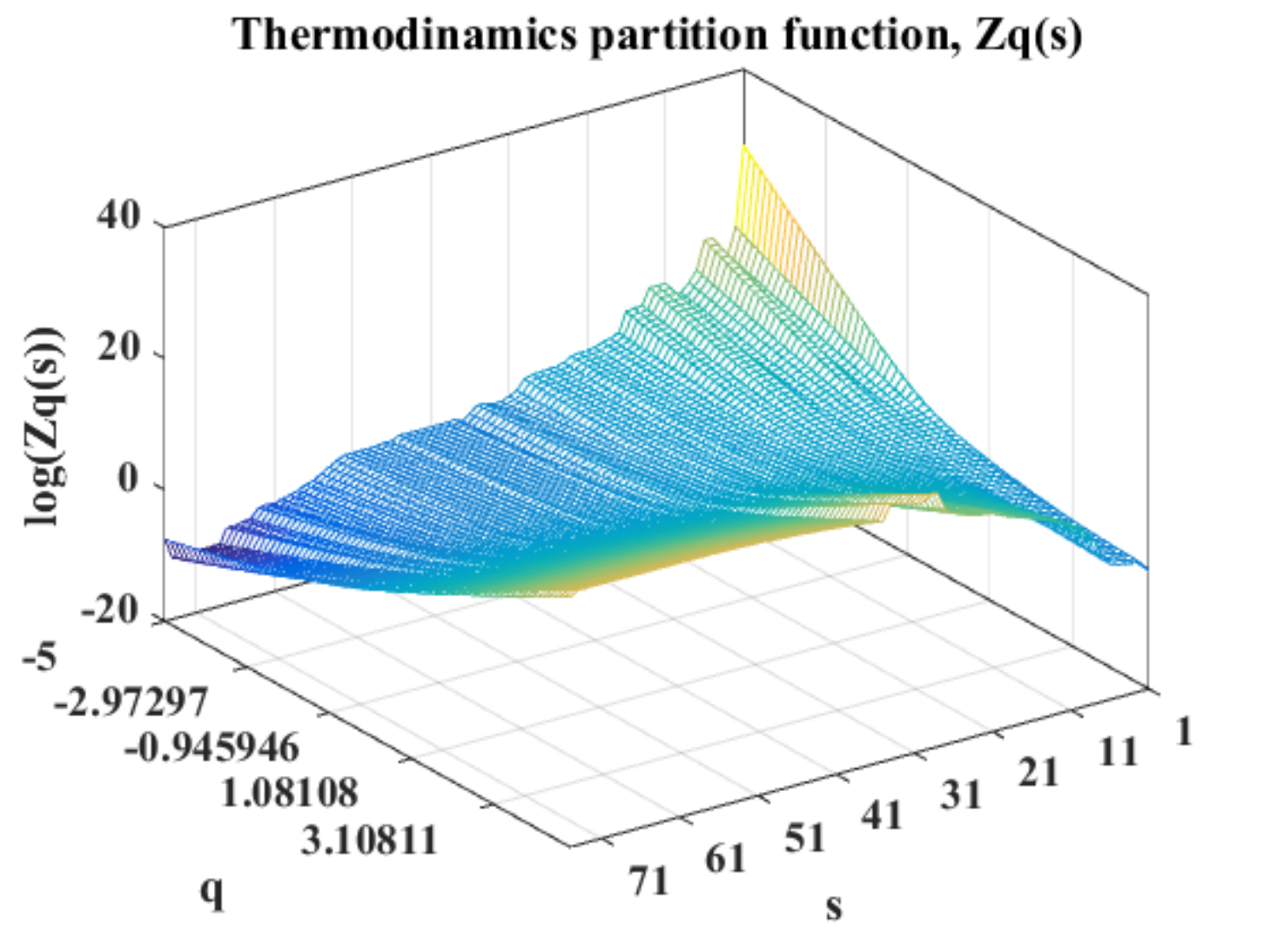} 
\includegraphics[scale=0.5]{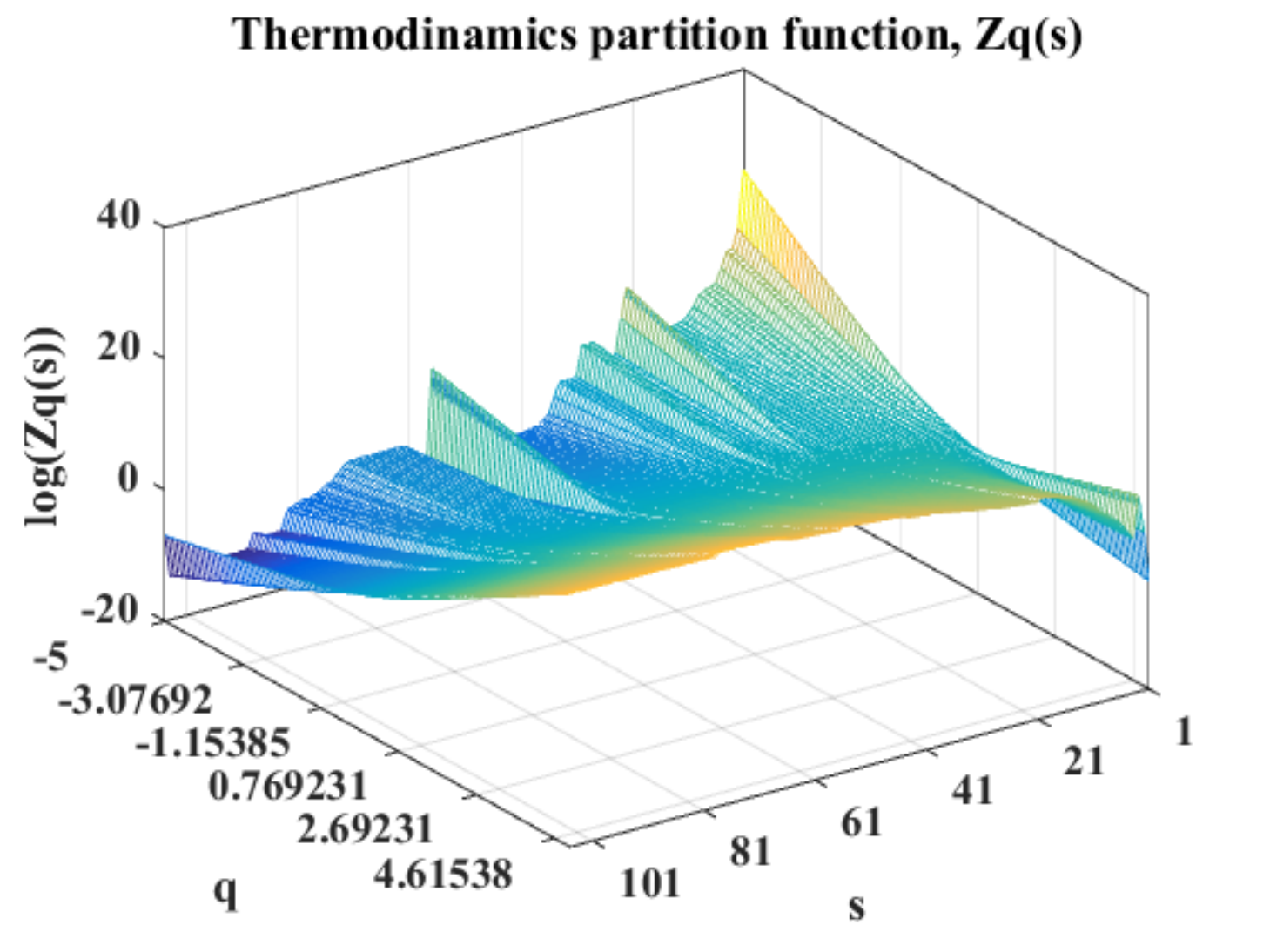} 

\caption{Upper panel: The thermodynamic partition functions at 22 GHz for the sources 3C 273 (left) and 3C 279 (right). Lower panel: The thermodynamic partition functions at 22 GHz for the sources 3C 345 (left) and 3C 454.3 (right).}
\label{fig4}
\end{figure*}

\begin{figure*}
\centering
\includegraphics[scale=0.5]{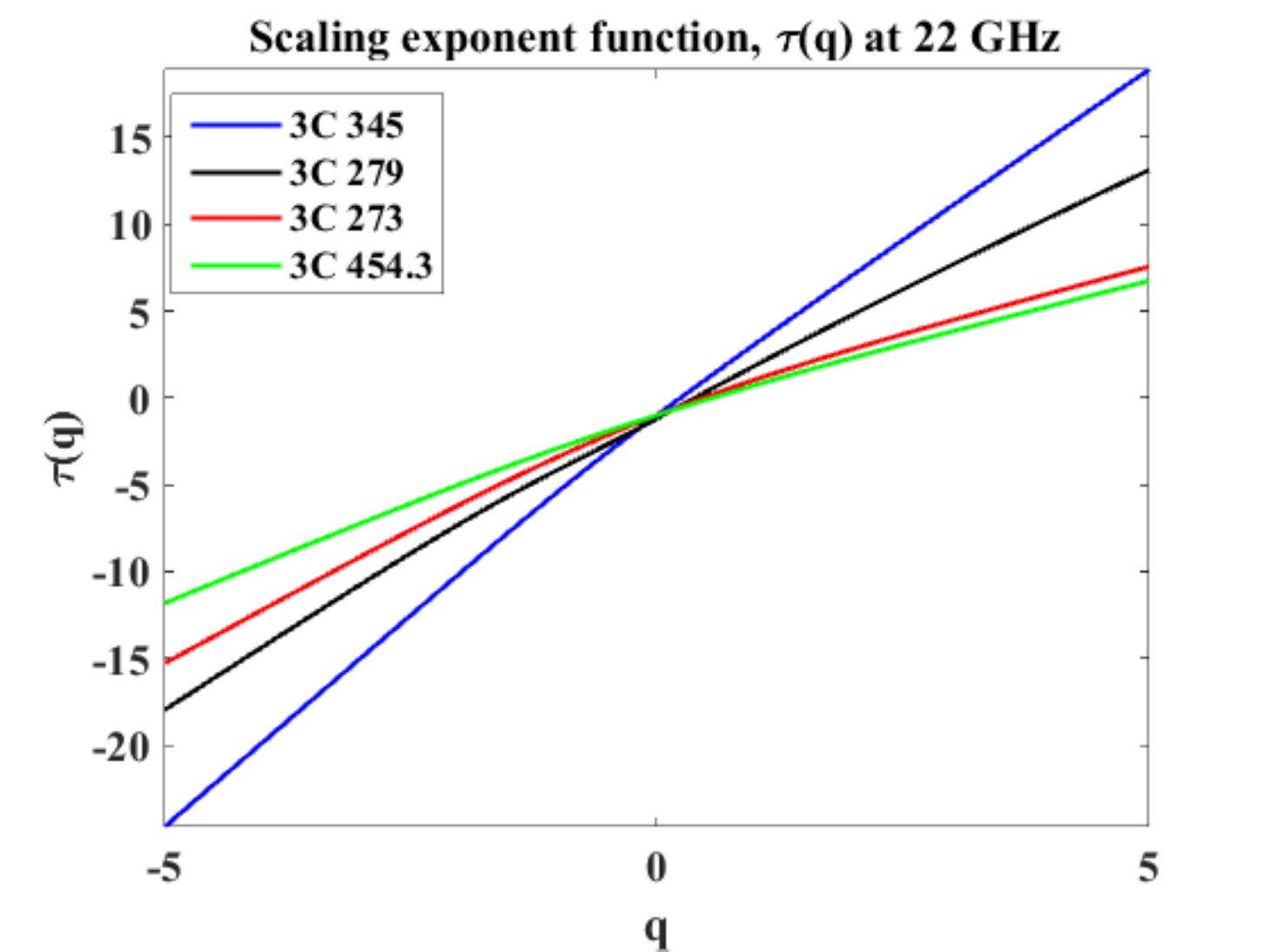}  
\includegraphics[scale=0.5]{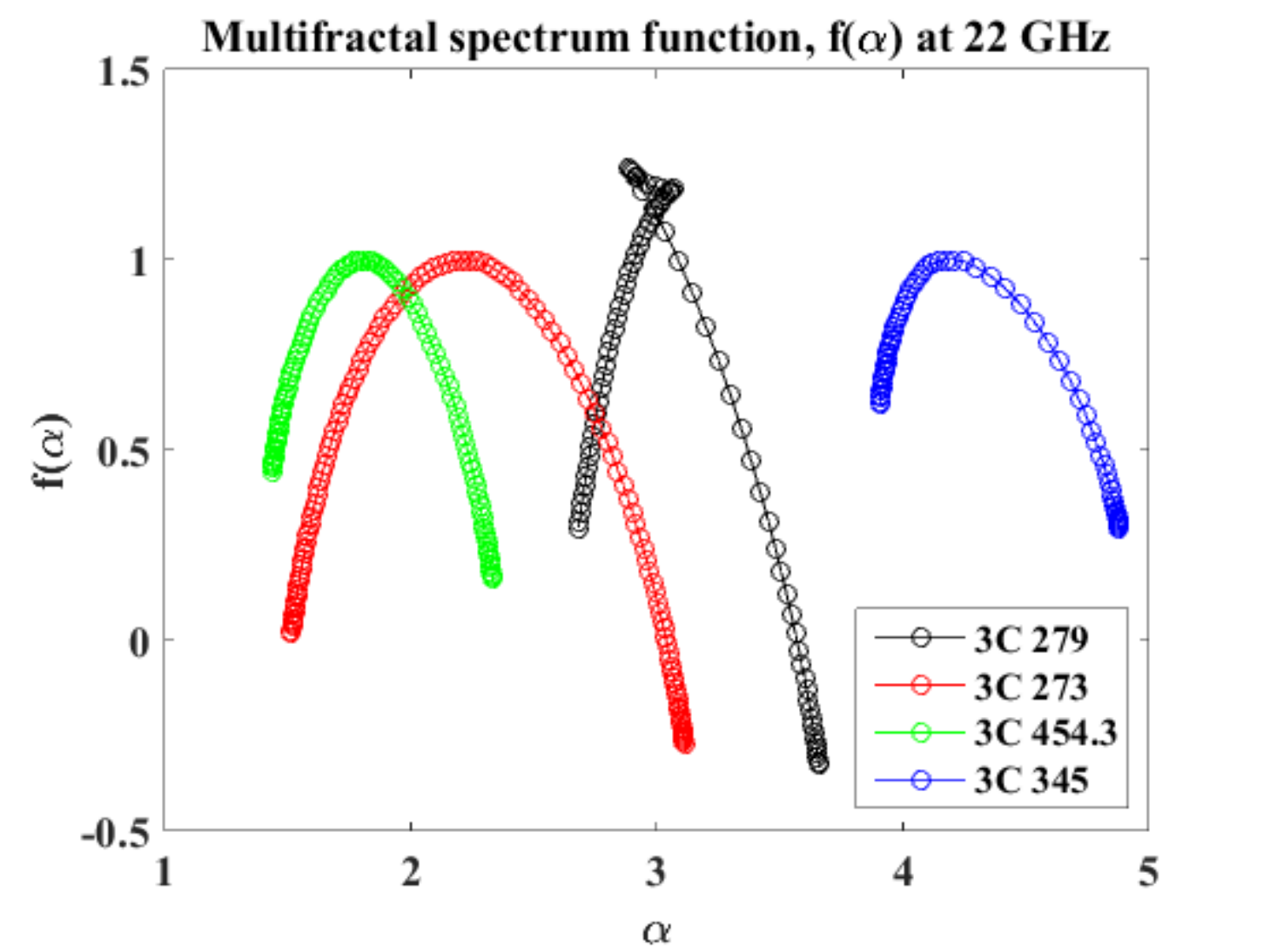}
\caption{The scaling exponent (left) and multifractal spectrum (right) at 22 GHz for the sources 3C 273, 3C 279, 3C 345, and 3C 454.3.}
\label{fig5}
\end{figure*}

\begin{figure*}
\centering
\includegraphics[scale=0.5]{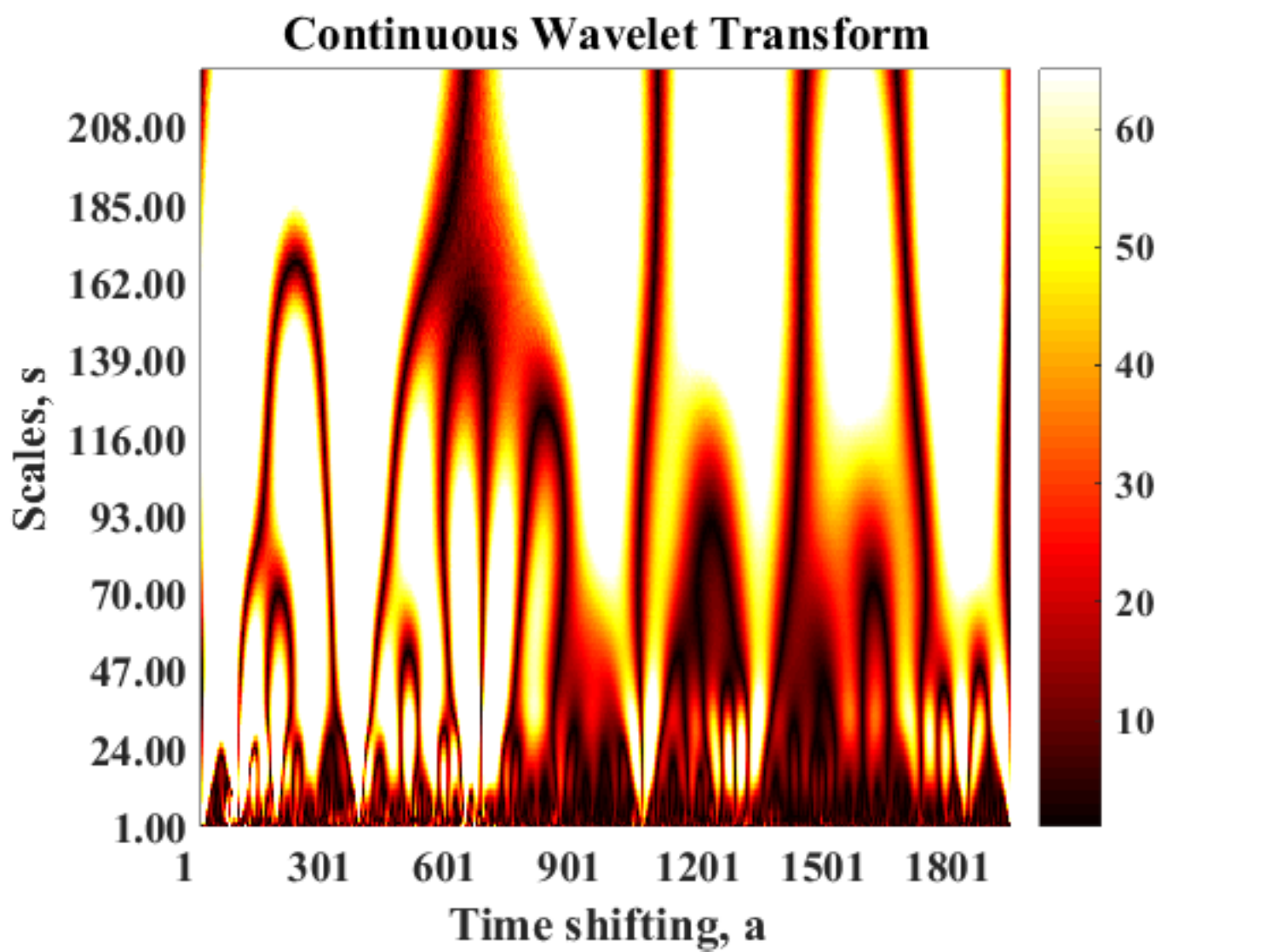}  
\includegraphics[scale=0.5]{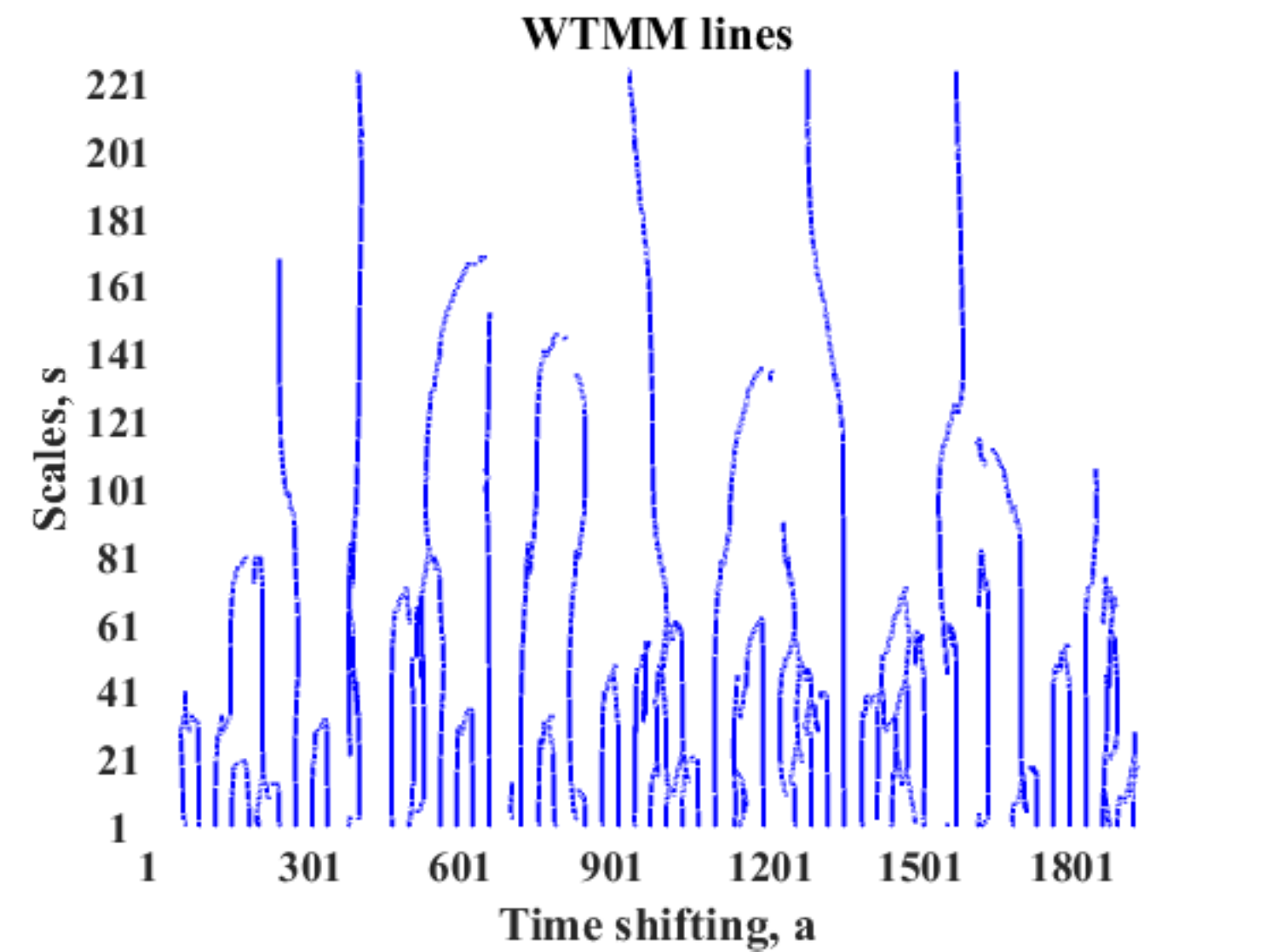}
\includegraphics[scale=0.5]{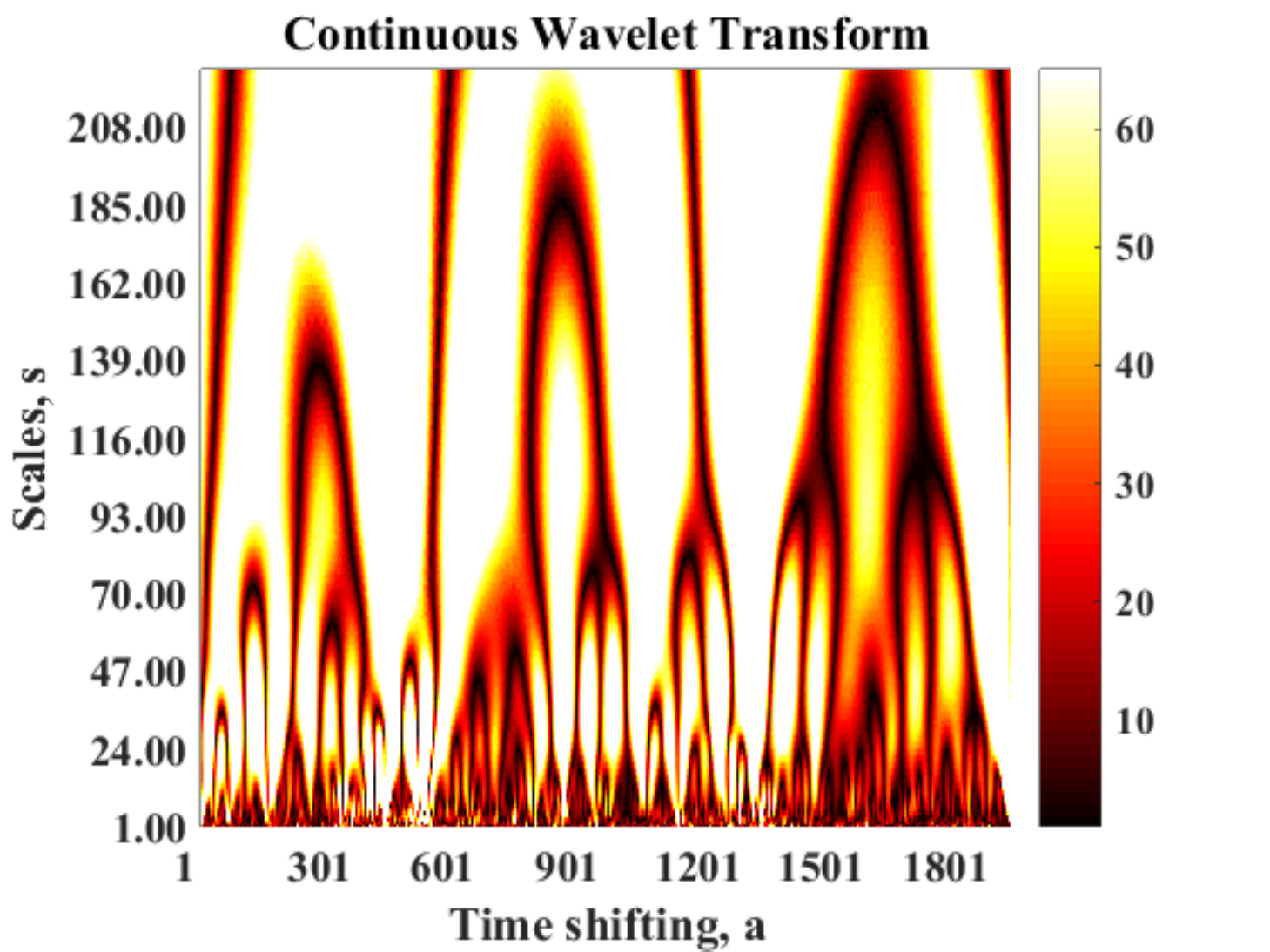}  
\includegraphics[scale=0.5]{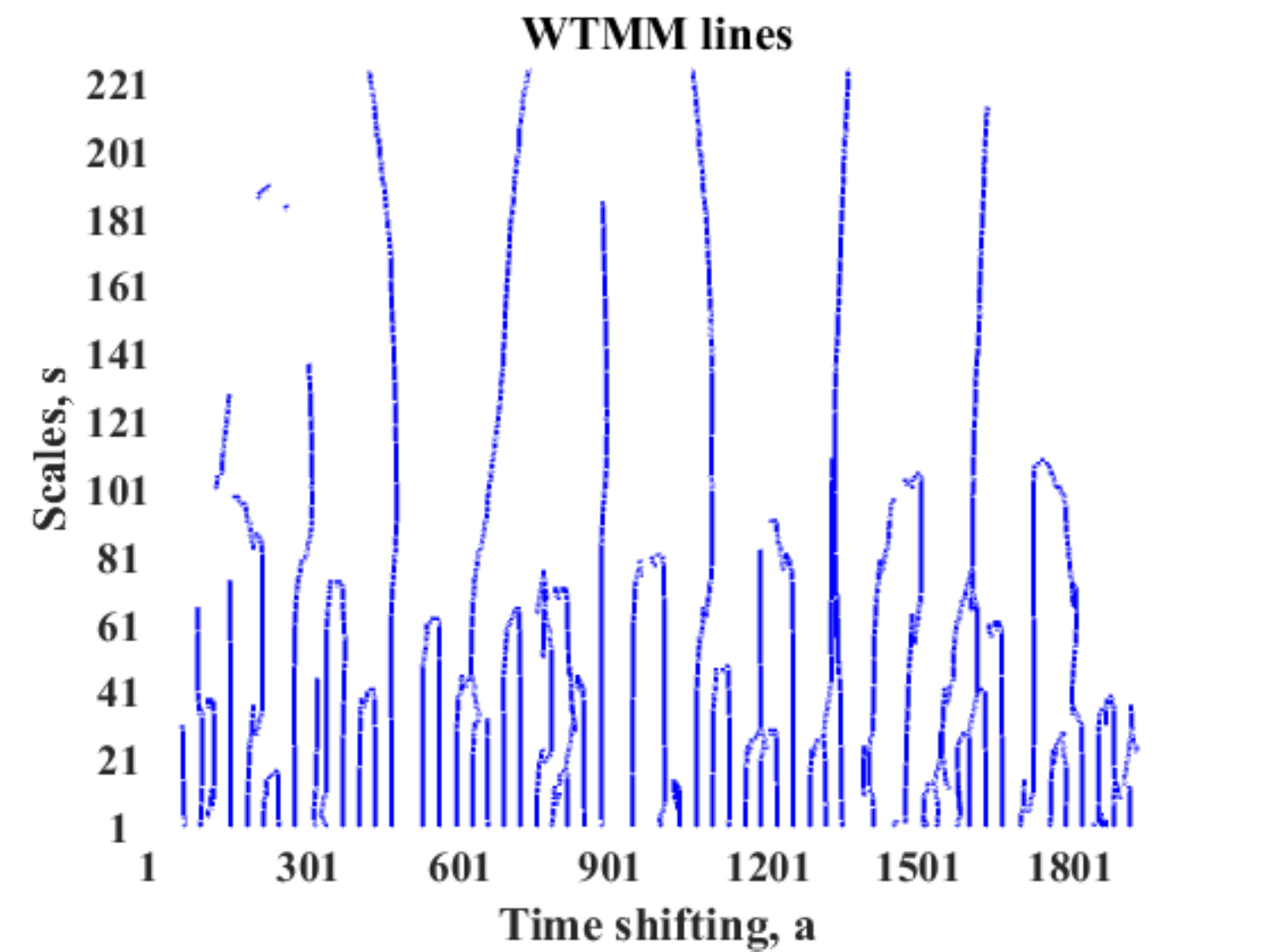}
\includegraphics[scale=0.5]{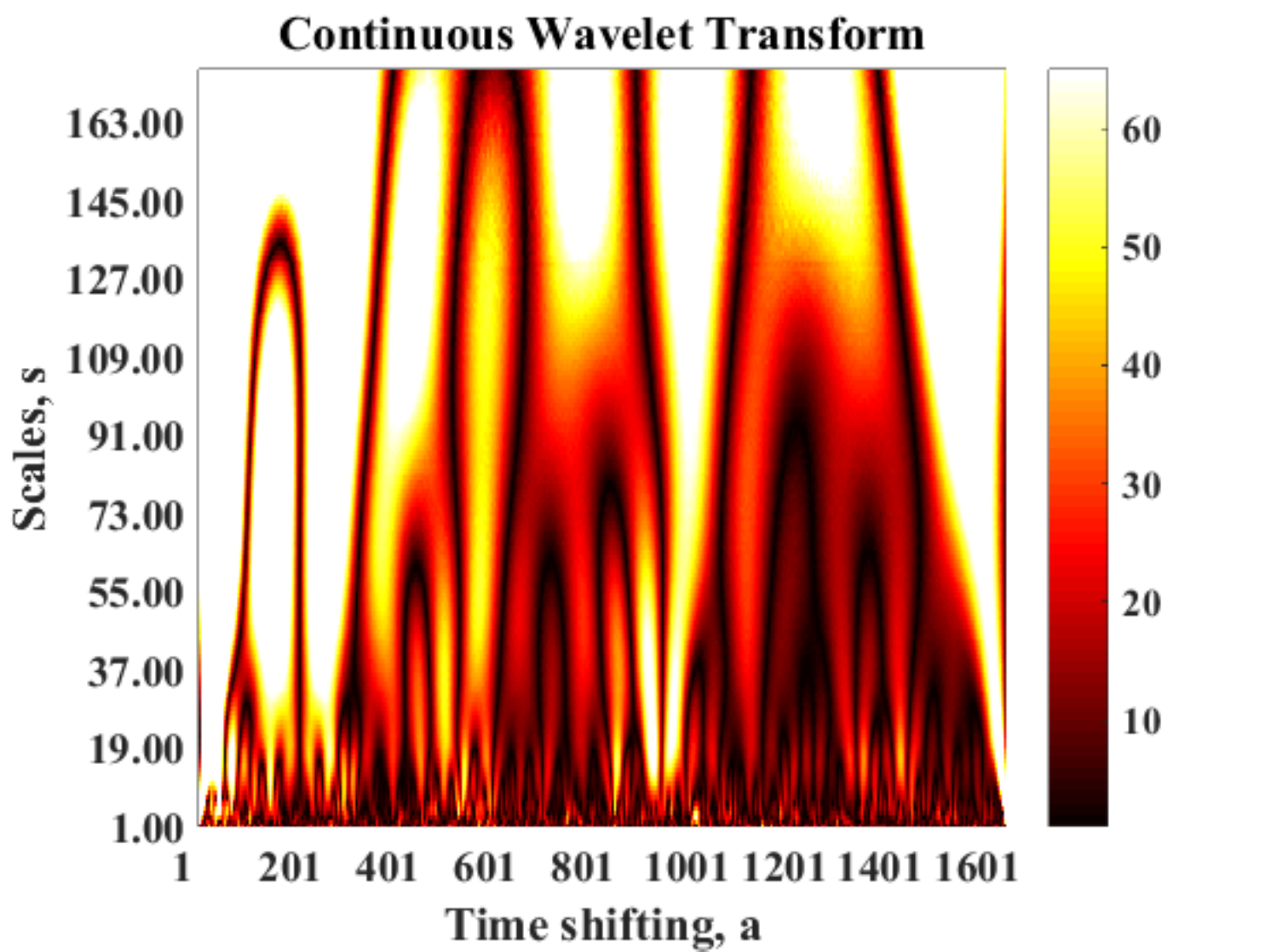} 
\includegraphics[scale=0.5]{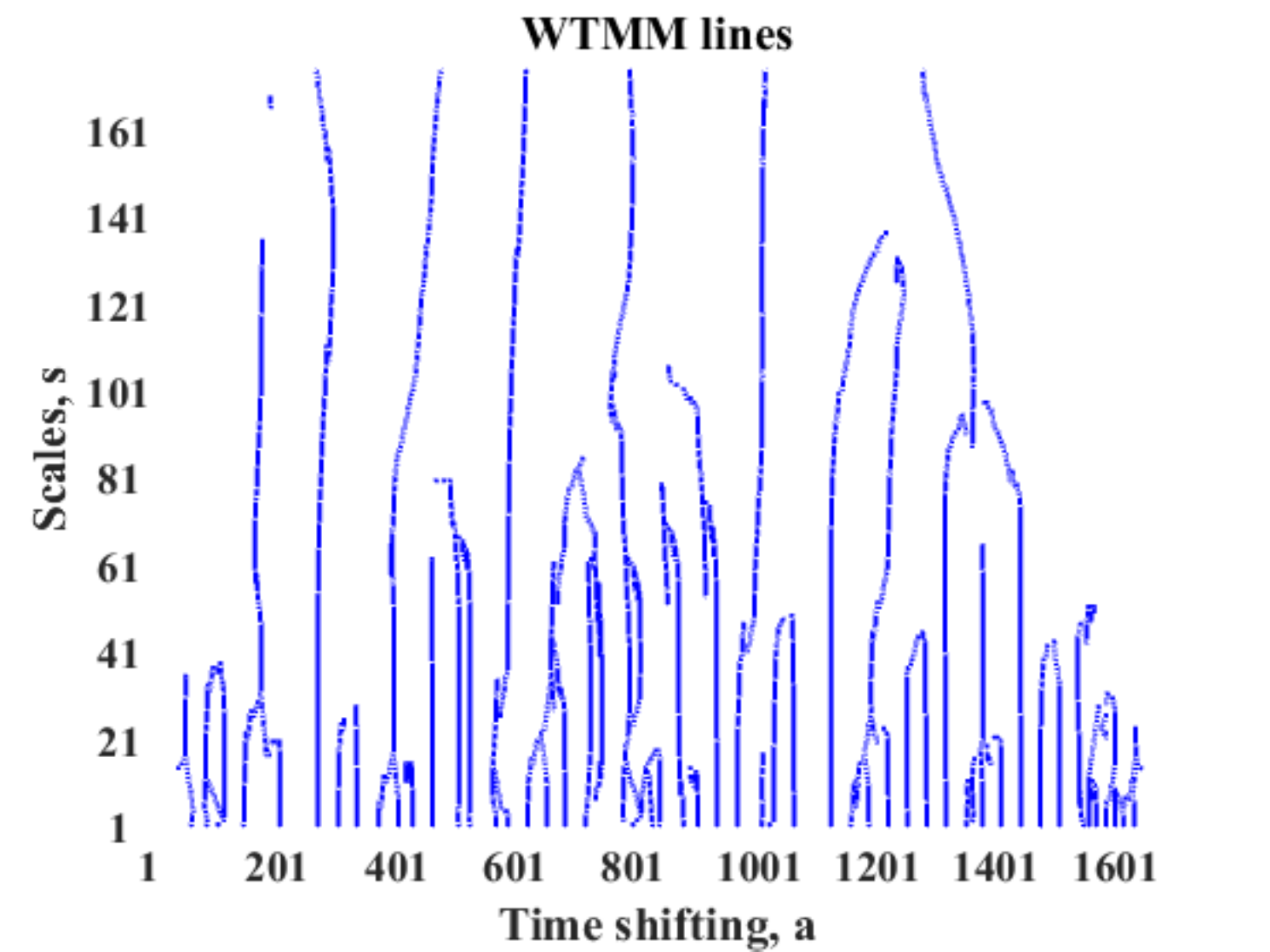}
\includegraphics[scale=0.5]{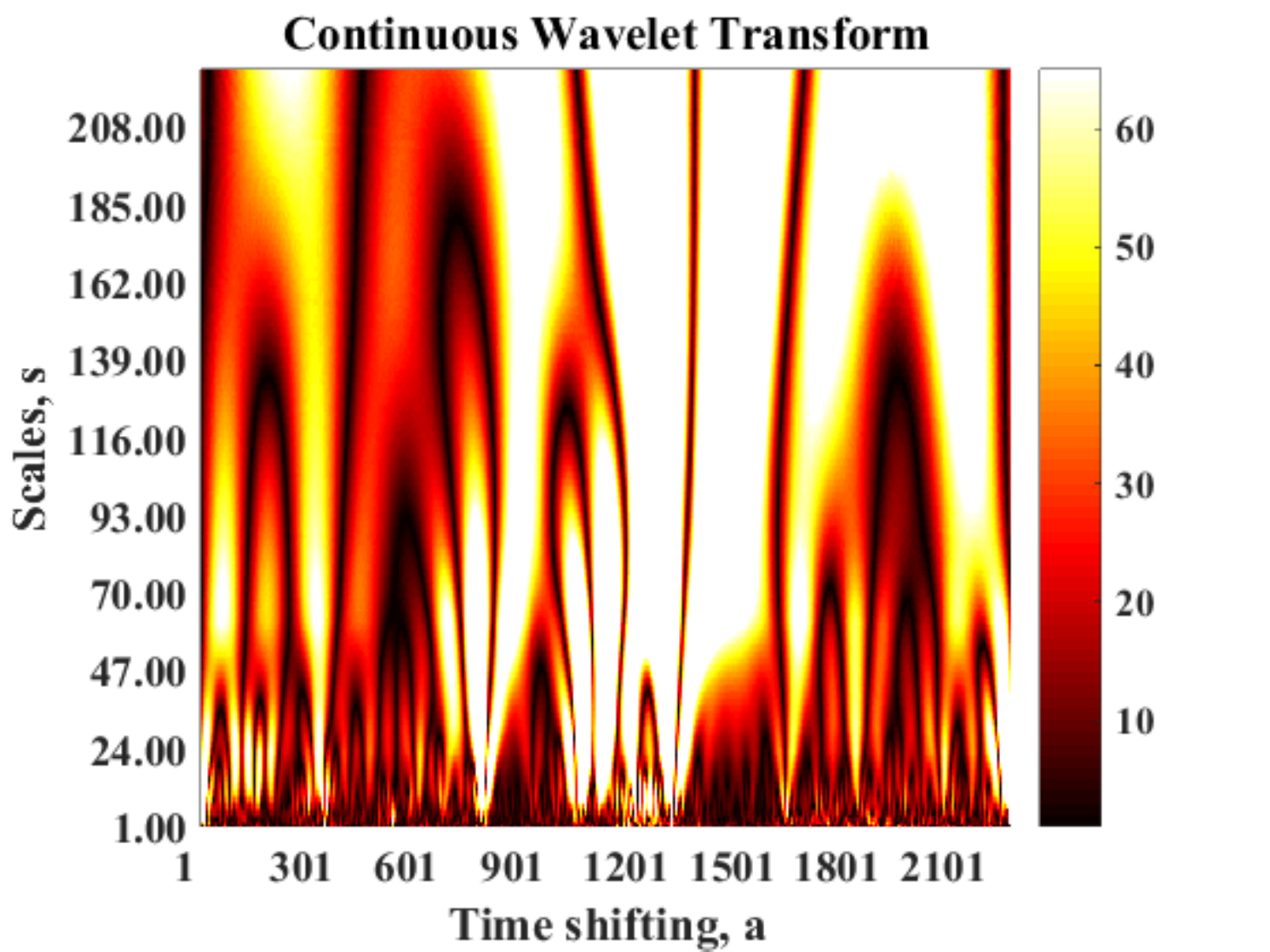}  
\includegraphics[scale=0.5]{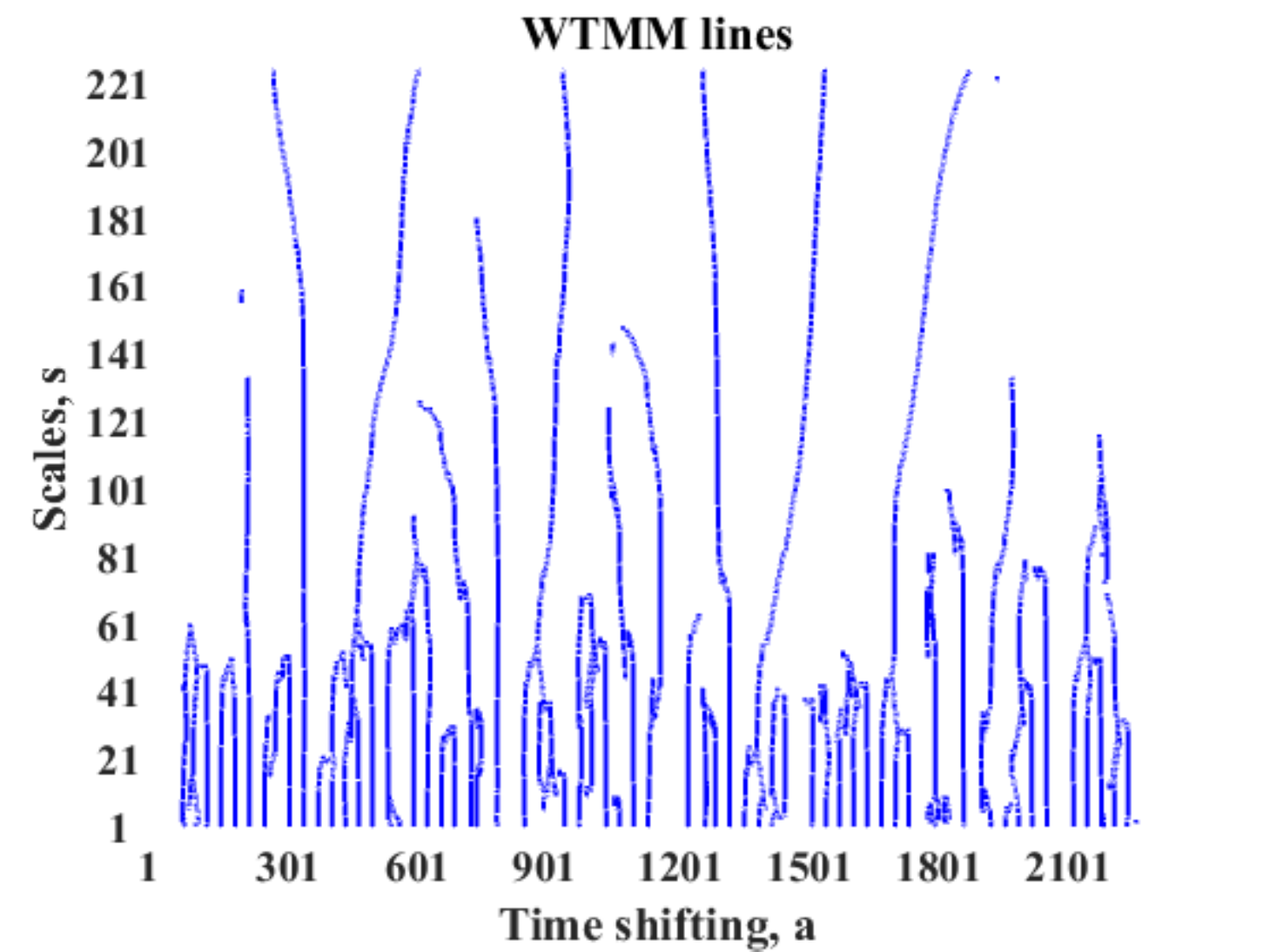}
\caption{The wavelet map and skeleton function (left to right) at 37 GHz for the sources 3C 273, 3C 279, 3C 345, and 3C 454.3 (top to bottom), respectively.}
\label{fig6}
\end{figure*}

Here, we analyze the multifractal behavior of the light curves in the observed frame at 22 and 37 GHz, Fig. \ref{fig1}. The scenario emerging for each light curve is discussed separately as follows. In Figs. \ref{fig3} and \ref{fig6}, we present the continuous wavelet transform maps and the constructed skeleton functions for all sources at 22 and 37 GHz, respectively. The wavelet maps present the calculated wavelet coefficients in their absolute values and colored as dark and light in accordance with the color map given. The dark and light colors show lower and higher absolute wavelet coefficients, respectively. From the wavelet coefficient matrix, we select local maxima lines, or maxima points at each scale. The collection of all local maxima lines, or maxima points, at each scale forms a function called the skeleton function. The need to construct the skeleton function is to consider only local maxima lines at each scale and simplify the multifractality analysis. In Figs. \ref{fig4} and \ref{fig7}, we present  the 3D plots of the thermodynamic partition functions, a function of two arguments - scaling parameter $s$ and moment $q$ - at 22 and 37 GHz for all the light curves to show how the thermodynamic function $Zq(s)$, the moment $q$, and the scaling parameter $s$ behave together. It is the thermodynamic partition function that connects the wavelet formalism to the multifractal formalism. To clearly see the scaling behavior between $Zq$($s$) and $s$, the scalability of the signals, we have created the log-log plots of $Zq$($s$) and $s$ using eq. \ref{eq12} for all the light curves and represented the calculated slope by the scaling exponent function $\tau$($q$) plots as shown in Figs. \ref{fig5} and \ref{fig8}. The observed nonlinearity between the scaling exponent $\tau(q)$ vesus the moment $q$  at both bands reveals the presence of nonlinear scaling behavior between the thermodynamic partition function $Zq$($s$) and the scale $s$ for all the light curves. Furthermore, the observed nonlinearity between $\tau$($q$) and $q$, which is the slope of log($Zq$(s)) against log($s$) plots, clearly indicates the presence of multifractal behavior in all the light curves though the degree of nonlinearity that varies between sources at both bands.

Based on the scaling exponent function $\tau$($q$),  Figs. \ref{fig5} and \ref{fig8}, we can see the similarity and difference in the degree of nonlinearity at the two bands for each source, Fig. \ref{fig9},  and between the sources,  Figs. \ref{fig5} and \ref{fig8}. After observing the presence of the multifractal signature based on the nonlinear scaling exponent function, the degree of multifractality is determined (i.e., how strong is the observed multifractal signature) at each band for all the sources. To this end, we estimate the multifractal spectrum function $f$($\alpha$) and calculate the width ($\Delta\alpha$ = $\alpha_{max}$ - $\alpha_{min}$) using eqs. \ref{eq13} and \ref{eq14}. The broader the spectrum (the larger the value of $\Delta\alpha$), the richer the multifractality is. The width value tells us how strong is the observed multifractal signature, which is an additional parameter one can use to see the difference in the degree of nonlinearity, or multifractality between signals. The calculated width $\Delta\alpha$  values for 3C 273, 3C 279, 3C 345, and 3C 454.3 at 22 and 37 GHz are 1.5961\slash 0.9763\slash 0.9745\slash 0.8951 and 1.5854\slash  0.9653\slash  0.9633\slash  8432, respectively (Table\ref{tab2}). Therefore, for all the sources considered, the width $\Delta\alpha$ value is not near zero at both frequencies, indicating the multifractal and intermittent nature of the sources at both bands, which is in agreement with the conclusion based on the nonlinear scaling exponent function in Figs. \ref{fig5} and \ref{fig8}. Additionally, the similarity and difference in the degree of the nonlinearity observed at 22 and 37 GHz in the scaling exponent function for each source and between the sources are further confirmed by the calculated width values of the multifractal spectrum function given in the same figures.\\
We have tested the stability of $\Delta\alpha$ by applying different fluctuations in the range of errors recorded for each observation to the original time series and found that our $\Delta\alpha$'s are stable with uncertainty $<$ 10 \%. The observed multifractality could be due to different physical mechanisms. It is the variation in the flux that results in the multifractal structure in a time series, and therefore, the observed multifractal signature could be due to turbulence in the radio radiation region (the blobs that propagate in the jet) since turbulent dynamics create multifractal and intermittent structures \citep{2013PhRvL.110t5002L, 2004AnGeo..22.2431Y}. Additionally, it has been indicated that a higher magnetic field strength will increase both Compton and synchrotron losses in blazars, which could result in an increase in variability at millimeter and longer wavelengths. Additionally, it has been shown that the change in Doppler factor resulting from the change in shock orientation could result in rapid flux variation \citep{1994ApJ...437...91S}, which may culminate in a multifractal structure.

\begin{figure*}
\centering
\includegraphics[scale=0.5]{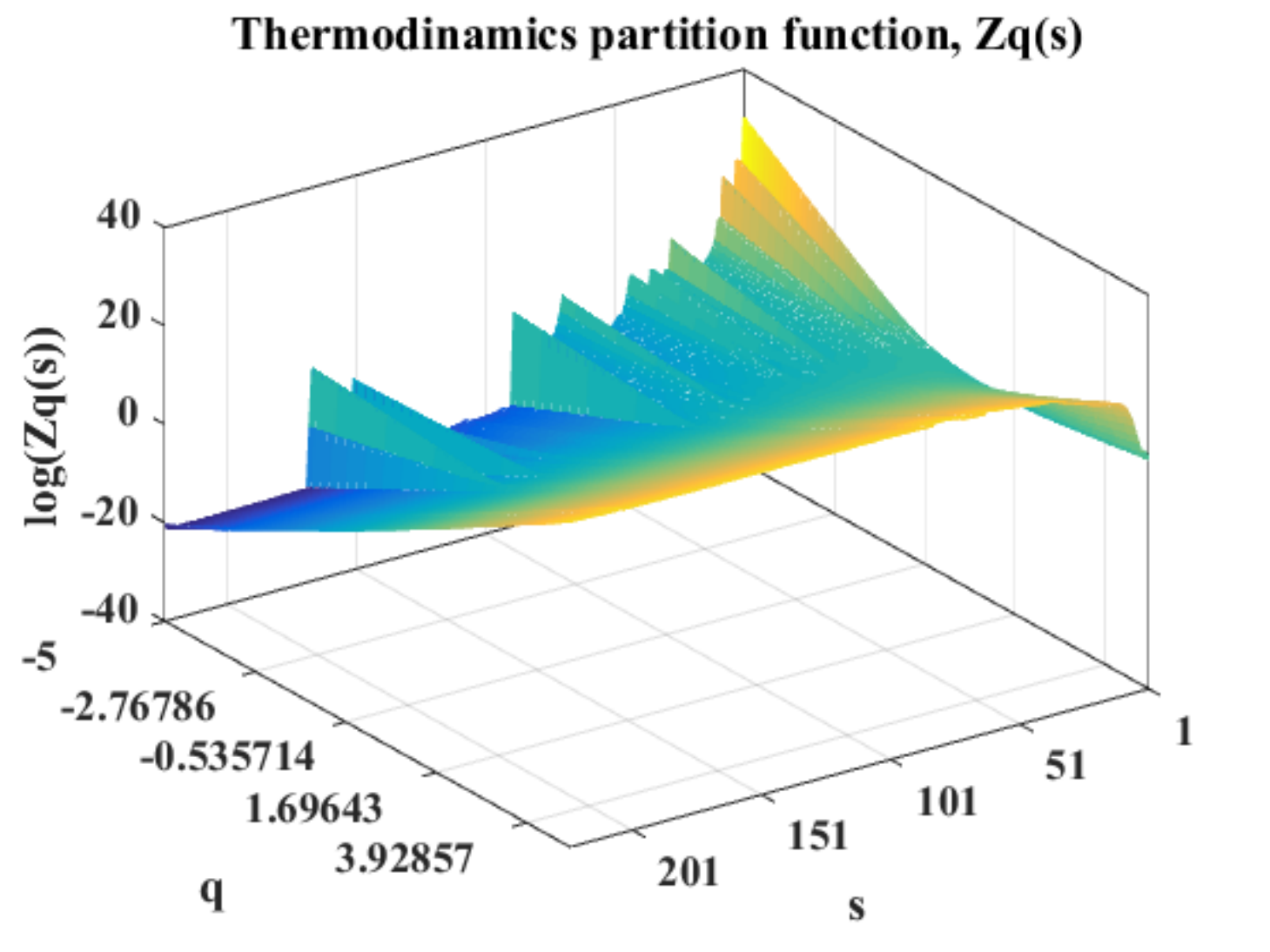} 
\includegraphics[scale=0.5]{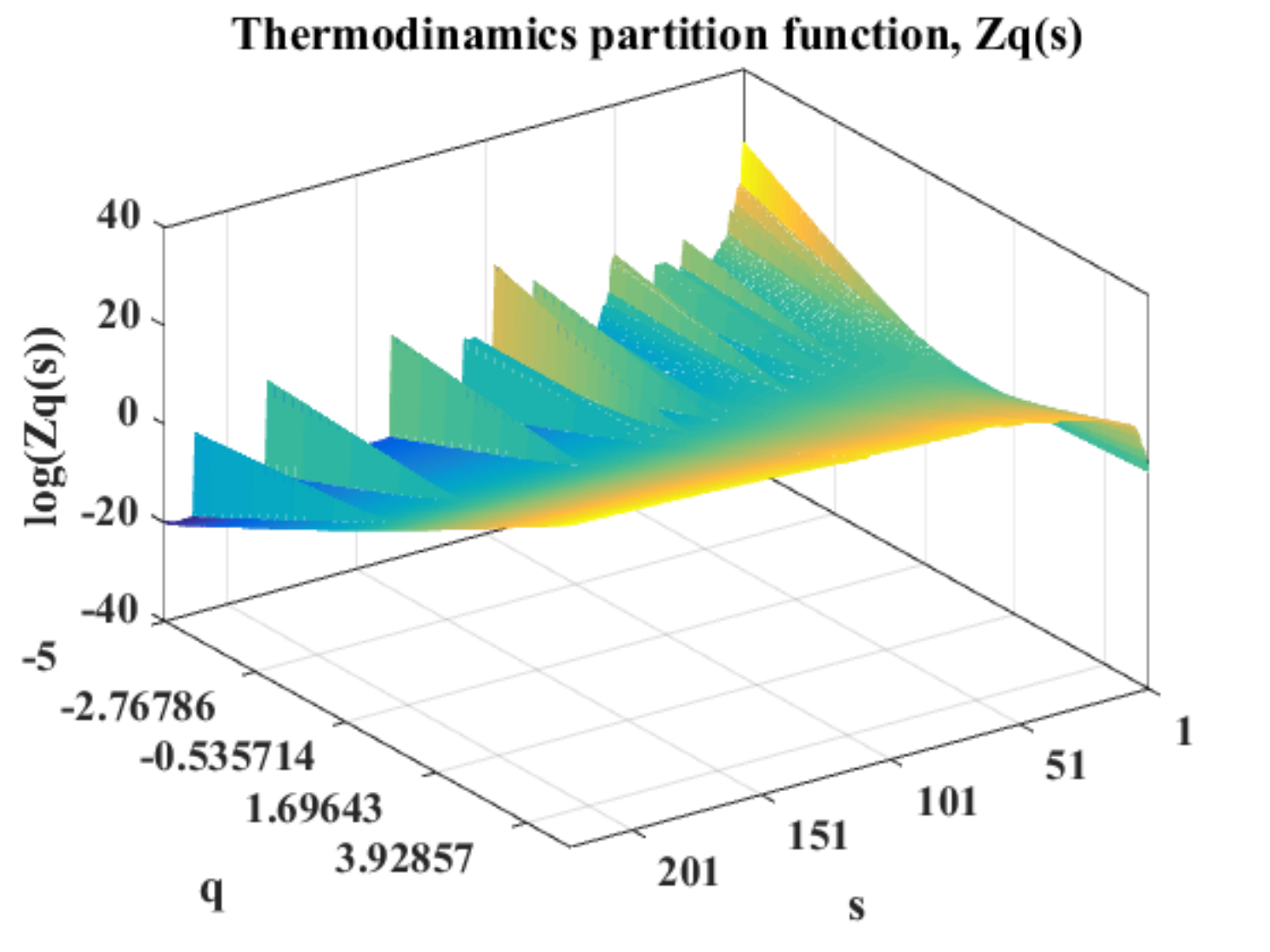} 
\includegraphics[scale=0.5]{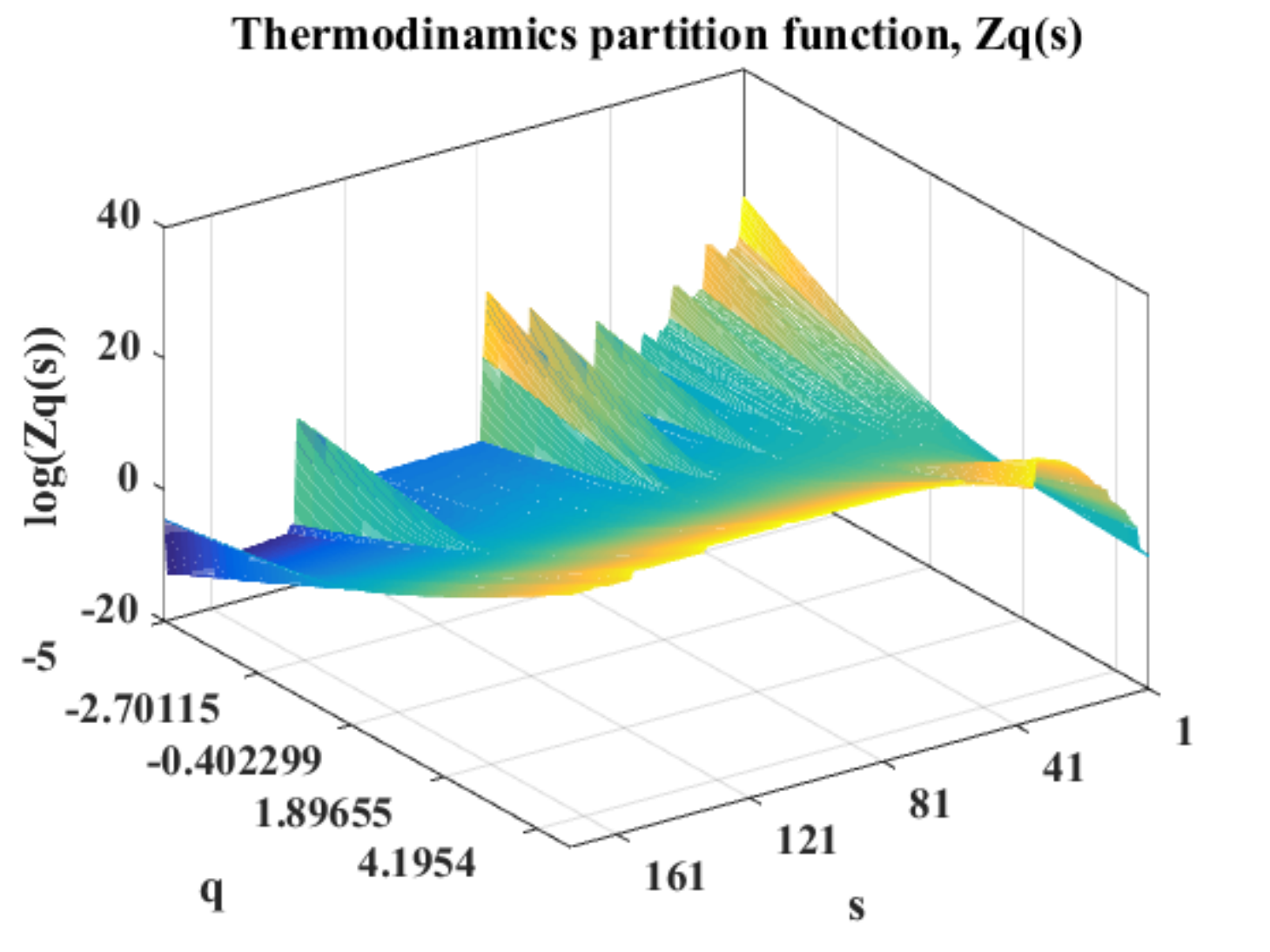} 
\includegraphics[scale=0.5]{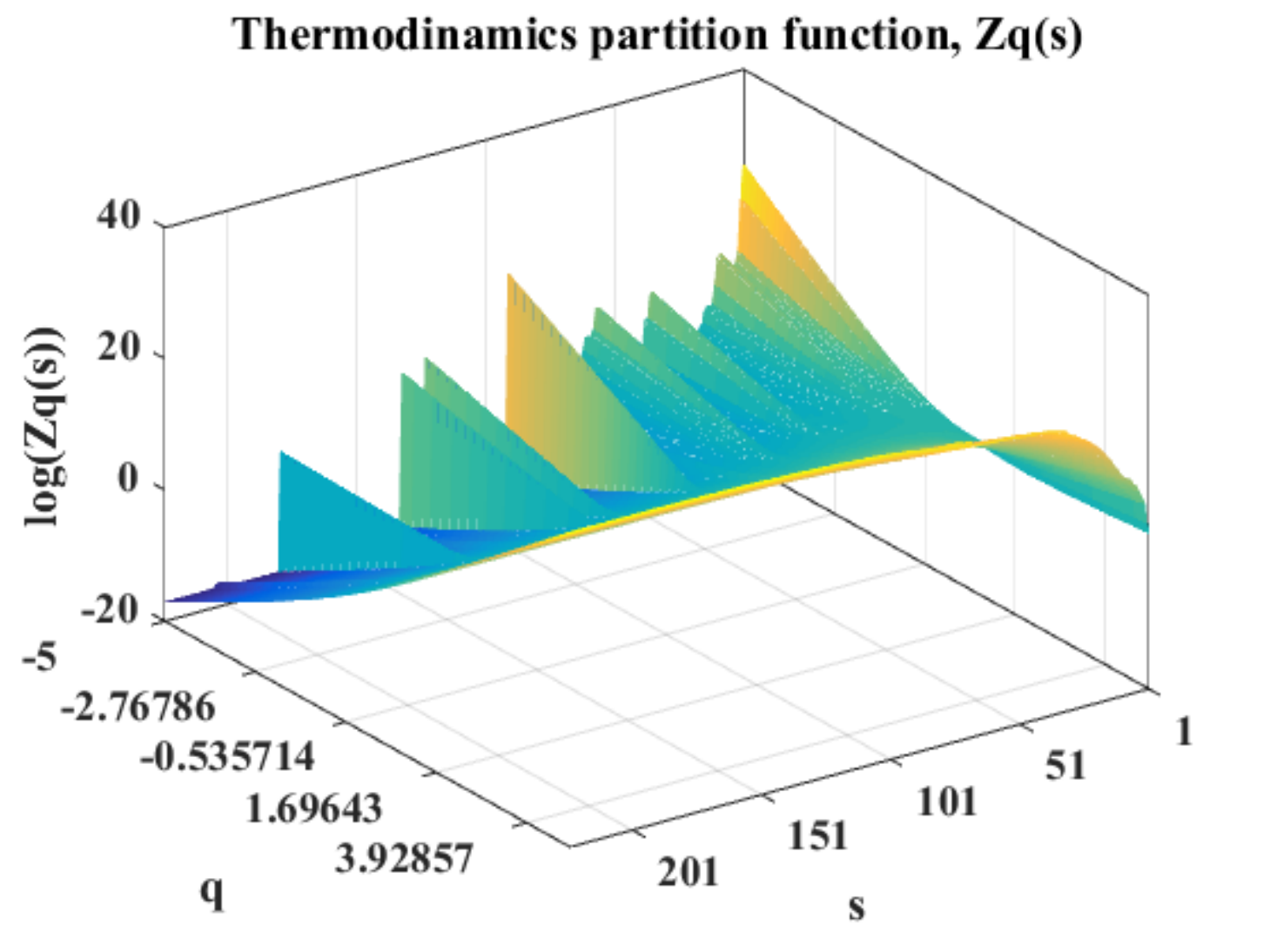} 
\caption{Upper panel: The thermodynamic partition functions at 37 GHz for the sources 3C 273 (left) and 3C 279 (right). Lower panel: The thermodynamic partition functions at 37 GHz for the sources 3C 345 (left) and 3C 454.3 (right).}
\label{fig7}
\end{figure*}

\begin{figure*}
\centering
\includegraphics[scale=0.5]{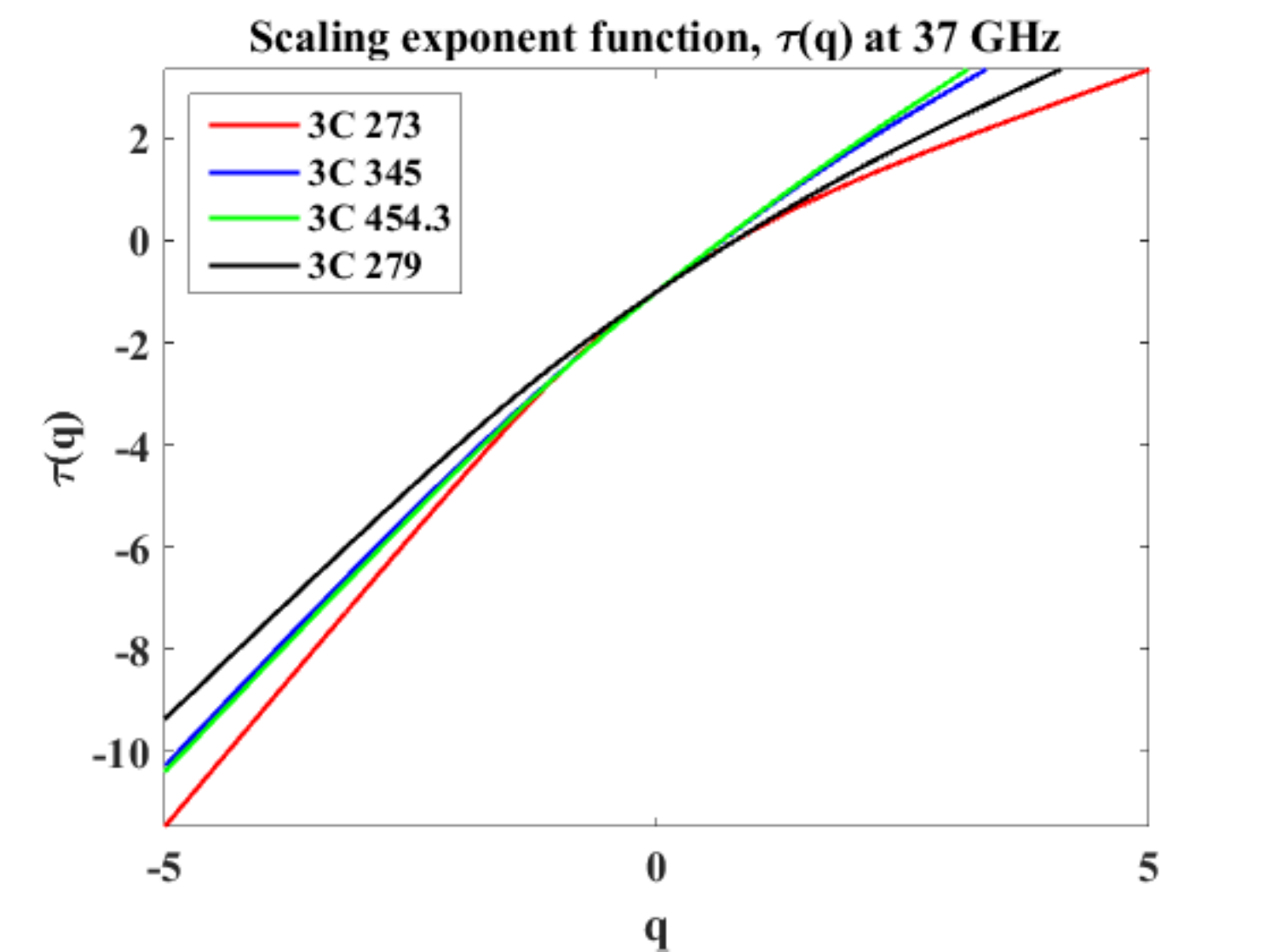}  
\includegraphics[scale=0.5]{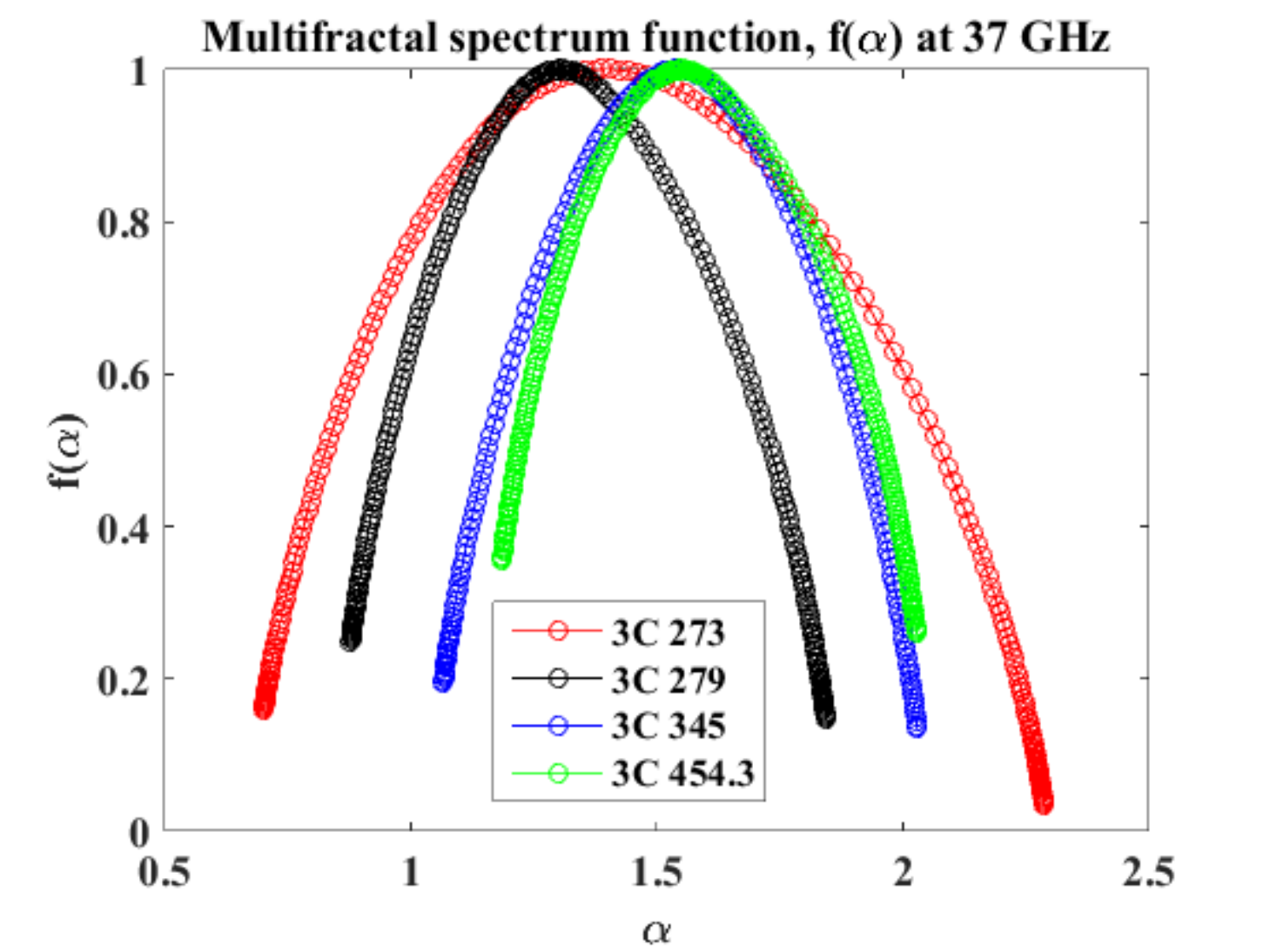}
\caption{The scaling exponent (left) and multifractal spectrum (right) at 22 GHz for the sources 3C 273, 3C 279, 3C 345, and 3C 454.3.}
\label{fig8}
\end{figure*}

The similarity and difference observed in the degree of nonlinearity at 22 and 37 GHz for each source and between the sources, respectively, provide us with valuable information about the radiation region and mechanisms of the sources considered. The similarity in the slope at those frequencies tells us that the signals of each source at 22 and 37 GHz fluctuate in a similar fashion.
The difference in the behavior of the light curves between the sources at both frequencies is clearly visible, and therefore, our finding of different degrees of nonlinearity or multifractality between the sources is somewhat expected.\\

\begin{figure*}
\centering
\includegraphics[scale=0.5]{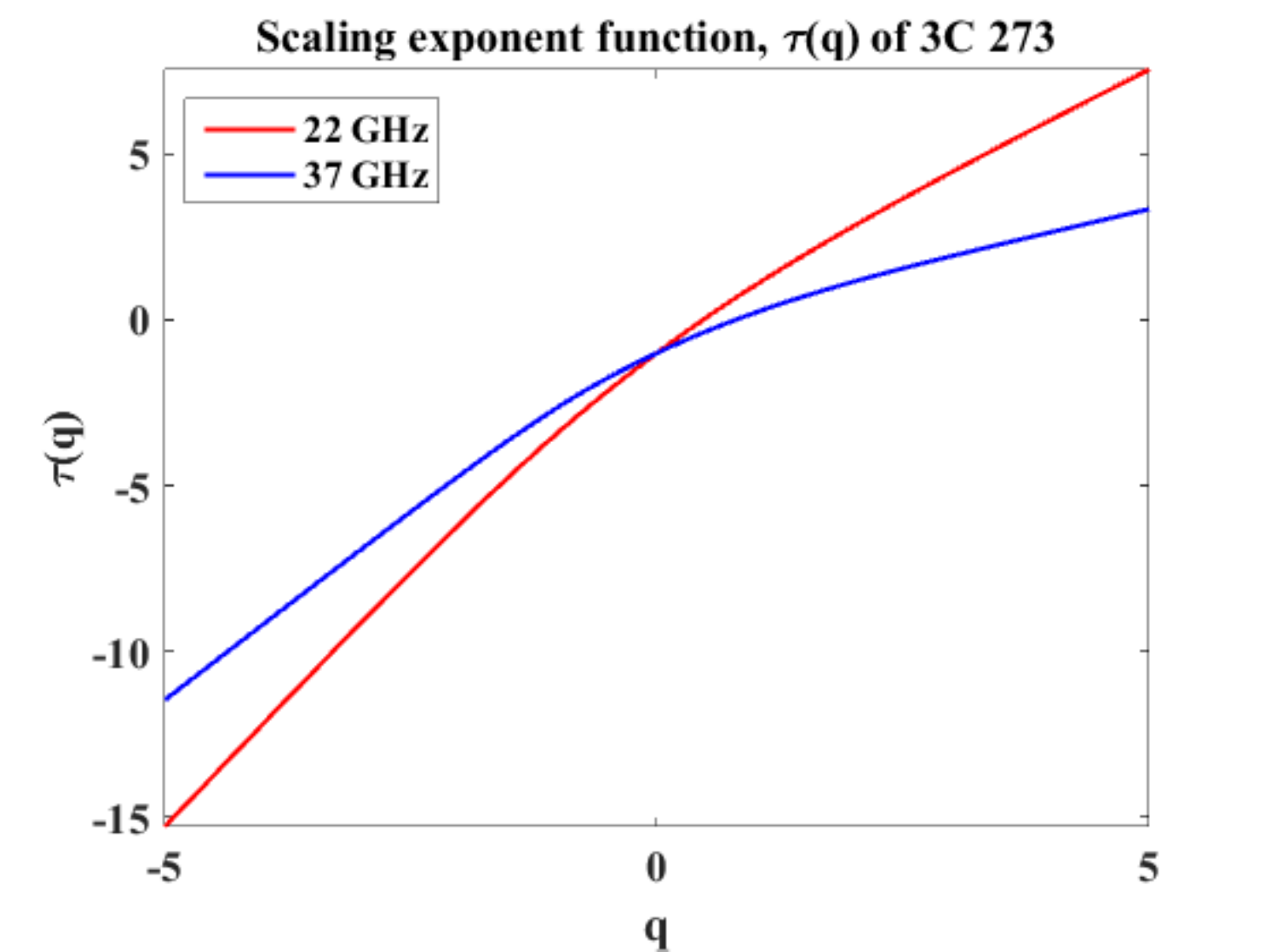}  
\includegraphics[scale=0.5]{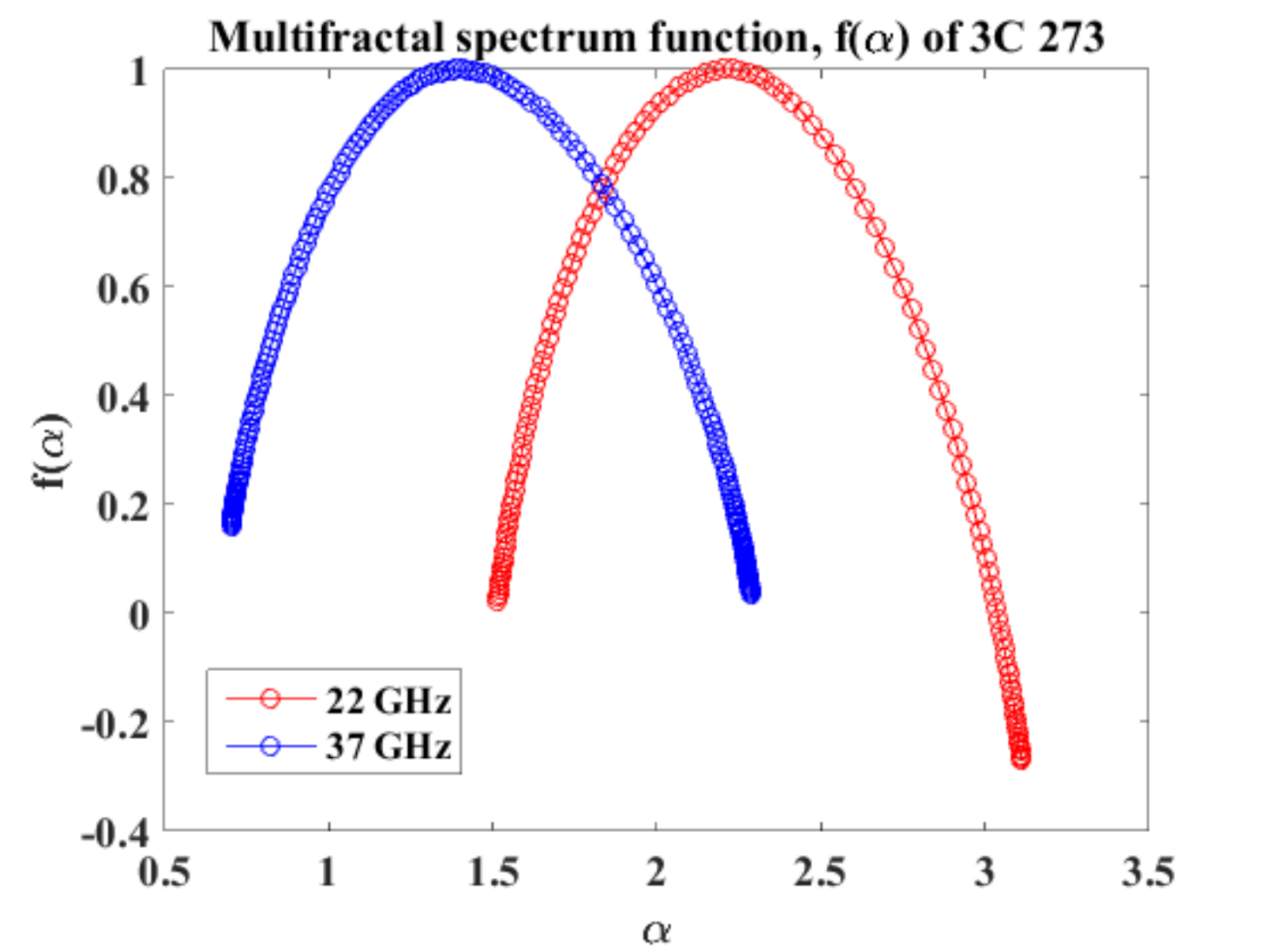}
\includegraphics[scale=0.5]{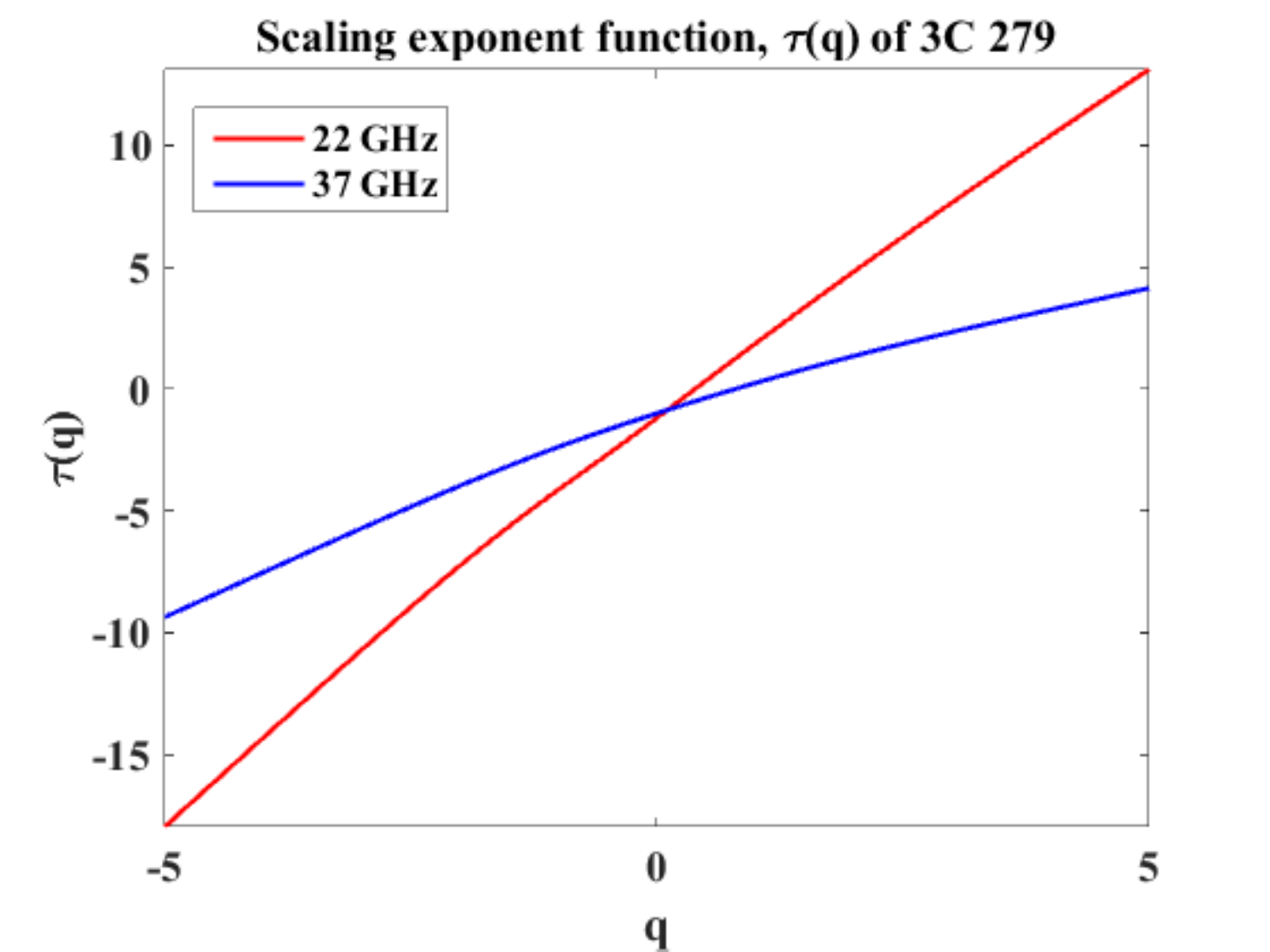}  
\includegraphics[scale=0.5]{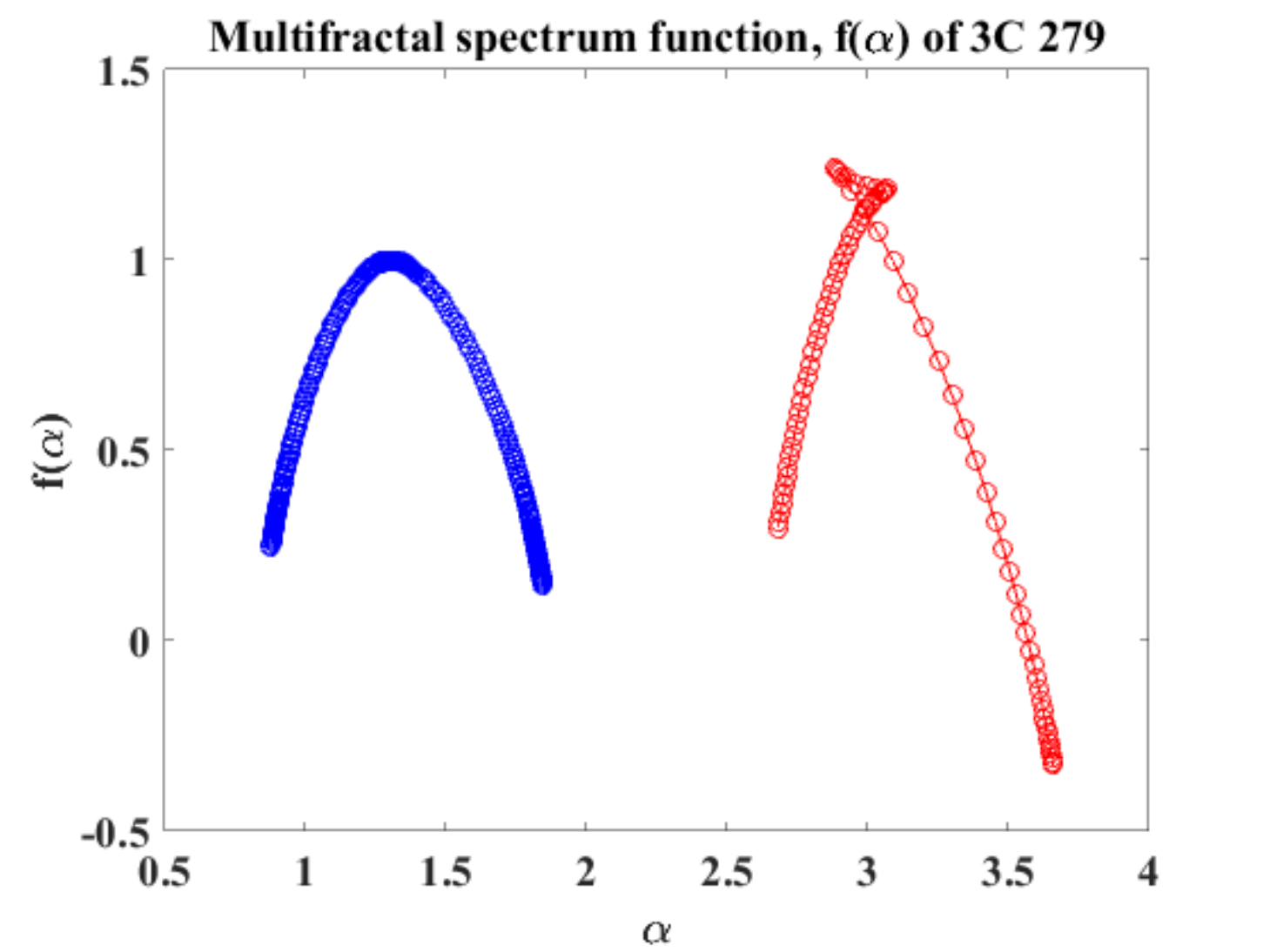}
\includegraphics[scale=0.5]{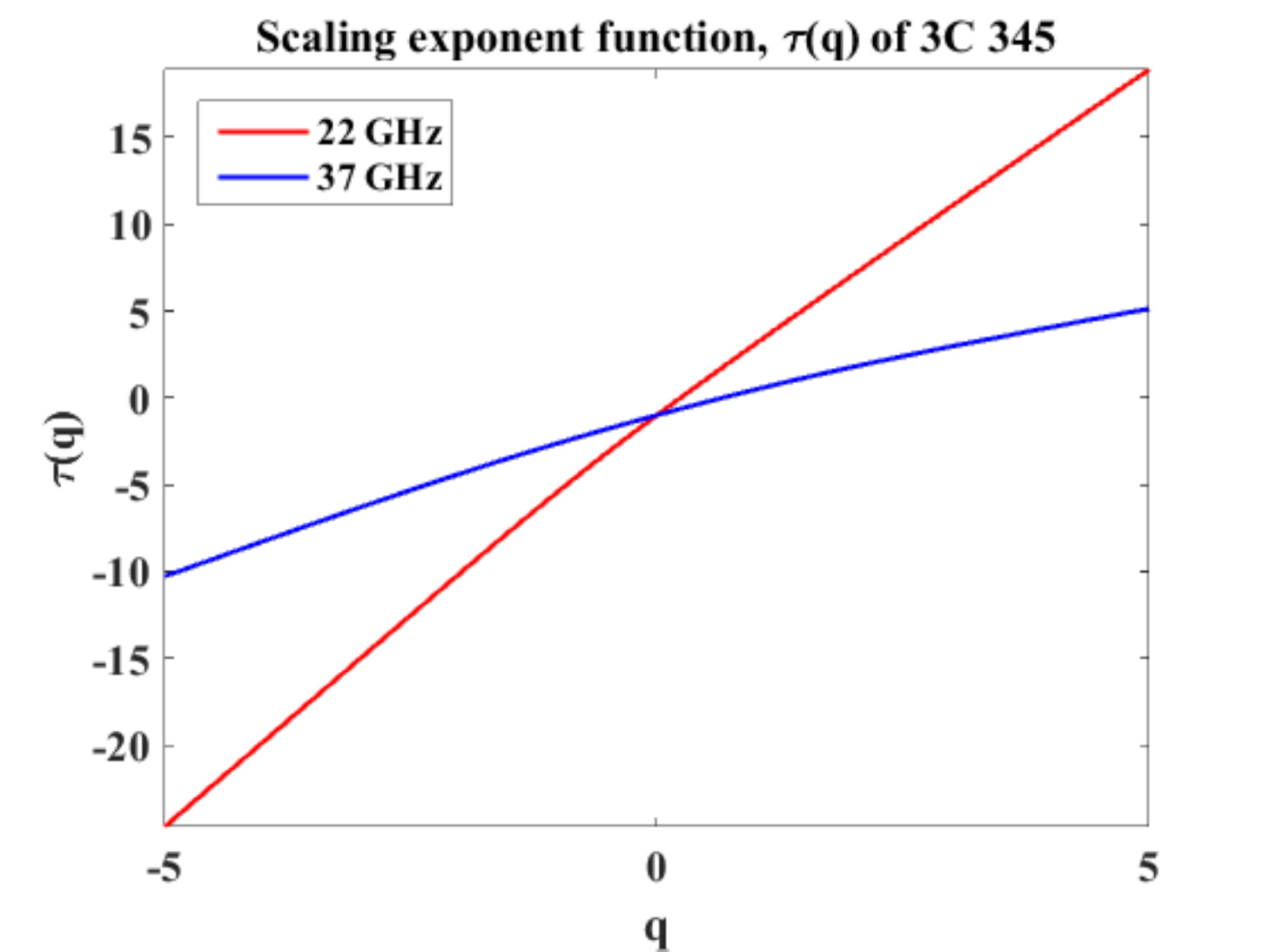}  
\includegraphics[scale=0.5]{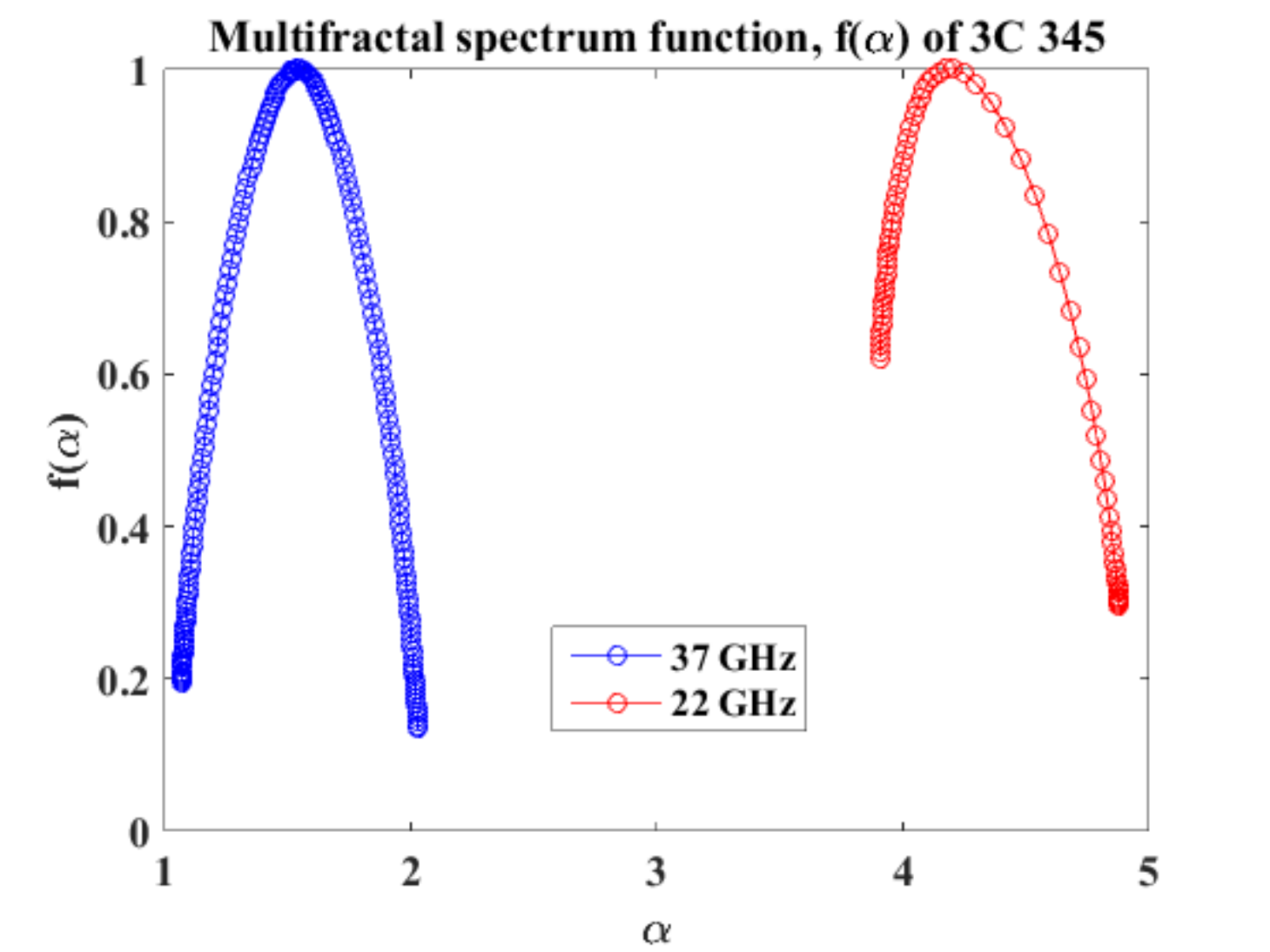}
\includegraphics[scale=0.5]{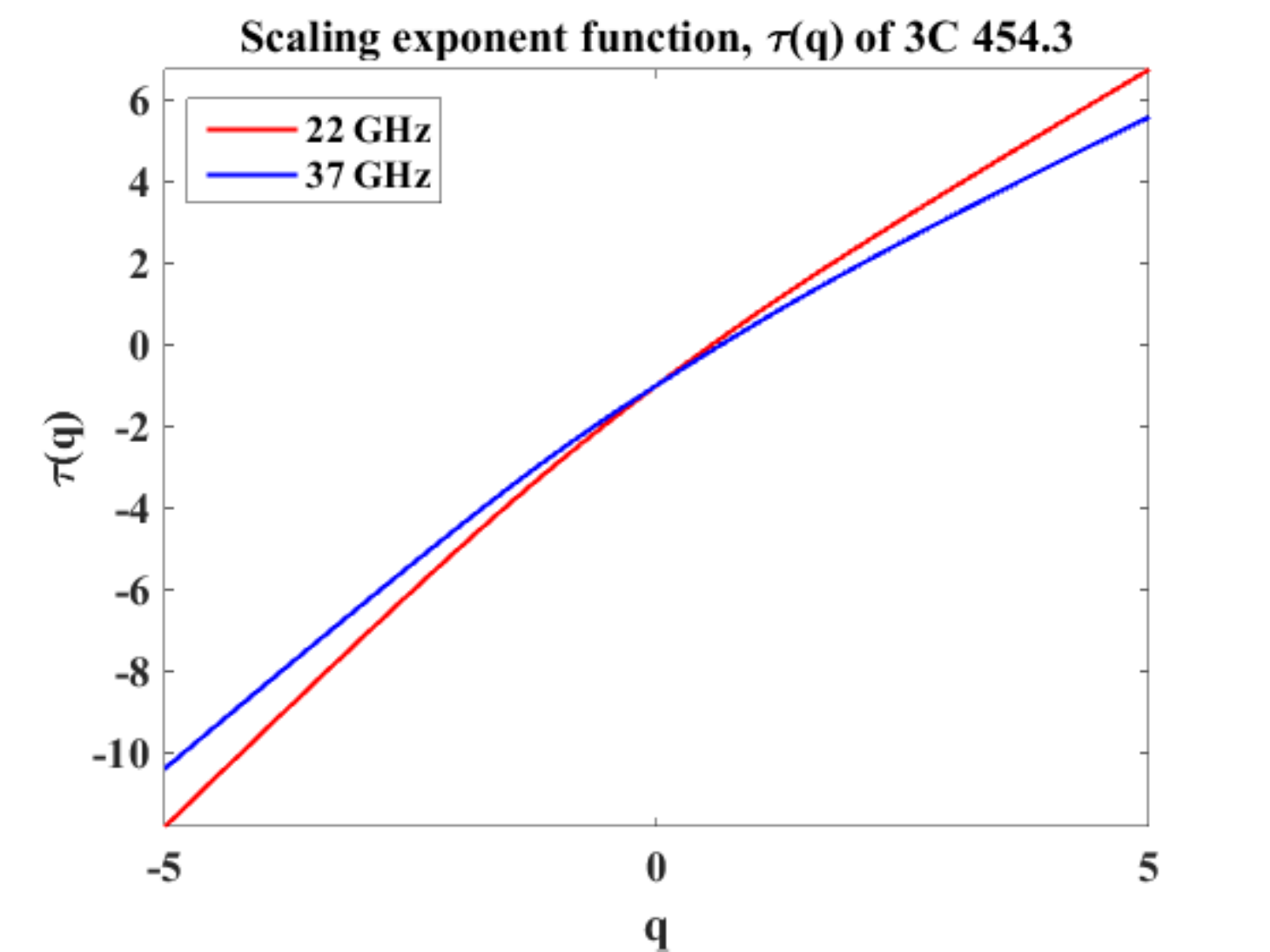}  
\includegraphics[scale=0.5]{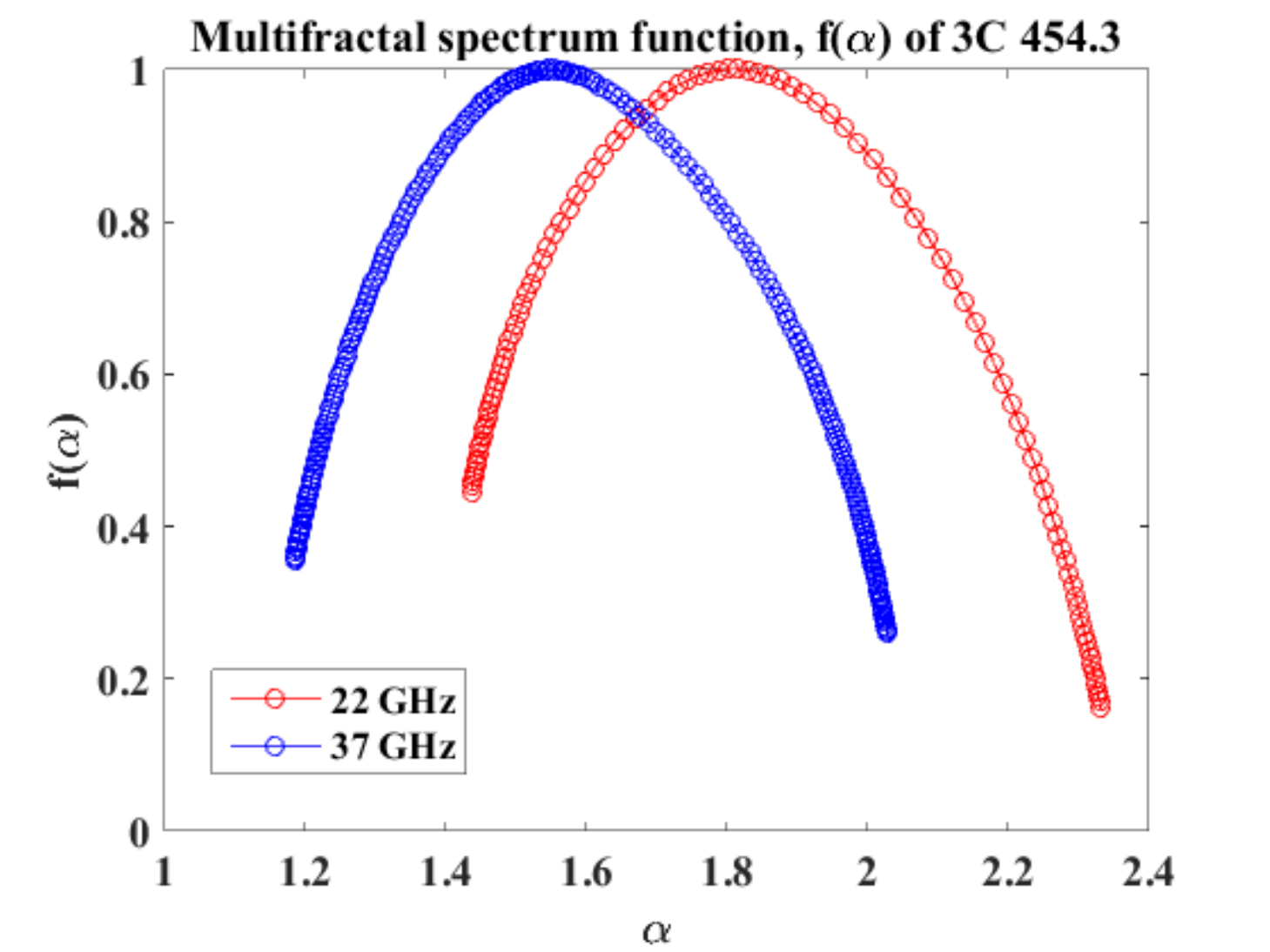}
\caption{The scaling exponent (left) and multifractal spectrum (right) at 22  and 37 GHz for the sources 3C 273, 3C 279, 3C 345, and 3C 454.3 (top to bottom), respectively.}
\label{fig9}
\end{figure*}
Indeed, the scale parameter for a time series of given length changes if the length of the time series changes, i.e., the choice of the scale parameter is dependent on the length of the time series. As seen from the calculated width values for each light curve at both bands, we found that the multifractality (nonlinearity) strength at 22 and 37 GHz is strongly similar though not the same for each source and differs from source to source. The similarity in the slope indicates that the sources have the same radiation region and radiation mechanism at the two frequencies as indicated by \citet{2010ChA&A..34..343W} for different light curves of 3C 273 and 3C 345. However, the data streams currently at hand are not of identical length, and the similarity observed at 22 and 37 GHz for each source might be affected by the difference in the data length at these two frequencies. Importantly, when we have full light curves (large data points) at 22 GHz, the scale parameters that we used to obtain better maxima lines for the present data may change based on the change in the singularity behavior of the light curves in time. Consequently, we may have different local maxima lines that affect the current relationship between the scaling exponent function $\tau$($q$) and the moment $q$. This in turn affects the behavior of the multifractality spectrum function $f$($\alpha$).
 Of course, still, there is difference in nonlinearity at 22 and 37 GHz for all the sources, though not strong, and this could be due to the difference in the number of data points. For the difference in nonlinearity between the sources at both bands, for example, we analyzed the light curves of the sources 3C 273 and 3C 279 at 37 GHz, where there are only 7 data points difference between them using the same scale parameter, and found strongly different results. For the time being, since we are considering scale parameters that give us better maxima lines, informative scale parameters, it is somewhat logical to perform a comparison between the sources in terms of the degree of multifractality at the two frequencies based on what we have on hand. In general, despite the difference in the number of data points (time series length) between the sources, all their light curves at 22 GHz span from 1980 - 2004 and from 1979/1980 -2018 at 37GHz, and therefore, the results obtained using the chosen informative scale parameters could inform us at least about how the multifractal (nonlinear) behavior of the light curves at the two radio bands behave throughout these observation periods.

\subsection{Analysis of light curves in the rest frame:}
Following the same procedures as applied for light curves in the observation frame, we have repeated the same multifractality analysis for the corresponding light curves in the rest frame, as shown in Fig. \ref{fig2}. We obtained the same results as shown in Figs. \ref{fig5} and \ref{fig8} for all the light curves. The similarity in the degree of multifractality (nonlinearity), $\Delta\alpha_{obs} = \Delta\alpha_{rest}$,  of the  light curves in the observation and rest frames implies that redshift-correction does not affect the multifractal behavior of quasars radio emissions, indicating that multifractality is an intrinsic behavior of quasars radio emissions. The redshift versus multifractality strength is given in Fig. \ref{fig7}.


\begin{table}
\caption{The calculated multifractal spectrum width ($\Delta\alpha$) for each source at 22 and 37 GHz.} 
\centering 
\begin{tabular}{l c c rrrrrrr} 
\hline\hline 
Source name   &  Red-shift (z)  &  Observation frame frequency&$\Delta$ $\alpha$
\\ [0.5ex]
\hline 
 & &22 GHz   &  1.5961  \\[-1ex]
\raisebox{1.5ex}{3C 273} & \raisebox{1.5ex}{0.158}&37 GHz
 &   1.5854 \\[1ex]
& &22 GHz & 0.9763  \\[-1ex]
\raisebox{1.5ex}{3C 279} & \raisebox{1.5ex}{0.536}& 37 GHz
& 0.9653 \\[1ex]
& &22 GHz& 0.9745 \\[-1ex]
\raisebox{1.5ex}{3C 345} & \raisebox{1.5ex}{0.595}& 37 GHz
& 0.9633  \\[1ex]
& &22 GHz& 0.8951  \\[-1ex]
\raisebox{1.5ex}{3C 454.3} & \raisebox{1.5ex}{0.859}& 37 GHz
& 0.8432 \\[1ex]
\hline 
\end{tabular}
\label{tab2}
\end{table}

\begin{figure*}
\centering
\includegraphics[scale=0.6]{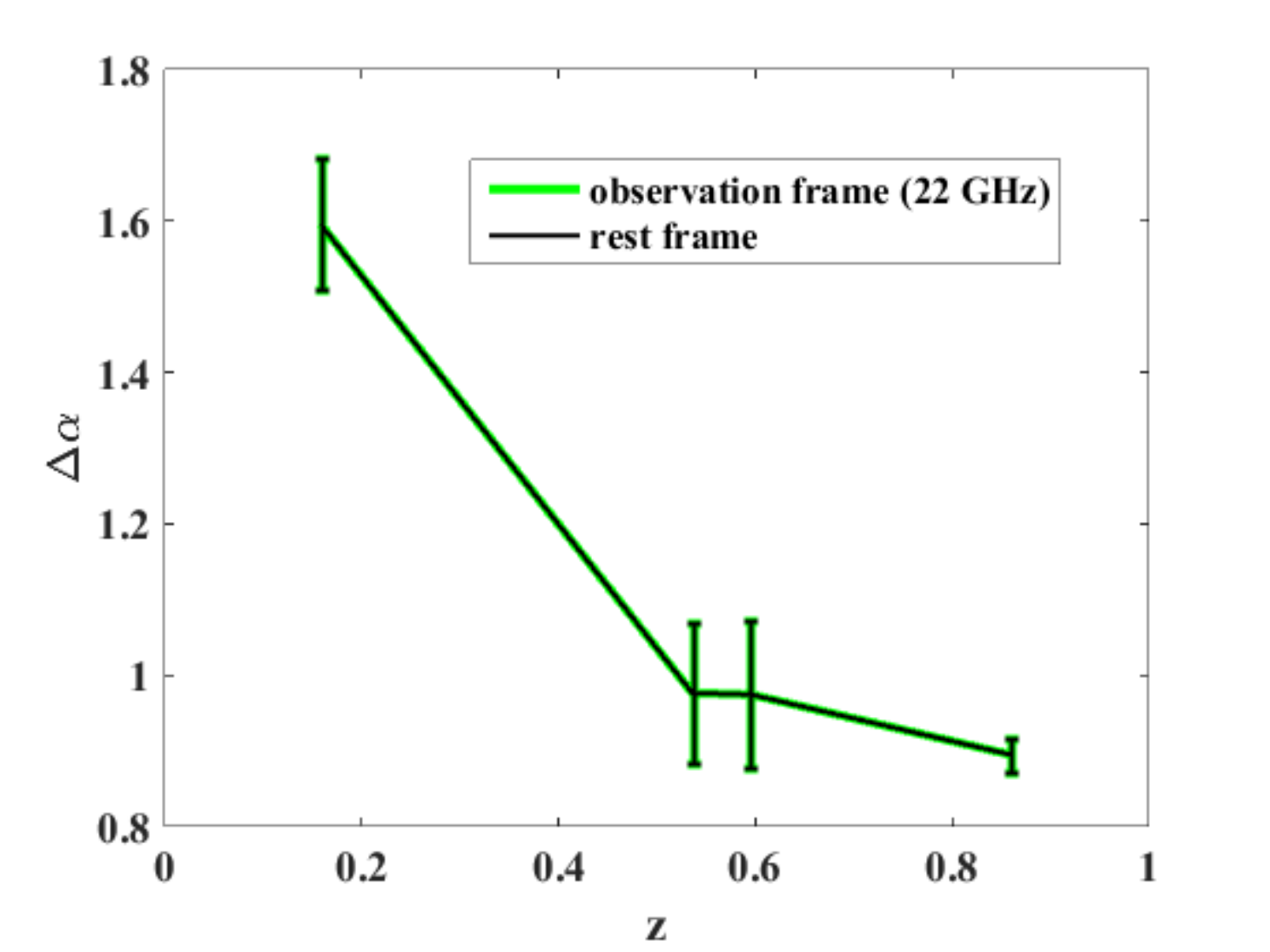}  
\includegraphics[scale=0.6]{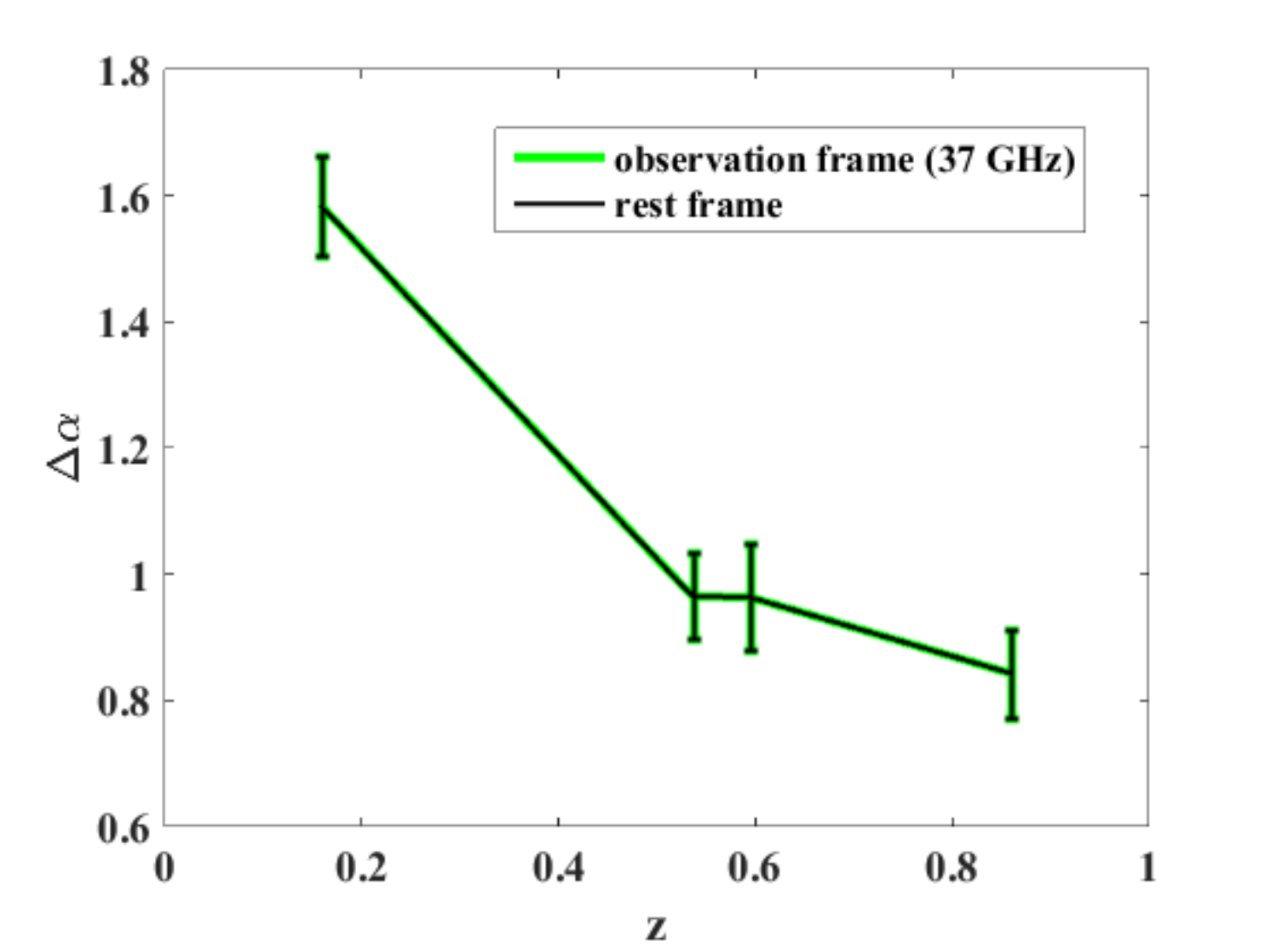}  
\caption{The degree of multifractality $\Delta\alpha$ in the observed, 22 GHz (left) and 37 GHz (right),  and rest frames for 3C 273 ($z$ = 0.158), 3C 279 ($z$ = 0.536), 3C 345 ($z$ = 0.595), and 3C 454.3 ($z$ = 0.859).}
\label{fig7}
\end{figure*}

\subsection{Analysis of the ARIMA(p,d,q) models}
\begin{figure*}
\centering
\includegraphics[scale=0.75]{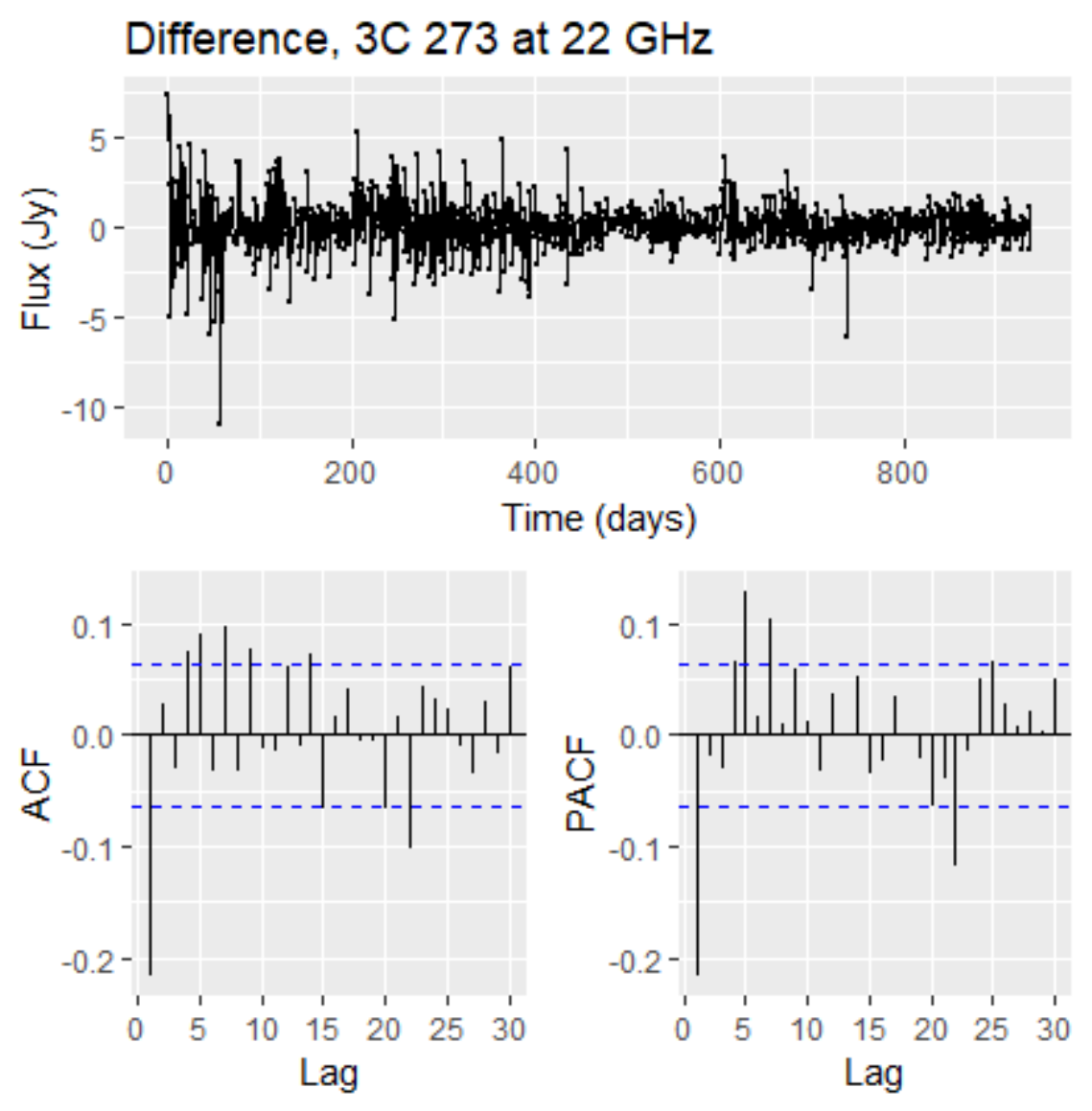}  
\includegraphics[scale=0.75]{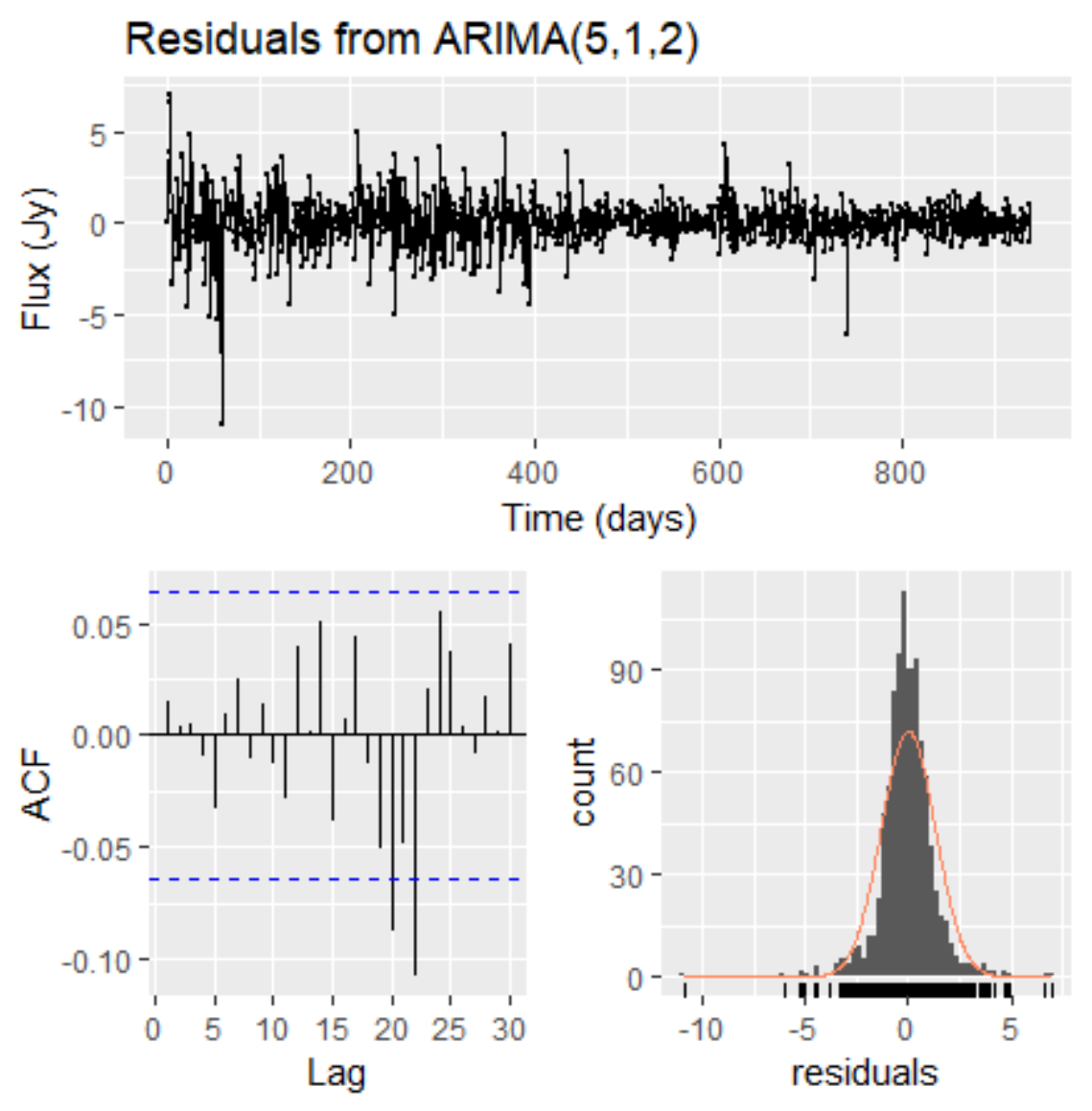}
\includegraphics[scale=0.75]{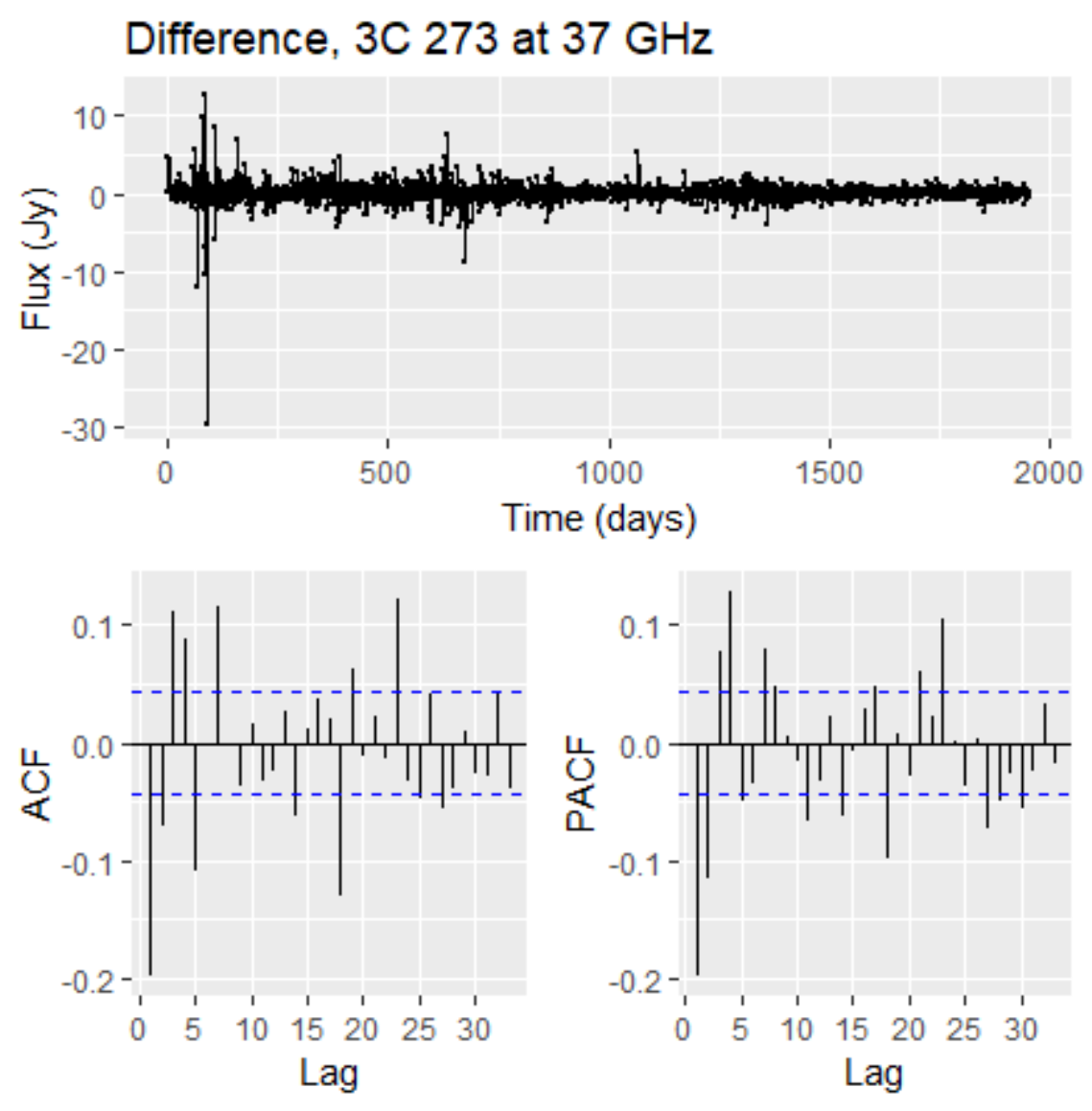}  
\includegraphics[scale=0.75]{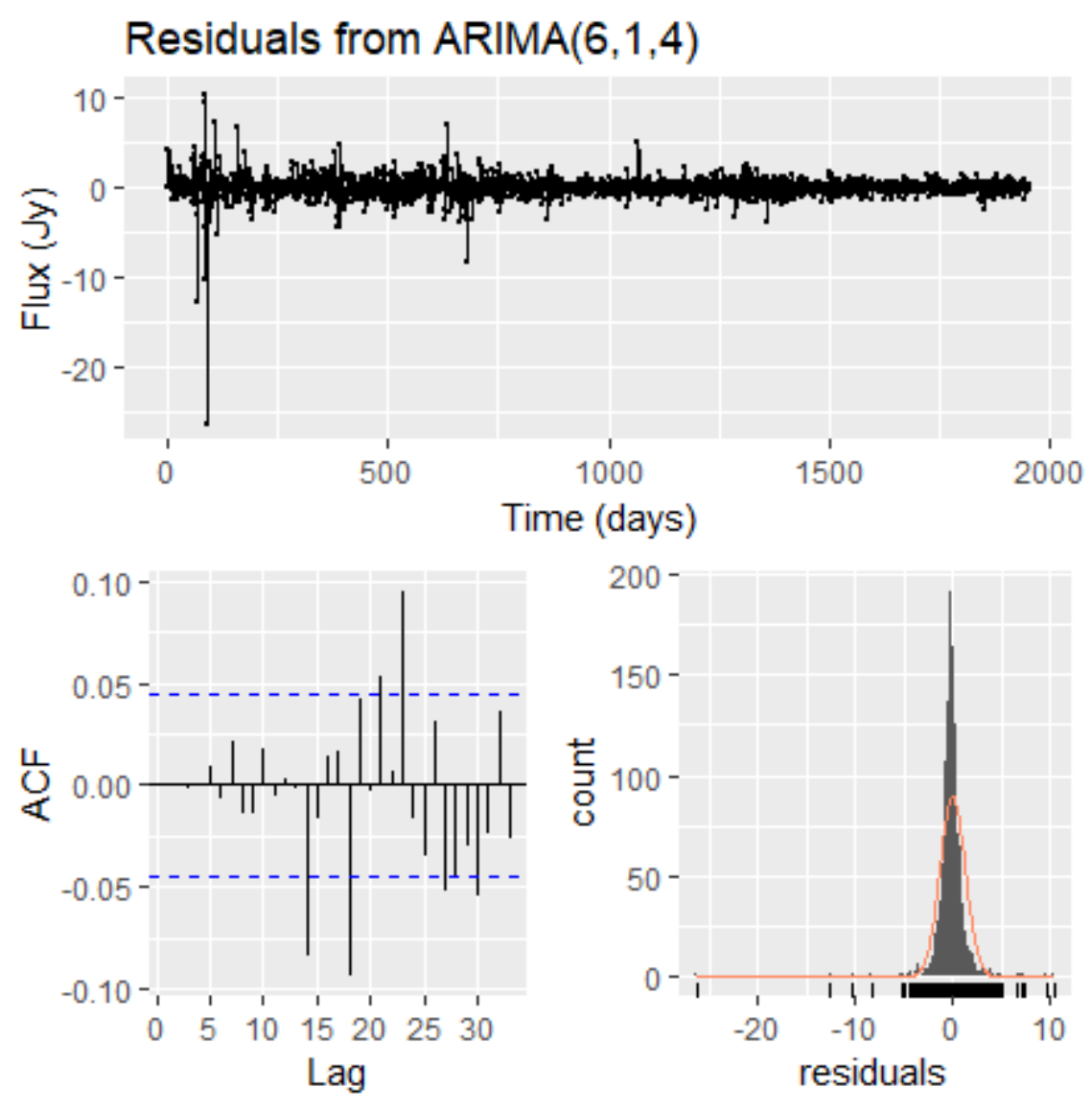}
\caption{Time plot of the differenced time series and its ACF and PACF plots (on the left of the first and second panels, respectively) and the corresponding residual plots (on the right of the first and second panels) for 3C 273 at 22 GHz . Similarly, time plot of the differenced time series and its ACF and PACF plots (on the left of the third and fourth panels, respectively) and the corresponding residual plots (on the right of the third and fourth panels) for 3C 273 at 37 GHz.}
\label{fig11}
\end{figure*}

\begin{figure*}
\centering
\includegraphics[scale=0.75]{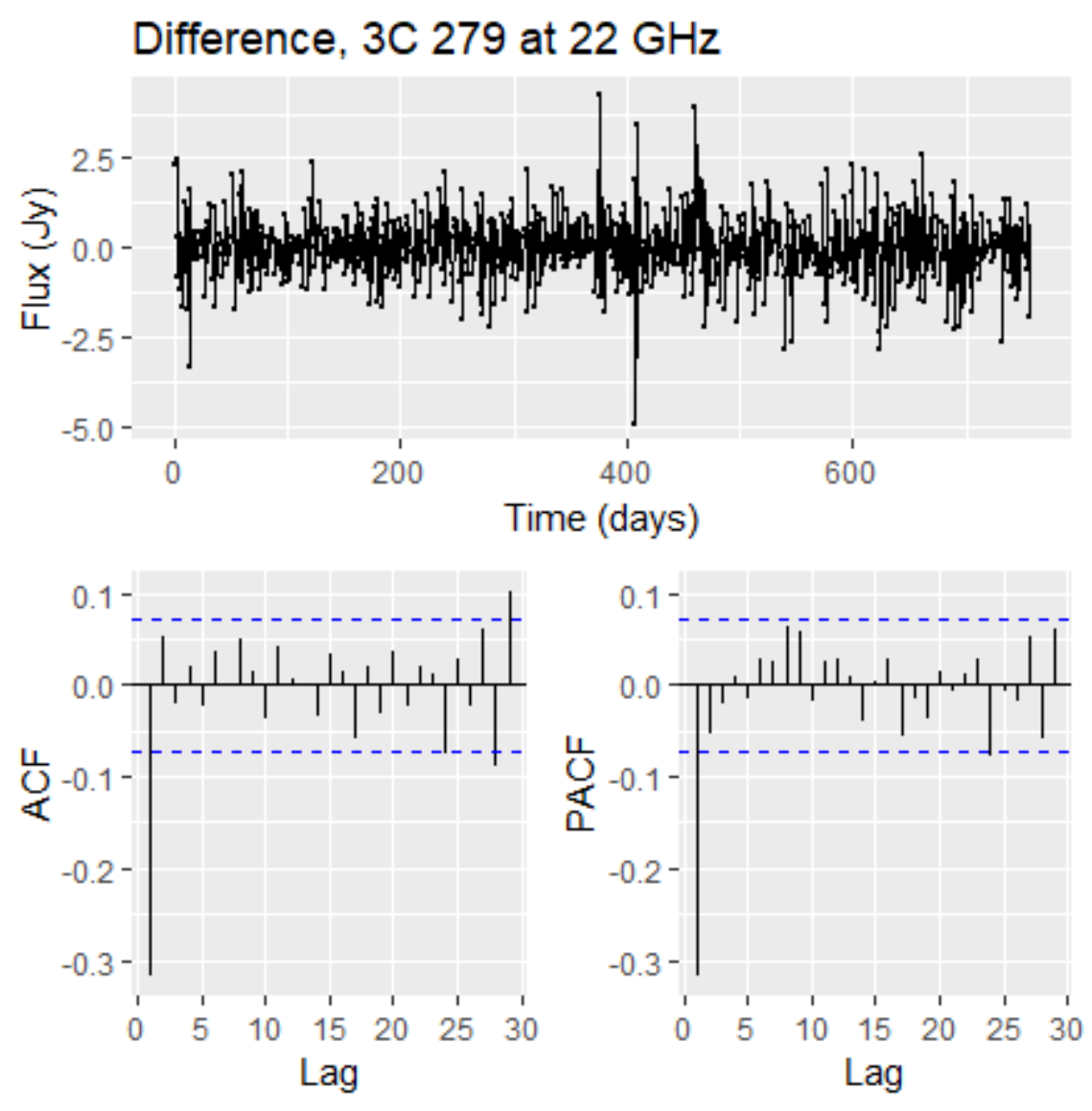}  
\includegraphics[scale=0.75]{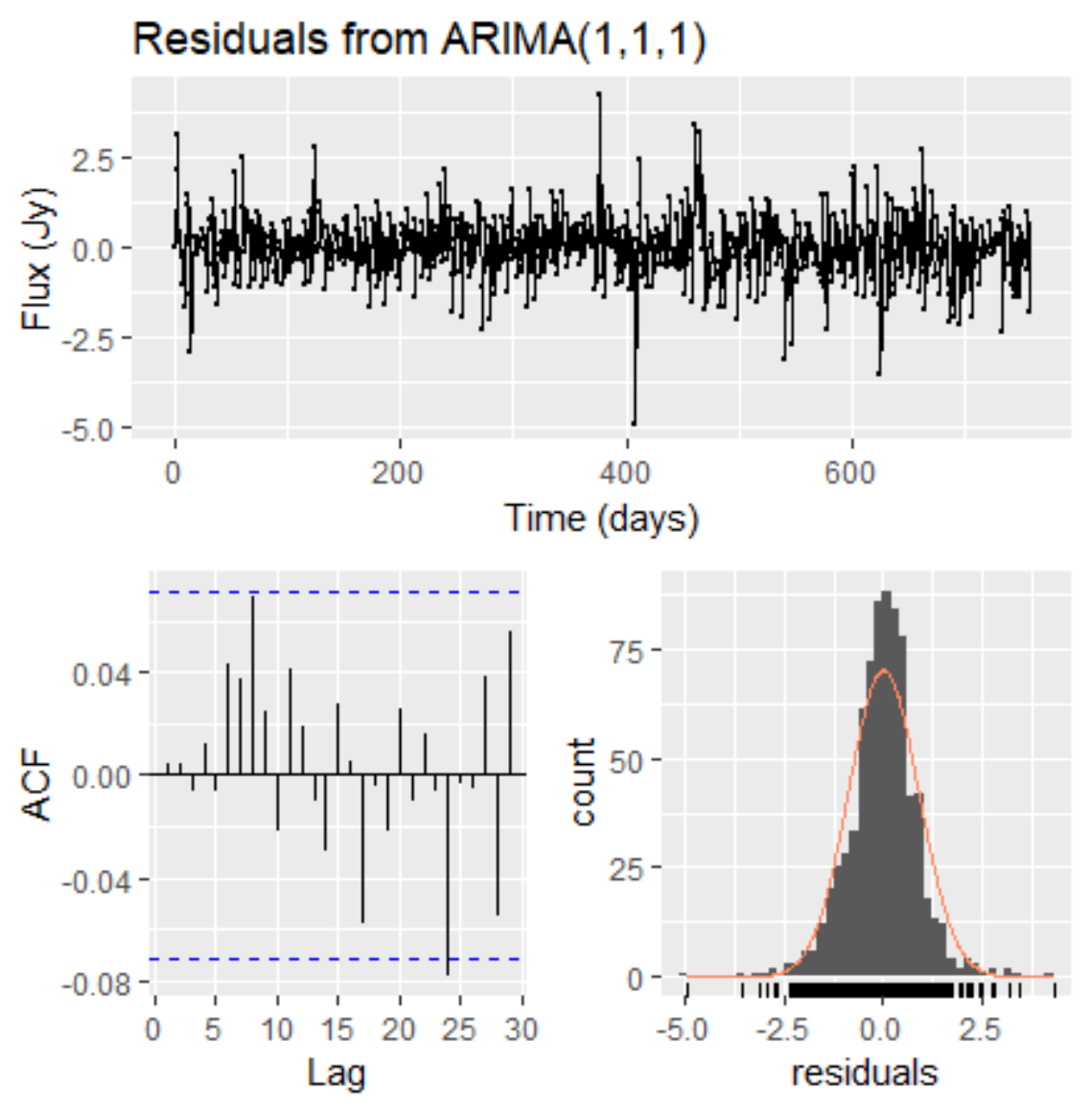}
\includegraphics[scale=0.75]{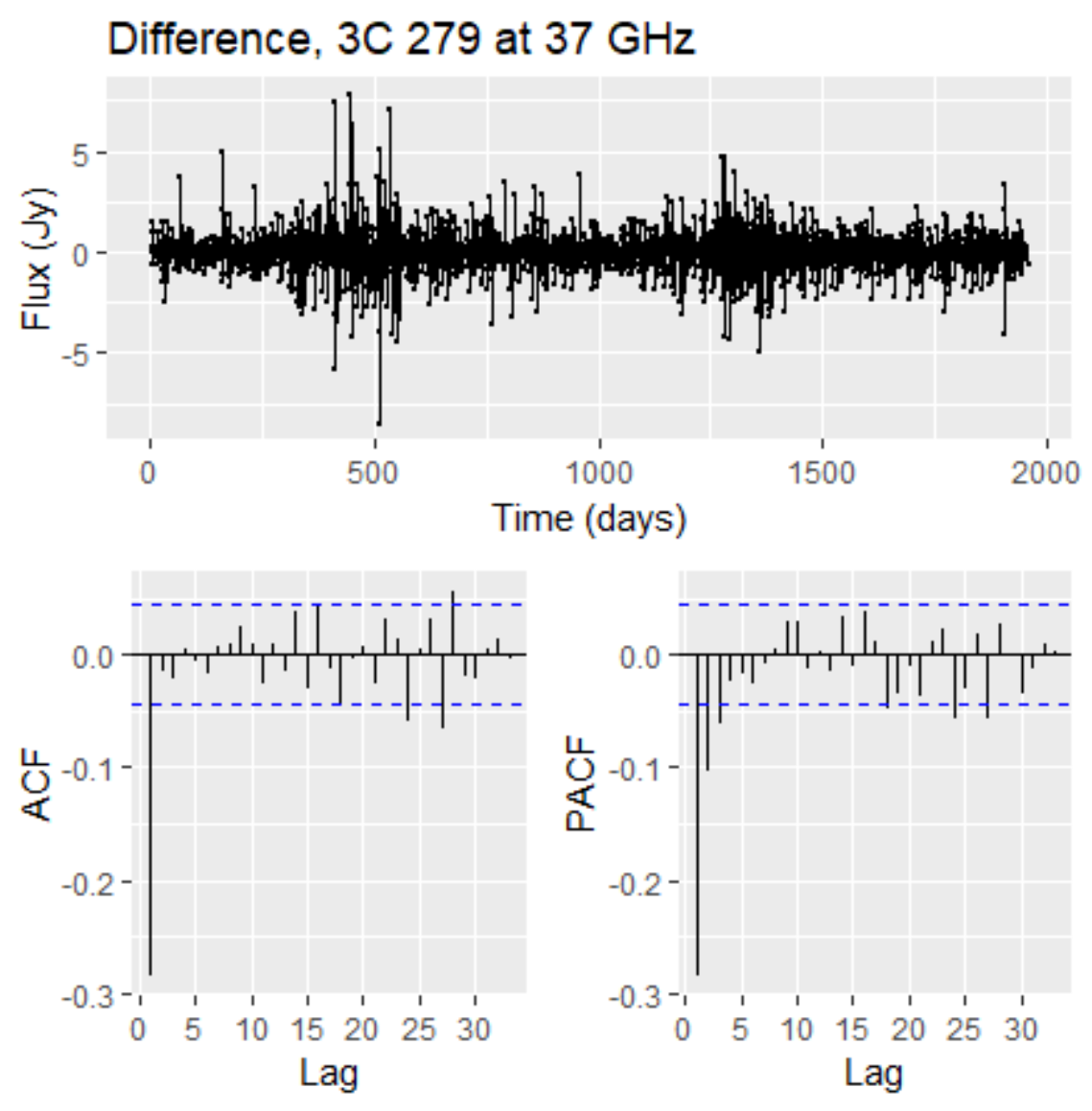}  
\includegraphics[scale=0.75]{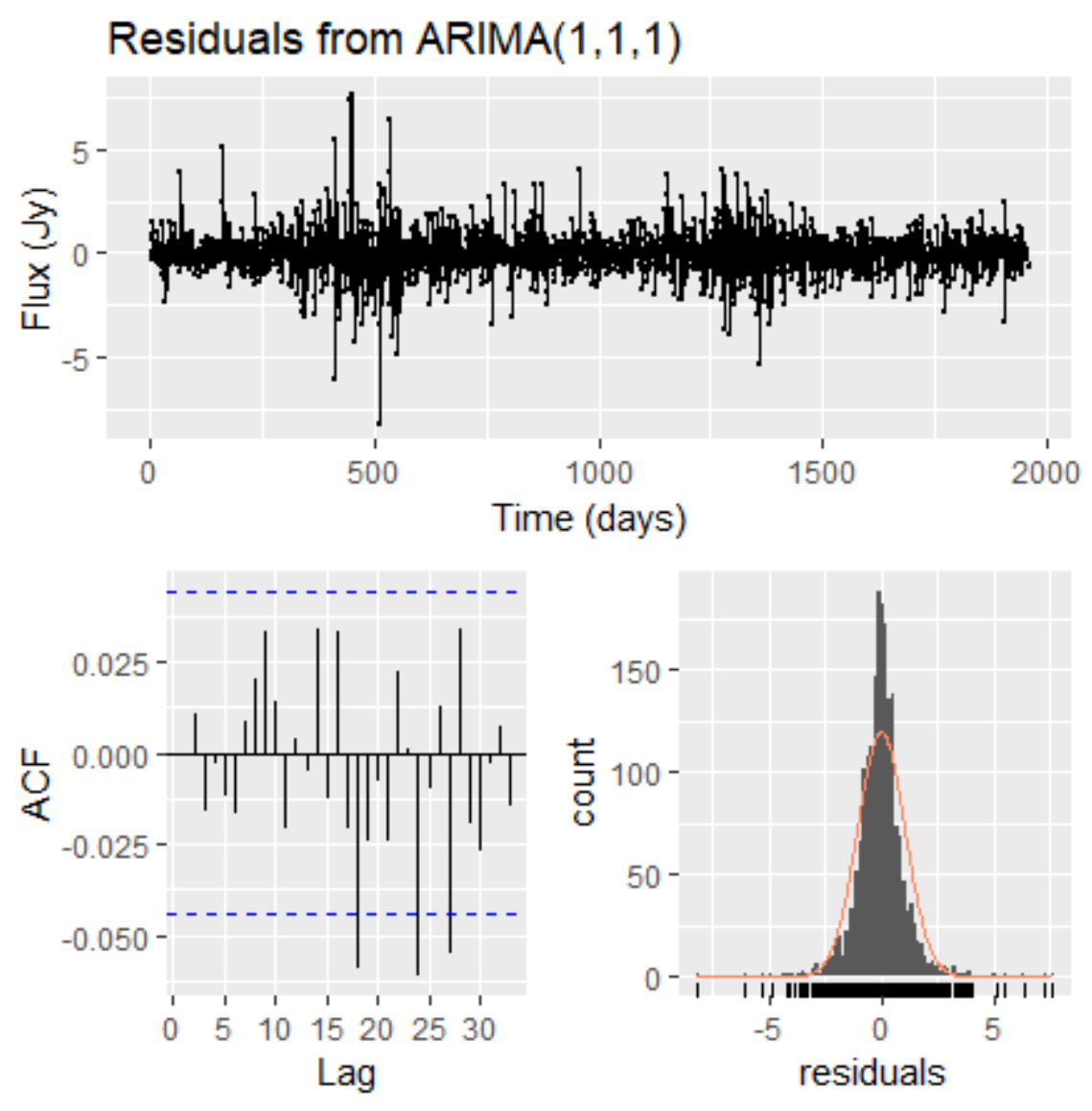}
\caption{Time plot of the differenced time series and its ACF and PACF plots (one the left of the first and second panels, respectively) and the corresponding residual plots (on the right of the first and second panels) for 3C 279 at 22 GHz . Similarly, time plot of the differenced time series and its ACF and PACF plots (one the left of the third and fourth panels, respectively) and the corresponding residual plots (on the right of the third and fourth panels) for 3C 279 at 37 GHz.}
\label{fig12}
\end{figure*}

\begin{figure*}
\centering
\includegraphics[scale=0.75]{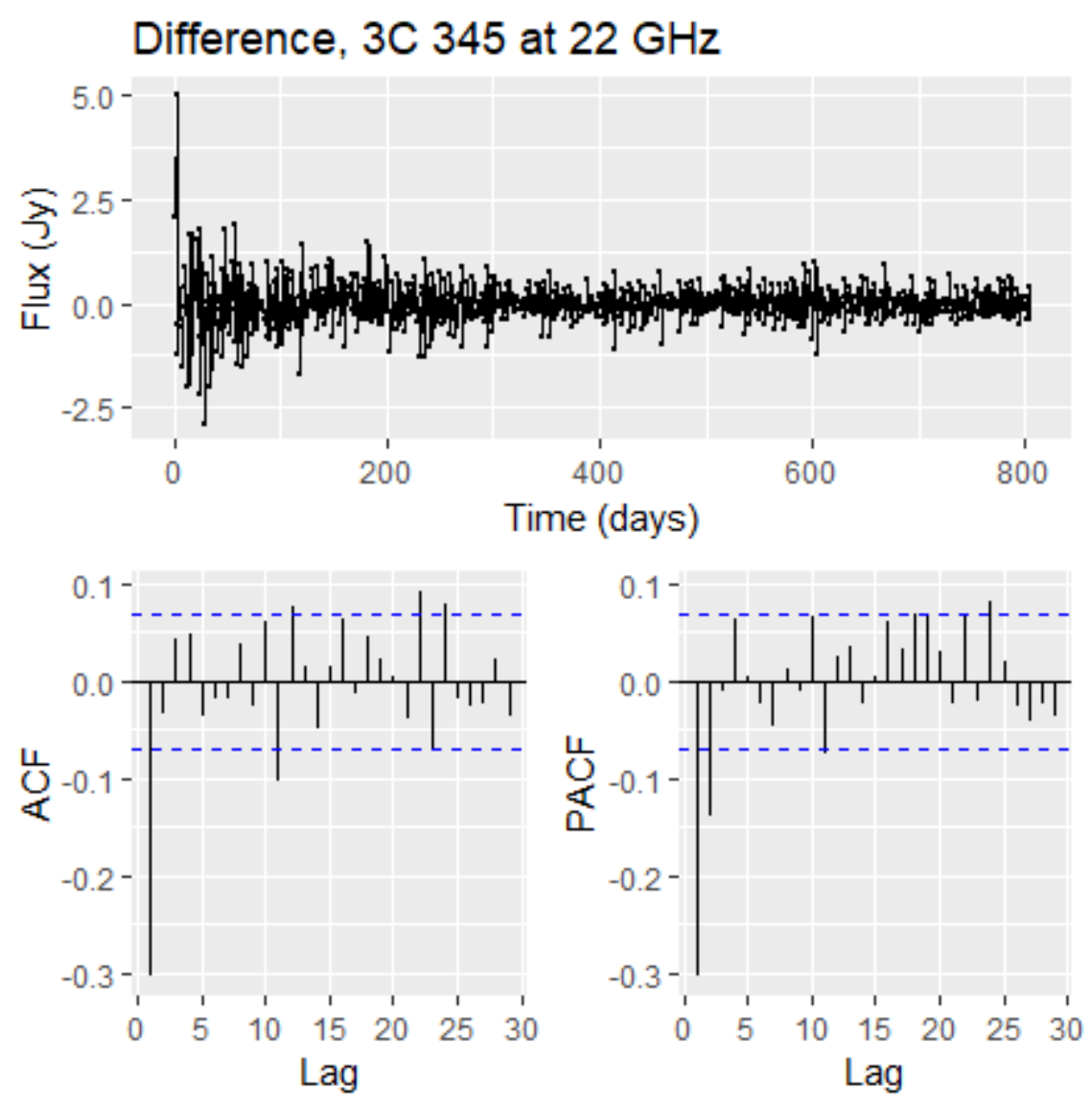}  
\includegraphics[scale=0.75]{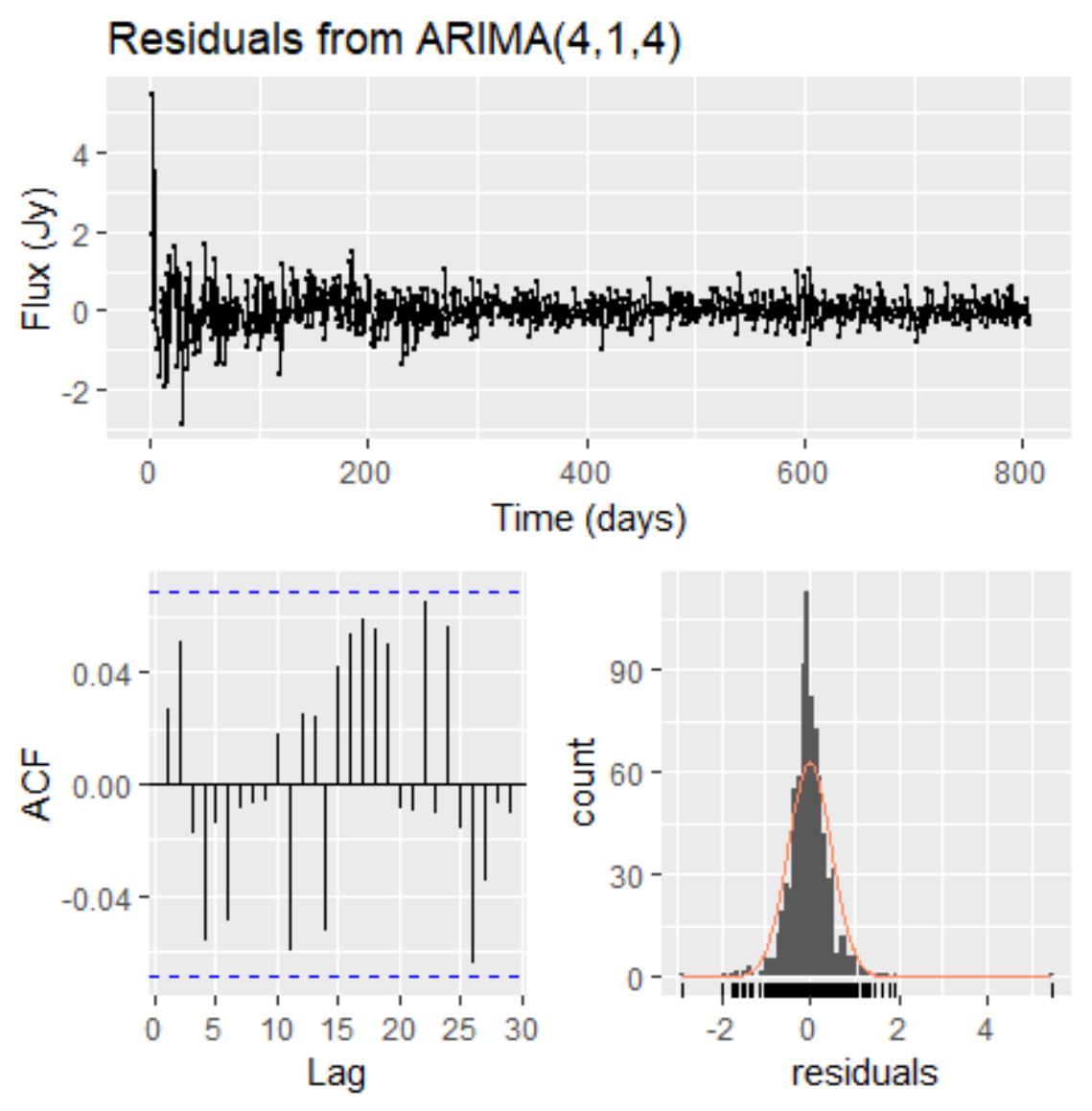}
\includegraphics[scale=0.75]{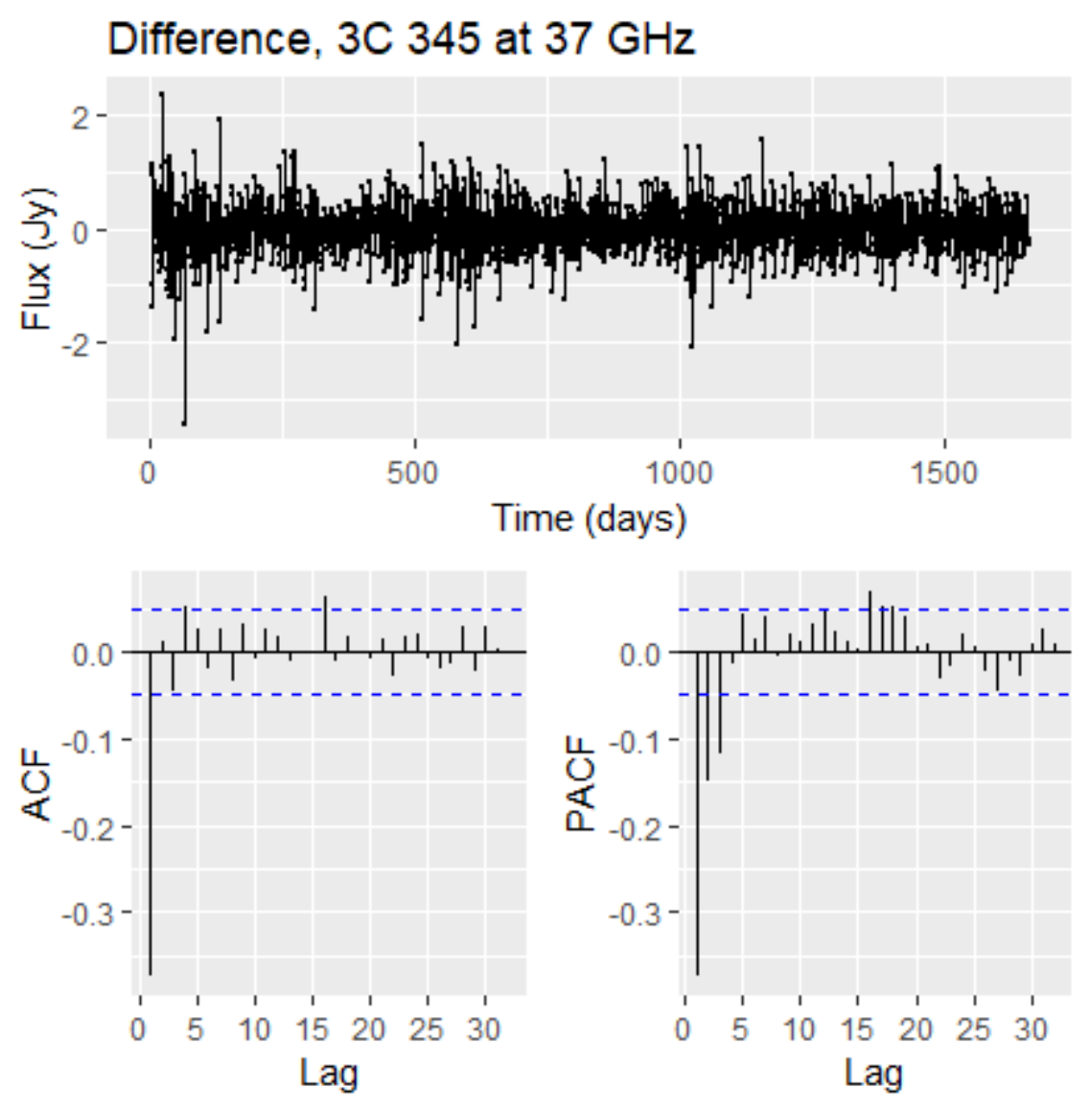}  
\includegraphics[scale=0.75]{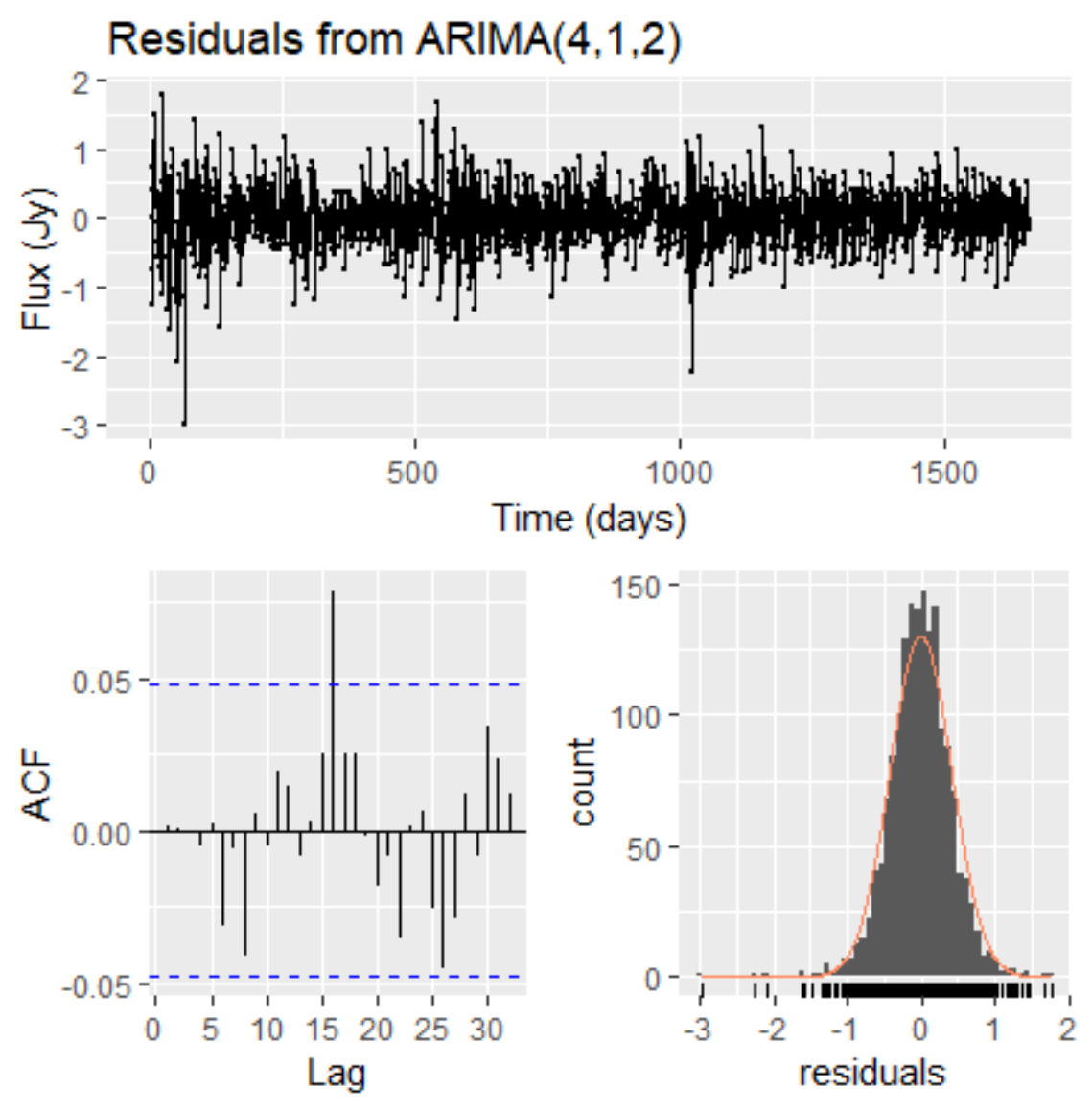}
\caption{Time plot of the differenced time series and its ACF and PACF plots (on the left of the first and second panels, respectively) and the corresponding residual plots (on the right of the first and second panels) for 3C 345 at 22 GHz . Similarly, time plot of the differenced time series and its ACF and PACF plots (on the left of the third and fourth panels, respectively) and the corresponding residual plots (on the right of the third and fourth panels) for 3C 345 at 37 GHz.}
\label{fig13}
\end{figure*}

\begin{figure*}
\centering
\includegraphics[scale=0.75]{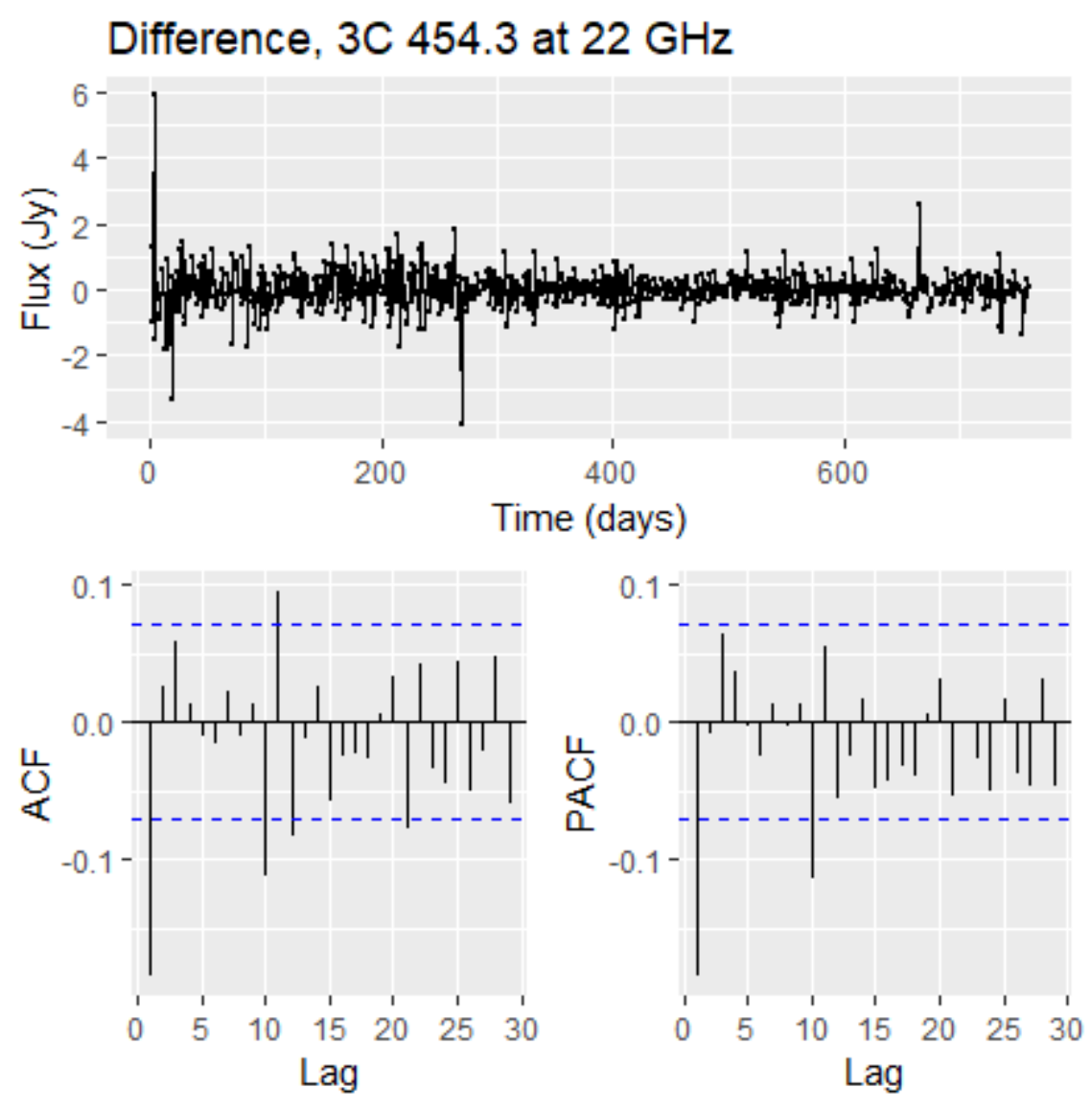}  
\includegraphics[scale=0.75]{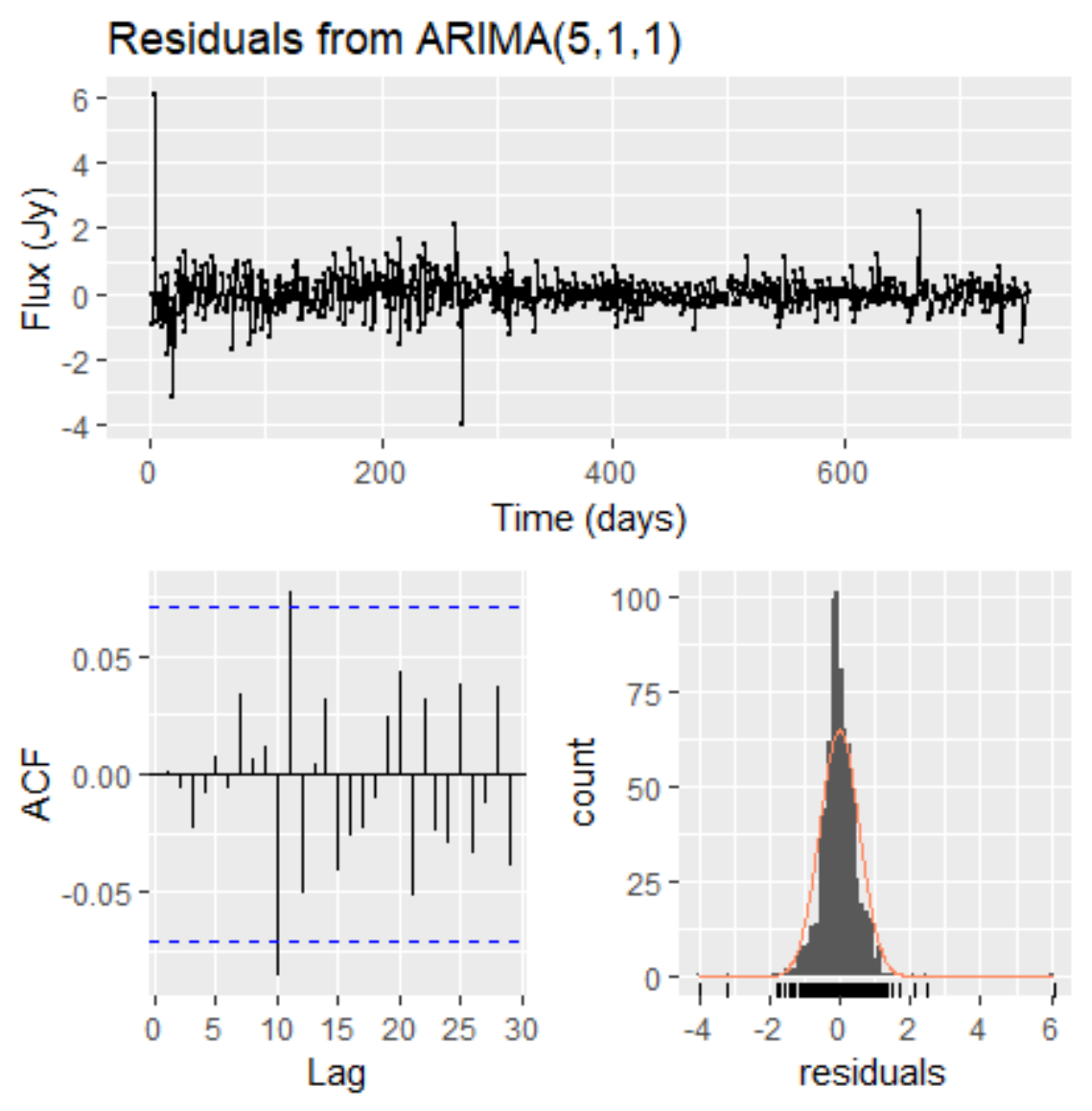}
\includegraphics[scale=0.75]{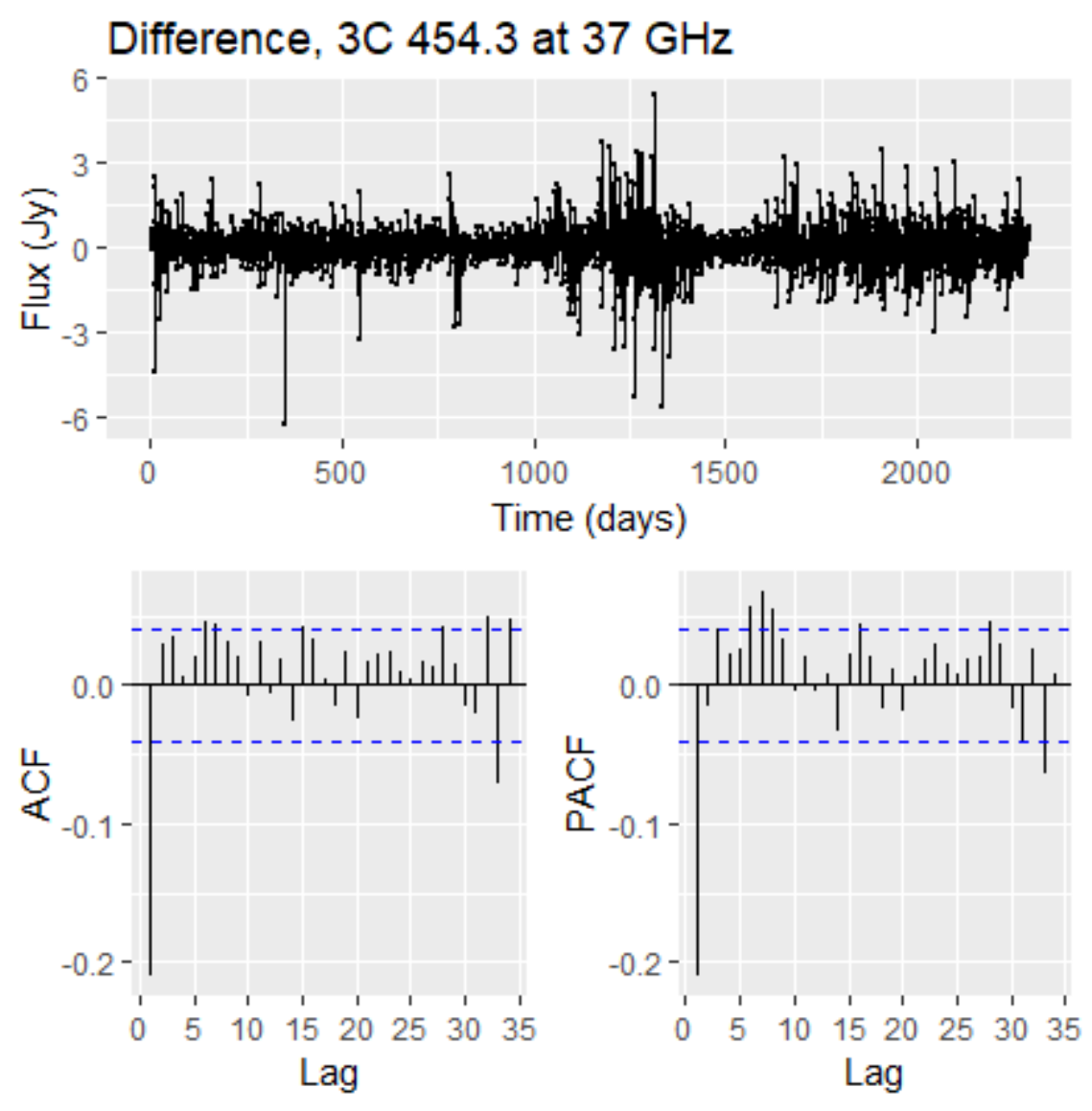}  
\includegraphics[scale=0.75]{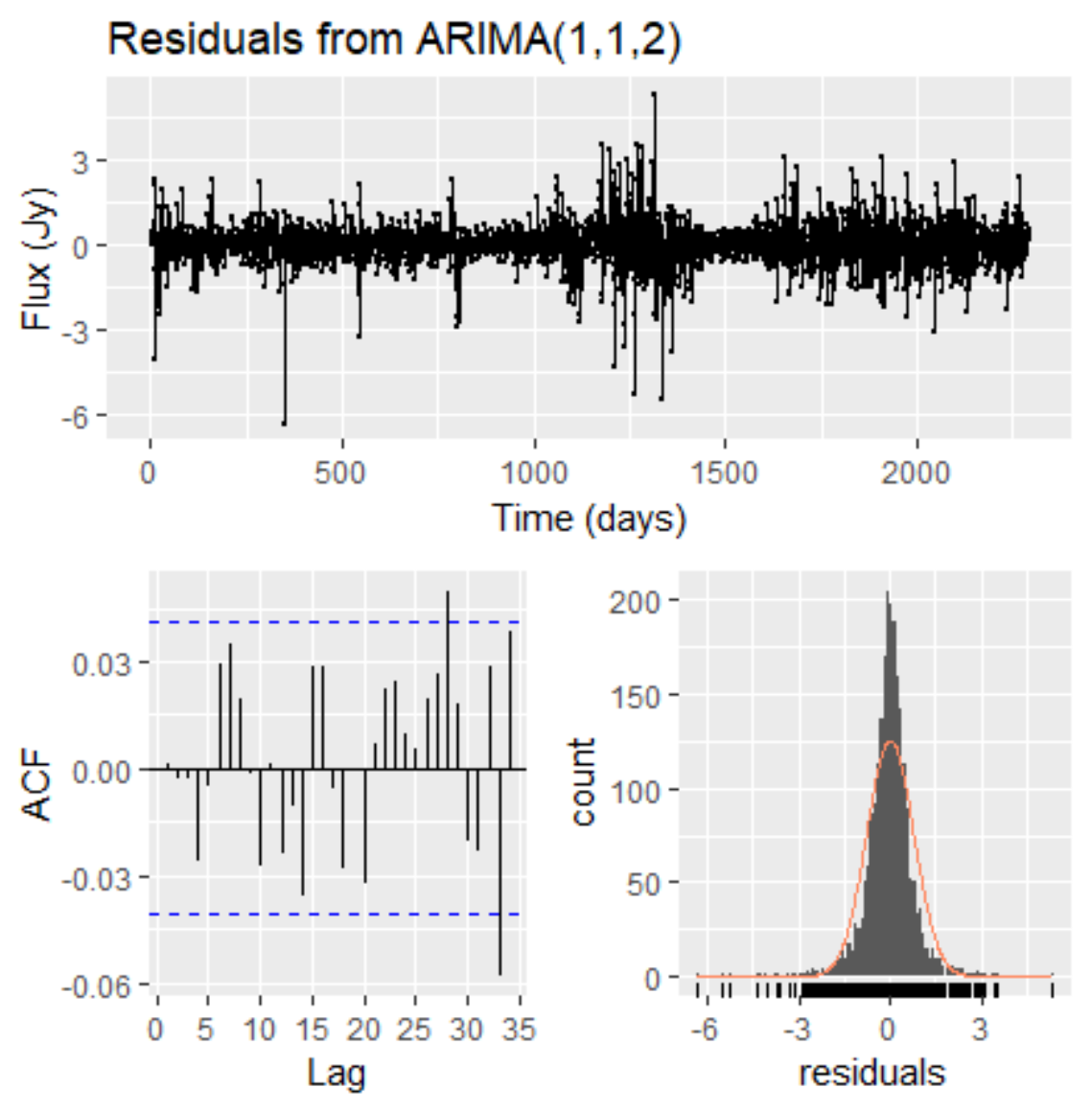}
\caption{Time plot of the differenced time series and its ACF and PACF plots (on the left of the first and second panels, respectively) and the corresponding residual plots (on the right of the first and second panels) for 3C 454.3 at 22 GHz . Similarly, time plot of the differenced time series and its ACF and PACF plots (one the left of the third and fourth panels, respectively) and the corresponding residual plots (on the right of the third and fourth panels) for 3C 454.3 at 37 GHz.}
\label{fig14}
\end{figure*}

In this section, the light curves shown in Fig. \ref{fig1} are analyzed using the ARIMA model. The scenario regarding the ARIMA model analysis of each light curve is discussed hereafter. The first step is verifying whether the time series are stationary or not. A visual inspection of the light curves show that they are not stationary, but one cannot be sure whether a time series is stationary or not only through visual inspection. There are different techniques to test for stationarity such as the augmented Dickey-Fuller unit root test. Though our light curves are shown to be nonstationary in the first subsection of section \ref{res}, we have checked the nonstationarity of each light curve based on their ACF plots, which are not included here. The ACF plot of a nonstationary time series decays slowly to zero, whereas stationary time series decay exponentially to zero. As a first step of our analysis, we have plotted ACF for all the time series considered here and found that all of them decay very slowly to zero, implying the presence of nonstationarity. Therefore, we proceed to our analysis directly by differencing, first differencing (d=1), the time series. The differenced time series and corresponding ACF and PACF plots are given in Figs. \ref{fig11}, \ref{fig12}, \ref{fig13}, and \ref{fig14} for 3C 273, 3C 279, 3C 345, and 3C 454.3, respectively.

As seen from the differenced time series plots, the differencing operator reduces the long-memory autocorrelations, although it left a few significant lags behind in most of the light curves except for 3C 273, where it does not reduce the autocorrelations that much. The next step is model selection and fitting. As a first attempt of best model selection, we estimate the parameters $p$ and $q$ based on the ACF and PACF of the differenced time series. Using estimated values of $p$ and $q$, we fit the time series to the ARIMA(p,d,q) model. The value of d is already known, d=1, since we differenced the time series only once. Additionally, we use the auto.arima() function in our best model selection. We select a model with the lowest log likelihood and the Akaike information criteria (AIC). If the models obtained based on the ACF and PACF plots and using auto.arima() do not fit our data, we try another model by changing the values of the parameters $p$ and $q$ and fit them to the ARIMA(p,d,q) model until we obtain the best one. The last step, in our case, is checking the diagnostic test to determine whether the residuals of the fitted models are white noise or not. We accept or reject the model selected based on two criteria: we accept the model if $p>0.05$ for the Ljung-Box test and the residual is white noise (randomly distributed), indicating no significant lags outside of the 95$\%$ limit, and we reject if these conditions are not satisfied. For the second condition, the residual should be white noise, and if it is not satisfied for all possible combinations of $p$ and $q$, we take the one that nearly provides a white-noise residual and $p>0.05$. Our interest here is not to forecast but only to determine whether ARIMA(p,d,q) models fit our data or not. Except for 3C 273 and 3C 279 at 37 GHz, where the fit to ARIMA models does not reduce the autocorrelations left behind more than 2 significant autocorrelations outside of the 95$\%$ limit, the ARIMA models fit sufficiently to reduce most of the autocorrelations for the rest of the time series. However, the models require a different number of coefficients, and as a result, the residuals of the ARIMA model fits are nearly white noise, leaving behind no more than two significant lags outside of the 95$\%$ limit, as shown from Figs. \ref{fig11}, \ref{fig12}, \ref{fig13}, and \ref{fig14} for 3C 273, 3C 279, 3C 345, and 3C 454.3, respectively. Except for 3C 345 at 22 GHz, the ARIMA(4,1,4) model fit reduces all significant autocorrelations leaving behind a white noise residual, indicating no single significant autocorrelation outside of the 95$\%$ limit as shown in  Fig. \ref{fig13}. Note that the ARIMA(p,d,q) models capture only short-memory components, and the observed failure could be due to the presence of long-memory components and/or nonlinear signatures in the time series that cannot be captured by the ARIMA models. The parametric autoregressive model, called the autoregressive fractionally integrated moving average (ARFIMA) model, where the differencing operator $d$ is a real number, is preferable to model time series with long-memory components. Of course, multifractality analyses as applied in this work are also capable of detecting both short and long memory trends.

\section{Summary and Conclusions}\label{concl}

In this study, we analyze the multiscaling signatures in the radio emissions of selected radio-loud quasars 3C 273, 3C 279, 3C 345, and 3C 454.3 both in the observed frame (at 22 and 37 GHz passbands) and the rest frame($f_{rest} = f_{obs}*(1+z)$) using a WTMM-based multifractality analysis approach. In addition, we fit the light curves of the sources to the ARIMA models. In our work, we first calculate the wavelet coefficients using a continuous wavelet transform and form a matrix of maxima lines (construct the skeleton function) by aggregating the absolute wavelet coefficients that only hold maxima lines. Second, using the collected local maxima lines and the constructed skeleton function, we determine the thermodynamics partition function. Third, by creating the log-log plots of the thermodynamics function $Zq(s)$ and the scale $s$, we estimate the slope using the least squares fitting method. The behavior of the estimated slopes are presented by the scaling exponent $\tau$($q$) versus $q$ plots in Fig. \ref{fig3} and  Fig. \ref{fig5} for each source in the observed frame at 22 and 37 GHz, respectively. Finally, we estimate the multifractal spectrum functions at each band for all the light curves and calculate the multifractality strength from the width $\Delta\alpha$ of the spectrum. Additionally, we analyze the corresponding light curves in the rest of the frame following the same procedures. Finally, we fit the time series in the observed frame to the ARIMA models. Our main conclusions are as follows. In this work, we have shown that (i) the scaling nature of quasars' radio emissions at 22 and 37 GHz is strongly multifractal and intermittent, (ii) the degree of nonlinearity or multifractality is similar for each source and strongly different between sources at both frequencies, (iii) the redshift correction does not affect the nonlinear or multifractal behavior of quasars' radio emission, and (iv) the ARIMA models fit most of the time series partially, except for 3C 345 at 22 GHz where the ARIMA(4,1,4) model fit obtains a white noise residual and for 3C 273 and 3C 279 at 37 GHz where the models are shown to be inadequate. The strong multifractal signature observed in quasars' radio emission further supports what has been previously posited: that quasars are intrinsically multifractal and complex systems that have nonlinear time series characterized by fractal behavior \citep{2018MNRAS.tmp.1264B, 1991ApJ...380..351V}. It is the physical mechanism that causes variation in the flux of radio emissions that possibly presents the detected multifractal signatures. It has been explained that the low-frequency emissions in general and radio emissions from radio-loud quasars/blazars in particular are due to synchrotron emissions from nonthermal electrons in a relativistic jet \citep{2011ApJ...743..104S, 2000A&A...361..850T, 1993MNRAS.262..249R,1988Natur.335..330C}. This means any physical process that causes a change in the dynamics of relativistic jets and presents variations in the flux of quasars' radio emissions is likely responsible for the existence of multifractal signatures in the radio observations of quasars. Therefore, our results play a significant role in providing valuable information for those working to develop models to better understand emissions in terms of synchrotron radiation from shocks propagating along relativistic jets and the dynamics of relativistic jets coming out from the center of radio-loud quasars, which in turn helps to constrain some physical properties of quasars in relation to the dynamics of their relativistic jets.

The observed similarity in the slope (degree of nonlinearity) or multifractality strength at 22 and 37 GHz for all the sources considered further supports the claim that the radiations of the sources at 22 and 37 GHz have the same emission region and mechanism, at least for 3C 273 and 3C 345 \citep{2010ChA&A..34..343W}. Despite that all our sources are flat-spectrum radio-loud quasars, the difference in the degree of multifractality (nonlinearity) between them at those radio bands provides very useful information about the physics of the sources of relativistic jets. This difference could be due to the different nature of turbulence in the accretion rate, internal shocks in the relativistic jets of the sources, fluctuations in the local magnetic fields and particle density, variation in the activity of the central engine, the difference in their black hole mass since it determines the mass accretion rate, which in turn affects the radio emissions, or due to the difference in any other variability mechanism not mentioned here. The other result we have found is that the degree of multifractality (nonlinearity) is the same both in the observed and rest frames of the sources, providing physically important information that multifractality is an intrinsic behavior of quasars' radio emissions. This finding further supports the conclusion that most of the long-term variations of quasars, in general, are intrinsic to the quasars themselves  \citep{2005AJ....129..615D}. In our recent work \citet{10.1093/mnras/stz203}, we have shown that extrinsic variations in relation to gravitational lensing, mainly microlensing effects, affect the multifractal behavior of quasars, and the microlensing effect increases the degree of multifractality. Therefore, the results obtained in this work, the sameness in the degree of multiractality in the observation and rest frames, possibly indicate the absence of extrinsic variations, mainly due to microlensing, in the light curves of the sources. A possible reason why the ARIMA(p,d,q) models do not fit most of the time series well could be due to the presence of memories and/or trends that cannot be easily captured by the models. This work provides valuable information mainly to model relativistic jet dynamics in particular and to understand the interior of quasars in general.

\section*{Acknowledgments}

This publication makes use of data obtained at the Mets\"ahovi Radio Observatory, operated by Aalto University in Finland.
Research activities of the Observational Astronomy Board of the Federal University of Rio Grande do Norte (UFRN) are supported by continuous grants from CNPq and FAPERN Brazilian agencies. We also acknowledge financial support from INCT INEspa\c{c}o/CNPq/MCT. 
AB acknowledges a CAPES PhD fellowship. ICL acknowledges a CNPq/PDE fellowship. We warmly thank the anonymous referee for the fruitful comments and suggestions that greatly improved this work.




\begin{thebibliography}{}

\bibitem[Agarwal (2016 )] {2016AGUFM.P11A1846A} Agarwal, S., Sordo, F. D., Wettlaufer, J. S.\ 2016, AGUFall Meeting Abstracts,
pp P11A–1846
\bibitem[Aliouane \& Ouadfeul(2013)] {2013EPSC....8...30A}  Aliouane, L., Ouadfeul, S.\ 2013, European Planetary Science Congress, 8,
EPSC2013
\bibitem[ Aller (1985 )] {1985ApJS...59..513A}  Aller, H. D., Aller, M. F., Latimer, G. E., Hodge, P. E.\ 1985, \apjs, 59, 513
\bibitem[Arneodo (1988 )] {1988PhRvL..61.2281A}  Arneodo, A., Grasseau, G., Holschneider, M.\ 1988, Physical Review Letters,
61, 2281
\bibitem[Arp(1987 )] {1987qrc..book.....A}  Arp, H.\ 1987, Berkeley, CA, Interstellar Media, 204p
\bibitem[Arp (1995 )] {1995PhLA..203..161A}  Arp, H. C.\ 1995, Physics Letters A, 203, 161
\bibitem[ Ashkenazy(2003 )] {2003GeoRL..30.2146A}  Ashkenazy, Y., Baker, D. R., Gildor, H., Havlin S.\ 2003, Geophys. Res. Lett.,
30, 2146

\bibitem[Bachev (2011 )] {2011A&A...528L..10B}  Bachev, R., et al.\ 2011, \aap, 528, L10
\bibitem[Barbieri (1990 )] {1990ApJ...359...63B}  Barbieri, C., Vio R., Cappellaro, E., Turatto, M.\ 1990, \apj, 359, 63
\bibitem[Barvainis (2005 )] {2005ApJ...618..108B}  Barvainis, R., Leh{\'a}r J., Birkinshaw, M., Falcke, H., Blundell, K. ~M.\ 2005,
\apj, 618, 108
\bibitem[Bewketu Belete et al (2019)]{10.1093/mnras/stz203} Bewketu Belete, A., Canto Martins, B L,  Leão, I C and De Medeiros, J R., et al.\ 2019, Monthly Notices of the Royal Astronomical Society, 484, 3552
\bibitem[Bewketu Belete et al (2018 )] {2018MNRAS.tmp.1264B}  Bewketu Belete, A., Bravo, J. P., Canto Martins, B. L.,  Le\~ao, I. C., De Araujo, J. M.,
De Medeiros, J. R., et al.\ 2018, \mnras, 478, 3976
\bibitem[Berton (2018)] {2018A&A...614A.148B}  Berton, M., et al.\ 2018, \aap, 614, A148
\bibitem[ B{\"o}ttcher (2007)] {2007ApJ...670..968B}  B{\"o}ttcher, M., et al.\ 2007, \apj, 670, 968
\bibitem[Brocksopp (2007)] {2007MNRAS.382.1019B}  Brocksopp, C., Kaiser C. R., Schoenmakers, A. P., de Bruyn, A. G.\ 2007,
\apj, 382, 1019
\bibitem[ Burbidge(1965 )] {1965ApJ...142.1674B}  Burbidge, E. M.\ 1965, \apj, 142, 1674

\bibitem[Chatterjee (2008 )] {2008ApJ...689...79C}  Chatterjee, R., et al.\ 2008, \apj, 689, 79
\bibitem[Cotton (1979)] {1979ApJ...229L.115C}  Cotton, W. D., et al.\ 1979, \apj, 229, L115
\bibitem[Courvoisier (2005 )] {2005A&A...444..417C}  Courvoisier,  T.~J.-L., T{\"u}rler, M.\ 2005, \aap, 444, 417
\bibitem[Courvoisier (1987 )] {1987A&A...176..197C}  Courvoisier, T.~J.-L., et al.\ 1987, \aap, 176, 197
\bibitem[Courvoisier (1988 )] {1988Natur.335..330C}  Courvoisier,  T.~J.-L., Robson, E. I., Blecha, A., Bouchet, P., Hughes, D. H.,
Krisciunas K., Schwarz H. E.\ 1988, Nature, 335, 330
\bibitem[Courvoisier et al. ( 1990)] {1990A&A...234...73C}  Courvoisier, T.~J.-L., et al.\ 1990, \aap, 234, 73

\bibitem[Dar (1991 )] {1991ApJ...382L...1D}  Dar, A.\ 1991, \apj, 382, L1
\bibitem[de Freitas  (2016 )] {2016ApJ...831...87D}  de Freitas, D. B., et al.\ 2016, \apj, 831, 87
\bibitem[Degaudenzi \& Arizmendi (1998 )] {1998adap.org..8004D}  Degaudenzi, M. E., Arizmendi, C. M.\ 1998, Advances in Astrophysics,
\bibitem[de Vries (2005)]{2005AJ....129..615D}{de Vries}, W.~H. and {Becker}, R.~H. and {White}, R.~L. and 
	{Loomis}, C.\ 2005, The Astronomical Journal, 129, 615-629

\bibitem[Dondi \& Ghisellini (1995 )] {1995MNRAS.273..583D}  Dondi L., Ghisellini, G.\ 1995, \mnras, 273, 583
\bibitem[Dong (2010 )] {2010ChA&A..34..357D}  Dong, F.-t., Zhang, H.-j., Mao, L.-s., Zhang, X., Zheng, Y.-g., Tang, L.\ 2010,
Chinese Astron. Astrophys., 34, 357
\bibitem[Dunlop \& Peacock (1990 )] {1990MNRAS.247...19D}  Dunlop, J. S., Peacock, J. A.\ 1990, \mnras, 247, 19

\bibitem[Enescu (2006 )] {2006GeoJI.164...63E}  Enescu, B., Ito, K., Struzik, Z. R.\ 2006, Geophysical Journal International,
164, 63

\bibitem[Fanti (1977 )] {1977A&A....61..487F}  Fanti, C., Fanti, R., Lari, C., Padrielli, L., van der Laan, H., de Ruiter, H.\ 1977,
\aap, 61, 487
\bibitem[Fernandes et al. (2017 )] {2017AAS...22925029F}  Fernandes, S., et al.\ 2017, in American Astronomical Society Meeting Abstracts
229,  250 - 29
\bibitem[Filippenko (2004 )] {2004mmu..symp..270F}  Filippenko, A. V.\ 2004, Measuring and Modeling the Universe, p. 270
\bibitem[Feigelson (2018)]{10.3389/fphy.2018.00080} Feigelson, Eric D. and Babu, G. Jogesh and Caceres, Gabriel A.\ 2018, Frontiers in Physics, 6, 80

\bibitem[Grandi ( 2004)] {2004Sci...306..998G}  Grandi, P., Palumbo, G. G. C.\ 2004, Science, 306, 998
\bibitem[Greenstein \&  Schmidt(1964 )] {1964ApJ...140....1G}  Greenstein, J. L., Schmidt, M.\ 1964, \apj, 140, 1
\bibitem[Gupta et al. (2017 )] {2017MNRAS.472..788G}  Gupta, A. C., et al.\ 2017, \mnras, 472, 788

\bibitem[Halsey (1986 )] {1986PhRvA..33.1141H}  Halsey, T. C., Jensen, M. H., Kadanoff, L. P., Procaccia, I., Shraiman, B. I.\
1986, Phys. Rev. A, 33, 1141
\bibitem[Hartman et al. (2001 )] {2001ApJ...558..583H}  Hartman, R. C., et al.\ 2001, \apj, 558, 583
\bibitem[ Hota \& Saikia  (2006 )] {2006MNRAS.371..945H}  Hota, A., Saikia, D. J.\ 2006, \mnras, 371, 945
\bibitem[Hovatta (2007 )] {2007A&A...469..899H}  Hovatta, T., Tornikoski, M., Lainela, M., Lehto, H. J., Valtaoja, E., Torniainen
I., Aller M. F., Aller H. D.\ 2007, \aap, 469, 899
\bibitem[Hovatta (2008 )] {2008A&A...488..897H}  Hovatta, T., Lehto, H. J., Tornikoski, M.\ 2008, \aap, 488, 897

\bibitem[ Hyndman and Athanasopoulos (2014)] {2014Hyndman}  Hyndman, R., Athanasopoulos, G.\ 2014, Forecasting: Principles and Practice, 2ND EDITION, Monash University, Australia

\bibitem[ Ihlen (2012 )] {2012Ihlen}  Ihlen, E. A. F.\ 2012, Frontiers in Physiology, 3, 141

\bibitem[Jagtap (2012 )] {2012SPIE.8222E..0FJ}  Jagtap, J., Ghosh, S., Panigrahi, P. K., Pradhan, A.\ 2012, in Dynamics and Fluctuations
in Biomedical Photonics IX. p. 82220F, doi:10.1117/12.907330
\bibitem[Jorstad et al. (2010 )] {2010ApJ...715..362J}  Jorstad, S. G., et al.\ 2010, \apj, 715, 362
\bibitem[Jorstad et al. (2013 )] {2013ApJ...773..147J}  Jorstad, S. G., et al.\ 2013, \apj, 773, 147

\bibitem[ Kasde (2016 )] {2016cosp...41E.946K}  Kasde, S. K., Gwal, A. K., Sondhiya, D. K.\ 2016, in 41st COSPAR Scientific
Assembly.
\bibitem[Kataoka (2002 )] {2002MNRAS.336..932K}  Kataoka, J., Tanihata, C., Kawai, N., Takahara, F., Takahashi, T., Edwards, P. G.,
  Makino F.\ 2002, \mnras, 336, 932
\bibitem[Katgert ( 1973)] {1973A&A....23..171K}  Katgert, P., Katgert-Merkelijn, J. K., Le Poole, R. S., van der Laan, H.\ 1973,
\aap, 23, 171
\bibitem[Kellermann (1966 )] {1966Natur.212..781K}  Kellermann, K. I., Pauliny-Toth, I. I. K.\ 1966, Nature, 212, 781
\bibitem[Kellermann (1989 )] {1989AJ.....98.1195K}  Kellermann, K. I., Sramek, R., Schmidt, M., Shaffer, D. B., Green, R.\ 1989,
\aj, 98, 1195
\bibitem[Kellermann (1994 )] {1994AJ....108.1163K}  Kellermann, K. I., Sramek, R. A., Schmidt, M., Green, R. F., Shaffer, D. B.\
1994, \aj, 108, 1163
\bibitem[Kharb (2016)] {2016MNRAS.459.1310K} Kharb, P., Srivastava, S., Singh, V., Gallimore, J. F., Ishwara-Chandra, C. H.,
Ananda, H.\ 2016, \mnras, 459, 1310
\bibitem[ Kidger (1989 )] {1989A&A...226....9K}  Kidger, M. R.\ 1989, \aap, 226, 9
\bibitem[Kollgaard (1989 )] {1989AJ.....97.1550K}  Kollgaard, R. I., Wardle, J. F. C., Roberts, D. H.\ 1989, \aj, 97, 1550

\bibitem[Lawson (1999)] {1999MNRAS.306..247L}  Lawson, A. J., McHardy, I. M., Marscher, A. P.\ 1999, \mnras, 306, 247
\bibitem[Le{\'o}n-Tavares (2011)] {2011A&A...532A.146L}  Le{\'o}n-Tavares, J., Valtaoja, E., Tornikoski, M., L{\"a}hteenm{\"a}ki, A., Nieppola, E.\
2011, \aap, 532, A146
\bibitem[Leonardis (2013 )] {2013PhRvL.110t5002L}  Leonardis, E., Chapman, S. C., Daughton, W., Roytershteyn, V., Karimabadi,
H.\ 2013, Physical Review Letters, 110, 205002
\bibitem[Lin \&  Sharif (2007 )] {2007EPJB...60..483L}  Lin, D. C., Sharif, A.\ 2007, European Physical Journal B, 60, 483
\bibitem[Lindfors (2005 )] {2005A&A...440..845L}  Lindfors, E. J., Valtaoja, E.,  T{\"u}rler M.\ 2005, \aap, 440, 845
\bibitem[Lindfors (2006 )] {2006A&A...456..895L}  Lindfors, E. J., et al.\ 2006, \aap, 456, 895
\bibitem[Lynden-Bell (1969)] {1969Natur.223..690L}  Lynden-Bell, D.\ 1969, Nature, 223, 690

\bibitem[Mallat \& Zhong (1992)]{mallat1992characterization} Mallat, Stephane and Zhong, Sifen.\ 1992, IEEE Transactions on Pattern Analysis \& Machine Intelligence, 7, 710-732
\bibitem[Mallat \& Hwang (1992)]{mallat1992singularity} Mallat, Stephane and Hwang, Wen Liang.\ 1992, IEEE transactions on information theory, 38, 617--643
\bibitem[Mandelbrot \& Whitrow(1983 )] {1983JBAA...93..238M}  Mandelbrot, B. B., Whitrow, G. J.\ 1983, Journal of the British Astronomical Association, 93, 238
\bibitem[ Maraschi (1992)] {1992ApJ...397L...5M}  Maraschi, L., Ghisellini, G., Celotti, A.\ 1992, \apj, 397, L5
\bibitem[ Maraschi et al. (1994 )] {1994ApJ...435L..91M}  Maraschi, L., et al.\ 1994, \apj, 435, L91
\bibitem[March{\~a} (2001 )] {2001MNRAS.326.1455M}  March{\~a}, M. J., Caccianiga, A., Browne, I. W. A., Jackson, N.\ 2001, \mnras,
326, 1455
\bibitem[Maruyama (2016 )] {2016JAsGe...5..301M}  Maruyama, F.\ 2016, NRIAG Journal of Astronomy and Geophysics, 5, 301
\bibitem[Maruyama (2017 )] {2017AdSpR..60.1363M}  Maruyama, F., Kai K., Morimoto, H.\ 2017, Advances in Space Research, 60,
1363
\bibitem[Miller (1990 )] {1990MNRAS.244..207M}  Miller, L., Peacock, J. A., Mead, A. R. G.\ 1990, \mnras, 244, 207
\bibitem[Moore \& Stockman (1984 )] {1984ApJ...279..465M}  Moore, R. L., Stockman, H. S.\ 1984, \apj, 279, 465
\bibitem[Muzy \& Arneodo (1991 )] {1991Muzy}  Muzy, J.F. B. E., Arneodo, A.\ 1991, Phys. Rev. Lett, 67, 3515
\bibitem[Muzy (1994)]{muzy1994multifractal} Muzy, Jean-Fran{\c{c}}ois and Bacry, Emmanuel and Arneodo, Alain.\ 1994, International Journal of Bifurcation and Chaos, 4, 245-302

\bibitem[Netzer et al. (1996 )] {1996MNRAS.279..429N}  Netzer H., et al.\ 1996, \mnras, 279, 429
\bibitem[Nurujjaman (2009 )] {2009PhPl...16j2307N}  Nurujjaman M., Narayanan R., Iyengar A. N. S.\ 2009, Physics of Plasmas,
16, 102307

\bibitem[ Ouahabi \& Femmam(2011 )] {Ouahabi:2011:WMA:2036797.2036809}  Ouahabi, A., Femmam, S.\ 2011, Analog Integr. Circuits Signal Process, 69, 3

\bibitem[Padovani et al. (2017 )] {2017A&ARv..25....2P}  Padovani, P., et al.\ 2017, \aapr, 25, 2
\bibitem[Page (2004 )] {2004MNRAS.349...57P}  Page, K. L., Turner, M. J. L., Done, C., O'Brien, P. T., Reeves, J. N., Sembay
S., Stuhlinger M.\ 2004, \mnras, 349, 57
\bibitem[Pan \& Coles (2000)] {2000MNRAS.318L..51P}  Pan, J., Coles, P.\ 2000, \mnras, 318, L51
\bibitem[Pati{\~n}o-{\'A}lvarez et al. (2017 )] {2017FrASS...4...47P}  Pati{\~n}o-{\'A}lvarez, V. M., et al.\ 2017, Frontiers in Astronomy and Space Sciences,
4, 47
\bibitem[ Petropoulou \&  Dimitrakoudis (2015 )] {2015MNRAS.452.1303P}  Petropoulou, M., Dimitrakoudis, S.\ 2015, \mnras, 452, 1303
\bibitem[Andrejs Puckovs \& Andrejs Matvejevs (2012 )] {WaveletTransformModulusMaximaApproachforWorldStockIndexMultifractalAnalysis}  Puckovs, A., Matvejevs, A.\ 2012, Information Technology and Management Science, 15, 76

\bibitem[Quirrenbach (1993)] {1993JAVSO..22...55Q} Quirrenbach, A.\ 1993, Journal of the American Association of Variable Star Observers (JAAVSO), 22, 55-63

\bibitem[Raiteri et al. (2008 )] {2008A&A...491..755R}  Raiteri, C. M., et al.\ 2008, \aap, 491, 755
\bibitem[Reynolds \& Begelman (1997 )] {1997ApJ...487L.135R}  Reynolds, C. S., Begelman, M. C.\ 1997, \apj, 487, L135
\bibitem[Robson (1993 )] {1993MNRAS.262..249R}  Robson, E. I., et al.\ 1993, \mnras, 262, 249

\bibitem[Saikia \& Jamrozy (2009 )] {2009BASI...37...63S}  Saikia, D. J., Jamrozy, M.\ 2009, Bulletin of the Astronomical Society of
India, 37, 63
\bibitem[Schinzel (2010 )] {2010arXiv1012.2820S}  Schinzel, F. K., Lobanov, A. P., Jorstad, S. G., Marscher, A. P., Taylor G. B.,
 Zensus J. A.\ 2010, preprint, (arXiv:1012.2820)
\bibitem[Shapiro (1966 )] {1966ApJ...143..598S}  Shapiro, I. I., Weinreb, S.\ 1966, \apj, 143, 598
\bibitem[Shimizu (2002 )] {2002SHIMIZU}  Shimizu, Y., Thurner, S., Ehrenberger, K.\ 2002, Fractals, 10, 103
\bibitem[Sikora (1994 )] {1994ApJ...421..153S}  Sikora, M., Begelman, M. C., Rees M. J.\ 1994, \apj, 421, 153
\bibitem[ Singal (2011 )] {2011ApJ...743..104S}  Singal, J., Petrosian, V., Lawrence, A., Stawarz, {\L}.\ 2011, \apj, 743, 104
\bibitem[Singh (2016 )] {2016ApJ...826..132S}  Singh, V., Ishwara-Chandra, C. H., Kharb, P., Srivastava, S., Janardhan, P.\
2016, \apj, 826, 132
\bibitem[Soldi et al. (2008 )] {2008A&A...486..411S}  Soldi, S., et al.\ 2008, \aap, 486, 411
\bibitem[Sramek (1980 )] {1980ApJ...238..435S}  Sramek, R. A., Weedman, D. W.\ 1980, \apj, 238, 435
\bibitem[Stevens (1994 )] {1994ApJ...437...91S}  Stevens, J. A., Litchfield, S. J., Robson, E. I., Hughes, D. H., Gear W. K.,
Terasranta H., Valtaoja E., Tornikoski M.\ 1994, \apj, 437, 91
\bibitem[Stocke (1992 )] {1992ApJ...396..487S}  Stocke, J. T., Morris, S. L., Weymann, R. J., Foltz C. B.\ 1992, \apj, 396, 487

\bibitem[ Tamhane (2015 )] {2015MNRAS.453.2438T}  Tamhane, P., Wadadekar, Y., Basu, A., Singh, V., Ishwara-Chandra, C. H.,
Beelen, A., Sirothia, S.\ 2015, \mnras, 453, 2438
\bibitem[Telesca (2004 )] {2004PCE....29..295T}  Telesca, L., Balasco, M., Colangelo, G., Lapenna, V., Macchiato, M.\ 2004,
Physics and Chemistry of the Earth, 29, 295
\bibitem[Ter{\"a}sranta (2005 )] {2005A&A...440..409T}  Ter{\"a}sranta, H., Wiren, S., Koivisto, P., Saarinen, V., Hovatta, T.\ 2005, \aap,
440, 409
\bibitem[Trevino \& Dal Negro (2012 )] {2012Trevino}  Trevino, J. L. S. N. H. C. H., Dal Negro, L.\ 2012, Optics Express, 20, 3015
\bibitem[ T{\"u}rler (1999 )] {1999A&A...349...45T}  T{\"u}rler, M., Courvoisier, T. J.-L., Paltani, S.\ 1999, \aap, 349, 45
\bibitem[ T{\"u}rler (2000 )] {2000A&A...361..850T}  T{\"u}rler, M., Courvoisier, T. J.-L., Paltani, S.\ 2000, \aap, 361, 850
\bibitem[ T{\"u}rler et al. (2006 )] {2006A&A...451L...1T}  T{\"u}rler, M., et al.\ 2006, \aap, 451, L1

\bibitem[Ulrich (1997 )] {1997ARA&A..35..445U}  Ulrich, M.-H., Maraschi, L., Urry C. M.\ 1997, \araa, 35, 445

\bibitem[Valtaoja (1991 )] {1991AJ....102.1946V}  Valtaoja, L., et al.\ 1991, \aj, 102, 1946
\bibitem[Valtaoja (1992a )] {1992A&A...254...71V}  Valtaoja, E., Terasranta, H., Urpo, S., Nesterov, N. S., Lainela, M., Valtonen,
M.\ 1992a, \aap, 254, 71
\bibitem[Valtaoja (1992b )] {1992A&A...254...80V}  Valtaoja, E., Terasranta, H., Urpo, S., Nesterov, N. S., Lainela, M., Valtonen,
M.\ 1992b, \aap, 254, 80
\bibitem[Valtaoja (1999 )] {1999ApJS..120...95V}  Valtaoja, E., L{\"a}hteenm{\"a}ki, A., Ter{\"a}sranta, H., Lainela, M.\ 1999, \apjs, 120,
95
\bibitem[Villata et al. (2006 )] {2006A&A...453..817V}  Villata, M., et al.\ 2006, \aap, 453, 817
\bibitem[Villata et al. (2009 )] {2009A&A...504L...9V}  Villata, M., et al.\ 2009, \aap, 504, L9
\bibitem[ Vio (1991 )] {1991ApJ...380..351V}  Vio, R., Cristiani, S., Lessi, O., Salvadori, L.\ 1991, \apj, 380, 351
\bibitem[ Vio (1992 )] {1992ApJ...391..518V}  Vio, R., Cristiani S., Lessi, O., Provenzale, A.\ 1992, \apj, 391, 518
\bibitem[Visnovsky (1992 )] {1992ApJ...391..560V}  Visnovsky, K. L., Impey, C. D., Foltz, C. B., Hewett, P. C., Weymann, R. J.,
 Morris, S. L.\ 1992, \apj, 391, 560

\bibitem[ Wang \& Yang (2010 )] {2010ChA&A..34..343W}  Wang, H.-t., Yang, J.-y.\ 2010, Chinese Astron. Astrophys., 34, 343
\bibitem[Wagner (1995)] {doi:10.1146/annurev.aa.33.090195.001115} Wagner, S. J. and Witzel, A.\ 1995, Annual Review of Astronomy and Astrophysics, 33, 163-197
\bibitem[Wehrle (1999 )] {1999APh....11..169W}  Wehrle, A. E.\ 1999, Astroparticle Physics, 11, 169
\bibitem[Wehrle (1998 )] {1998ApJ...497..178W}  Wehrle, A. E., et al.\ 1998, \apj, 497, 178
\bibitem[Willott (1998 )] {1998MNRAS.300..625W}  Willott, C. J., Rawlings, S., Blundell, K. M., Lacy, M.\ 1998, \mnras, 300,
625
\bibitem[Willott (2001 )] {2001MNRAS.322..536W}  Willott, C. J., Rawlings, S., Blundell, K. M., Lacy, M., Eales, S. A.\ 2001,
\mnras, 322, 536

\bibitem[Yordanova (2004 )] {2004AnGeo..22.2431Y}  Yordanova, E., Grzesiak, M.,Wernik, A., Popielawska, B., Stasiewicz, K.\ 2004,
Annales Geophysicae, 22, 2431
\bibitem[Yuan (2012 )] {2012ApJ...744...84Y}  Yuan, Z., Wang, J.\ 2012, \apj, 744, 84

\bibitem[Zhang (2018 )] {2018RAA....18...40Z}  Zhang, H.-M., Zhang, J., Lu R.-J., Yi T.-F., Huang, X.-L., Liang, E.-W.\ 2018,
Research in Astronomy and Astrophysics, 18, 040
\bibitem[Zheng \& Yang(2016 )] {2016MNRAS.457.3535Z}  Zheng, Y. G., Yang, C. Y.\ 2016, \mnras, 457, 3535

\end{thebibliography}
\end{document}